\documentclass[a4paper,11pt]{article}
\pdfoutput=1 
\usepackage{jinstpub} 
\let\ifpdf\relax
\usepackage{float}
\usepackage{color}
\usepackage[normalem]{ulem}
\usepackage{lineno}
\usepackage{graphicx}
\usepackage{subcaption}
\usepackage{mwe}
\usepackage{amsmath}
\usepackage{amssymb}
\usepackage{siunitx}
\usepackage{newtxmath,newtxtext}

\newcommand{\longitudinal}{{\mkern3mu\vphantom{\perp}\vrule depth 0pt\mkern2mu\vrule depth 0pt\mkern3mu}}
\newcommand{\transverse}{\perp}
\newcommand{\sigmaT}{\sigma_{\transverse}}
\newcommand{\sigmaTi}[1]{\sigma_{#1\transverse}}
\newcommand{\sigmaL}{\sigma_{\longitudinal}}
\newcommand{\sigmaLi}[1]{\sigma_{#1\longitudinal}}

\newcommand{\apriori}{\textit{a priori}\xspace}

\newlength{\figwidth}
\newlength{\fighalfwidth}
\collaboration{MicroBooNE Collaboration}

\title{\boldmath \center \LARGE Ionization Electron Signal Processing \\
  in Single Phase LArTPCs \\
  I. Algorithm Description and \\ Quantitative Evaluation with MicroBooNE Simulation}

\author[i]{C.~Adams}
\author[j]{R.~An}
\author[c]{J.~Anthony}
\author[aa]{J.~Asaadi}
\author[a]{M.~Auger}
\author[h]{L.~Bagby}
\author[ee]{S.~Balasubramanian}
\author[h]{B.~Baller}
\author[p]{C.~Barnes}
\author[s]{G.~Barr}
% ...

\affiliation[a]{Universit{\"a}t Bern, Bern CH-3012, Switzerland}
% ...
\affiliation[c]{University of Cambridge, Cambridge CB3 0HE, United Kingdom}
% ...
\affiliation[h]{Fermi National Accelerator Laboratory (FNAL), Batavia, IL 60510, USA}
\affiliation[i]{Harvard University, Cambridge, MA 02138, USA}
\affiliation[j]{Illinois Institute of Technology (IIT), Chicago, IL 60616, USA}
% ...
\affiliation[p]{University of Michigan, Ann Arbor, MI, 48109, USA}
% ...
\affiliation[s]{University of Oxford, Oxford OX1 3RH, United Kingdom}
% ...
\affiliation[aa]{University of Texas, Arlington, TX, 76019, USA}
% ...
\affiliation[ee]{Yale University, New Haven, CT, 06520, USA}

\emailAdd{microboone\_info@fnal.gov}

\abstract{
  We describe the concept and procedure of drifted-charge extraction developed
  in the MicroBooNE experiment, a single-phase liquid argon time projection
  chamber (LArTPC). This technique converts the raw digitized TPC waveform to
  the number of ionization electrons passing through a wire plane at a given time.
  A robust recovery of the number of ionization electrons from both induction and
  collection anode wire planes will augment the 3D reconstruction, and is 
  particularly important for tomographic reconstruction algorithms.
  A number of building blocks of the overall procedure are
  described. The performance of the signal processing is quantitatively
  evaluated by comparing extracted charge with the true charge through
  a detailed TPC detector simulation taking into account position-dependent
  induced current inside a single wire region and across multiple wires.
  Some areas for further improvement of the performance of the charge
  extraction procedure are also discussed. }
 

\begin{document}
\maketitle
\flushbottom


\section{Introduction}\label{sec:introduction}


The liquid argon time projection chamber (LArTPC)~\cite{rubbia77,Chen:1976pp,willis74,Nygren:1976fe} 
is an innovative detector technology being actively developed worldwide.
Several features of the LArTPC make it well adapted to the study
of neutrinos and other rare processes. Argon is readily available
commercially ($\sim$\SI{1}{\percent} by volume, the most abundant noble
gas in the atmosphere). Free electrons have high mobility, low
diffusion~\cite{Li:2015rqa}, and very high survival time in pure liquid
argon (LAr), making it an attractive material for a TPC.
In addition, LAr has a relatively high
density and a high scintillation light yield. In the near 
term, the Short-Baseline Neutrino Program~\cite{Antonello:2015lea} will utilize three LArTPCs
(MicroBooNE, SBND, and ICARUS) at Fermi National Accelerator Laboratory (Fermilab) to search for eV-scale sterile neutrino(s) and measure neutrino-argon
interaction cross sections. In the long term, the long-baseline Deep Underground Neutrino Experiment
(DUNE)~\cite{Acciarri:2015uup} is planning to use four large 10 kilotons LArTPC modules as far detectors to search for leptonic CP violation,
determine the neutrino mass hierarchy, test the standard three-neutrino paradigm, search for
proton decays, and potentially observe supernova neutrino bursts. 
The development of high-quality and fully-automated event reconstruction algorithms for LArTPC neutrino detectors is crucial to the success of the 
short-baseline and long-baseline physics programs and is an area of significant activity~\cite{uboone-pandora, uboone-dl, uboone-MCS, wirecell}. 
The robust recovery of the ionization signals from the LArTPC images is the critical first stage of LArTPC reconstruction.

The MicroBooNE detector~\cite{Acciarri:2016smi} is the first LArTPC in the Short-Baseline Neutrino Program~\cite{Antonello:2015lea} to be operational.
It is a single-phase LArTPC built to observe interactions of neutrinos from the on-axis
Booster~\cite{bnb} and off-axis NuMI~\cite{numi} beams at Fermi National Accelerator Laboratory in Batavia, IL.
The TPC LAr volume is 2.56 m $\times$ 2.3 m $\times$ 10.4 m with about 90 metric tons active mass, housed in a foam-insulated evacuable cryostat vessel. At the anode end of the 2.56 m drift distance,
there are three parallel wire readout planes~\cite{Acciarri:2017wp}. The first wire plane facing the cathode is labeled "U", and the second and third plane are labeled "V" and "Y", respectively. The wire pitch and the gap between two adjacent
wire planes are both 3 mm. The 3456 wires in the Y plane are oriented vertically. The U and V planes each
contain 2400 wires oriented $\pm$60$^\circ$ with respect to vertical. Behind the wire planes and external
to the TPC, there is an array of 32 photomultiplier tubes~\cite{Conrad:2015xta} to detect scintillation light
for triggering, timing, and other purposes. 

\begin{figure}[!h!tbp]
  \includegraphics[width=1.12\textwidth]{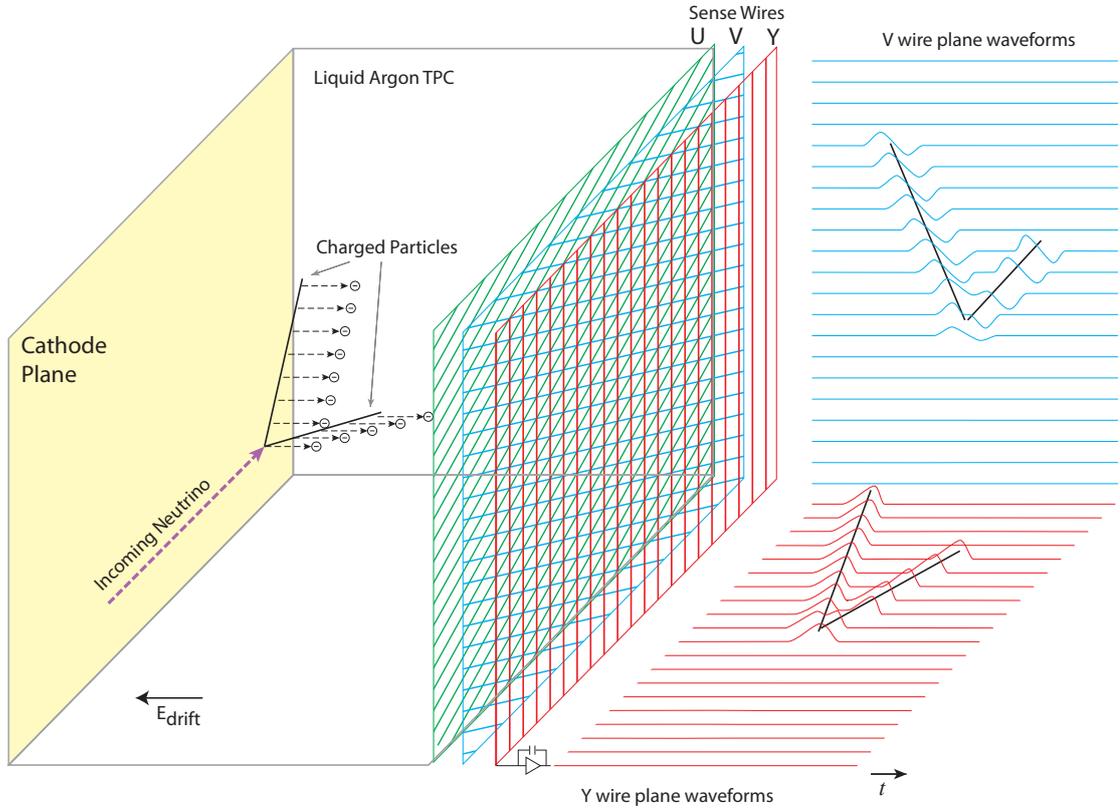}
\caption[TPC Basics]{Diagram illustrating the signal formation in a
  LArTPC with three wire planes~\cite{Acciarri:2016smi}. The signal on each
  plane produces a 2D image of the event. For simplicity,
  the signal in the U induction plane is omitted from the diagram. }
\label{fig:tpccartoon}
\end{figure}


The TPC signal formation in MicroBooNE is illustrated in figure~\ref{fig:tpccartoon}.
At a drift field of \SI{273}{\V/\cm} corresponding to a cathode high
voltage of \SI{-70}{\kV},
the ionization electrons drift through the
LAr detector volume along the electric field lines at a nominal speed of
about \SI{1.10}{\mm/\us}
toward the anode wire planes. Different bias voltages, \SI{-110}{\V}, \SI{0}{\V}, and \SI{+230}{\V} are
applied to the U, V, and Y wire planes, respectively, to ensure all ionization electrons
pass through the U and V planes before being collected by the Y
plane~\cite{Bunemann_Cranshaw_Harvey_1949} at the nominal MicroBooNE operating voltage of \SI{-70}{\kV}. As ionization electrons drift towards and then past
the wires of the U and V planes, currents with bipolar shape are induced on the U and V planes.
In contrast, a unipolar-shaped current is induced on a wire of the Y plane as
all nearby ionization charge
is collected. The U and V wire planes are commonly referred to as the
induction planes. Although also measuring induced current,
the Y wire plane is commonly referred to as the collection plane.

While the collection plane signal is mostly unipolar and large in amplitude with a Gaussian time profile, the induction plane signal is bipolar and small in amplitude with a complex time profile. The latter is due to the overlapping of many bipolar signal shapes as a distribution of drifting charge passes near the wires. Despite complications in the induction plane signal, the combination of the induction and collection wire planes is essential for tomographic event reconstruction~\cite{wirecell} in single-phase LArTPCs. Leveraging the induction signal in combination with the collection signal is important to fully exploit single-phase LArTPC capabilities.

The implementation of the induction and collection wire readout planes is a unique feature of the current single-phase LArTPC detectors. An alternative readout scheme, 2D pixel readout, would not suffer from ailments introduced from the complexity of the induction plane signal. However, this alternative scheme for large detectors is not actionable at present,
though good progress has been made~\cite{pixel-readout}. If the wire readout in MicroBooNE were replaced by a full 2D pixel readout, the total number of channels would be 2.7 million instead 
of 8,256, resulting in a significant increase in the cost of electronics. Furthermore, the power consumption of these electronics inside LAr would be a serious concern. 
Given the wire readout technology employed in the single-phase LArTPCs, reconstruction of the charge passing through the induction wire plane improves the correlation of signal 
between the multiple anode plane views and helps resolve degeneracies inherent in a projective wire geometry.

The successful reconstruction of a 3D event topology generally requires robust signal extraction in multiple 2D projection views. Since the ionization electrons are not collected on any of the induction wire planes, they naturally provide additional non-destructive views of the ionization electrons from the charged particle tracks. A successful extraction of the ionization electron information from the complicated induction plane signals is essential for 3D event reconstruction in single-phase LArTPCs using tomographic reconstruction and is expected to further enhance 3D reconstruction for techniques that match the image in different 2D projection views.


This paper is organized as follows. In section~\ref{sec:signal_form}, we
review the process of TPC signal formation including the induced current
generation, signal amplification and shaping, as well as the impact of noise.
In section~\ref{sec:charge_extract}, we describe the principle and the
algorithm implemented to extract ionization charge from the TPC signal. The
performance of the TPC signal processing chain is then evaluated with a
detailed TPC simulation in section~\ref{sec:evaluation}. The assessment of
this signal processing technique on MicroBooNE data is provided in a dedicated
accompanying paper~\cite{SP2_paper}. A discussion of some identified limitations of the current techniques is in section~\ref{sec:discussion}.
A summary and prospects for future improvements are presented in
section~\ref{sec:summary}.

\section{LArTPC signal formation} \label{sec:signal_form}

The formation of the TPC signal consists of three parts:
i) the electric field response to the drifting of a point ionization
charge leading to induced currents on the sense wires,
ii) the electronics response to the induced current waveform input to
each channel in terms of amplification and shaping, and
iii) the initial distribution of the ionization charge in the bulk of
the detector, how this charge drifts in the applied electric field and
how it undergoes diffusion and absorption as it drifts.
In the following sections, we describe each part in detail.


\subsection{Field response}\label{sec:field_resp}

\begin{figure}
  \centering
  \begin{subfigure}[b]{0.475\textwidth}
    \centering
    \includegraphics[width=1.0\textwidth]{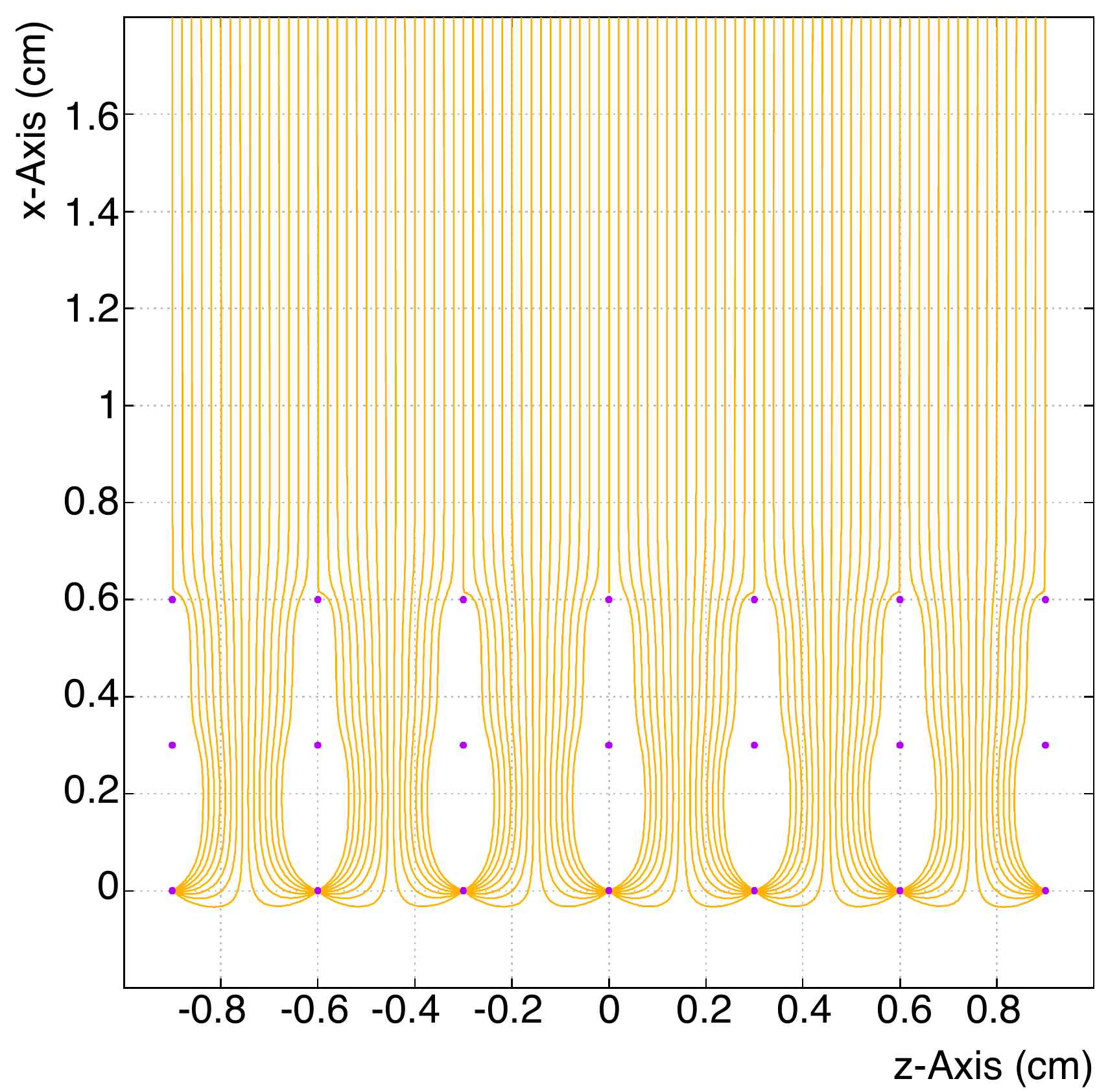}
    \caption[]%
    {{\small Electron drift paths.}}
    \label{fig:efield}
  \end{subfigure}
  \hfill
  \begin{subfigure}[b]{0.475\textwidth}  
    \centering 
    \includegraphics[width=1.0\textwidth]{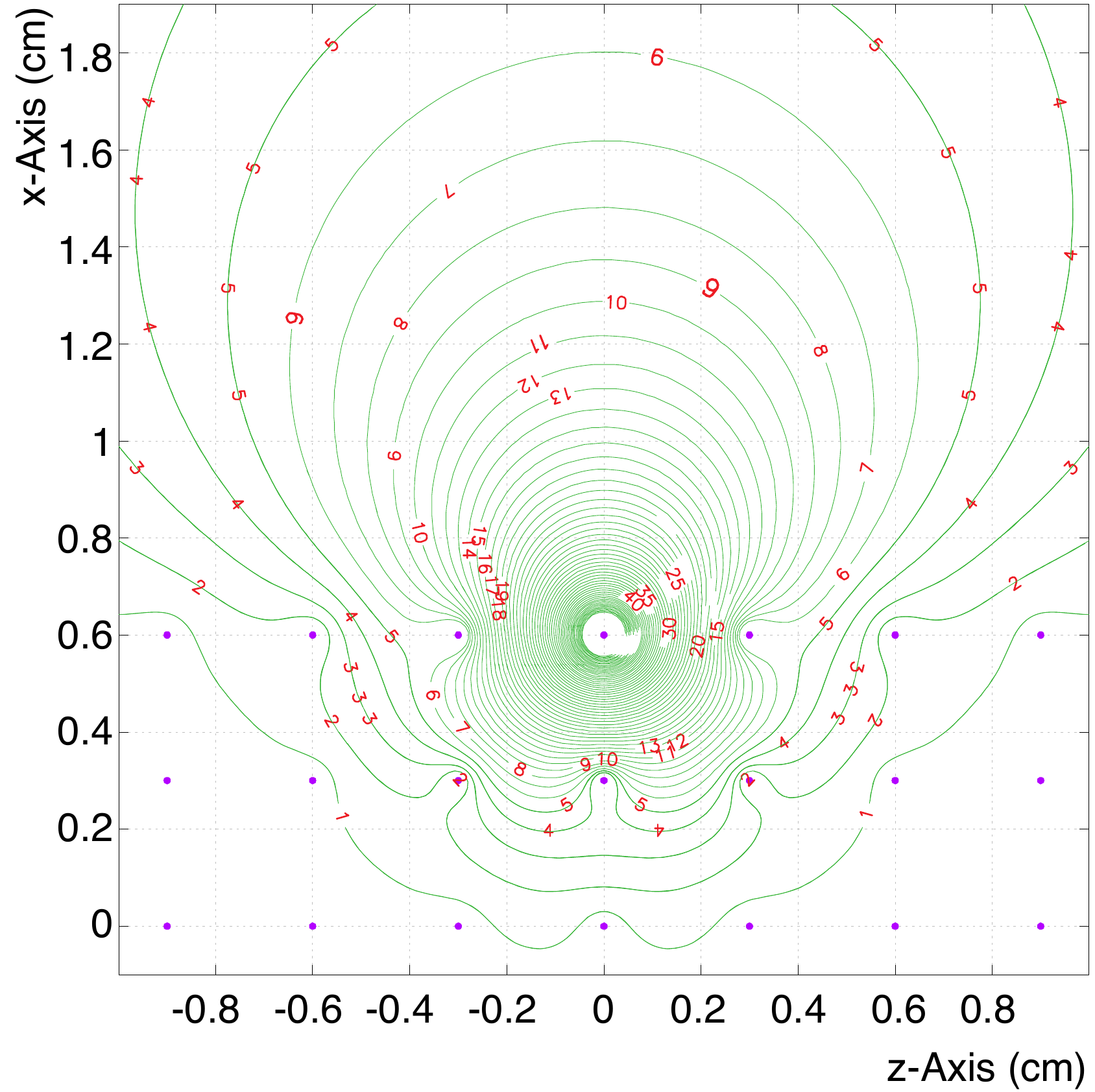}
    \caption[]%
    {{\small Weighting potential on a U wire.}}    
    \label{fig:u_weigthing}
  \end{subfigure}
  \vskip\baselineskip
  \begin{subfigure}[b]{0.475\textwidth}   
    \centering 
    \includegraphics[width=1.0\textwidth]{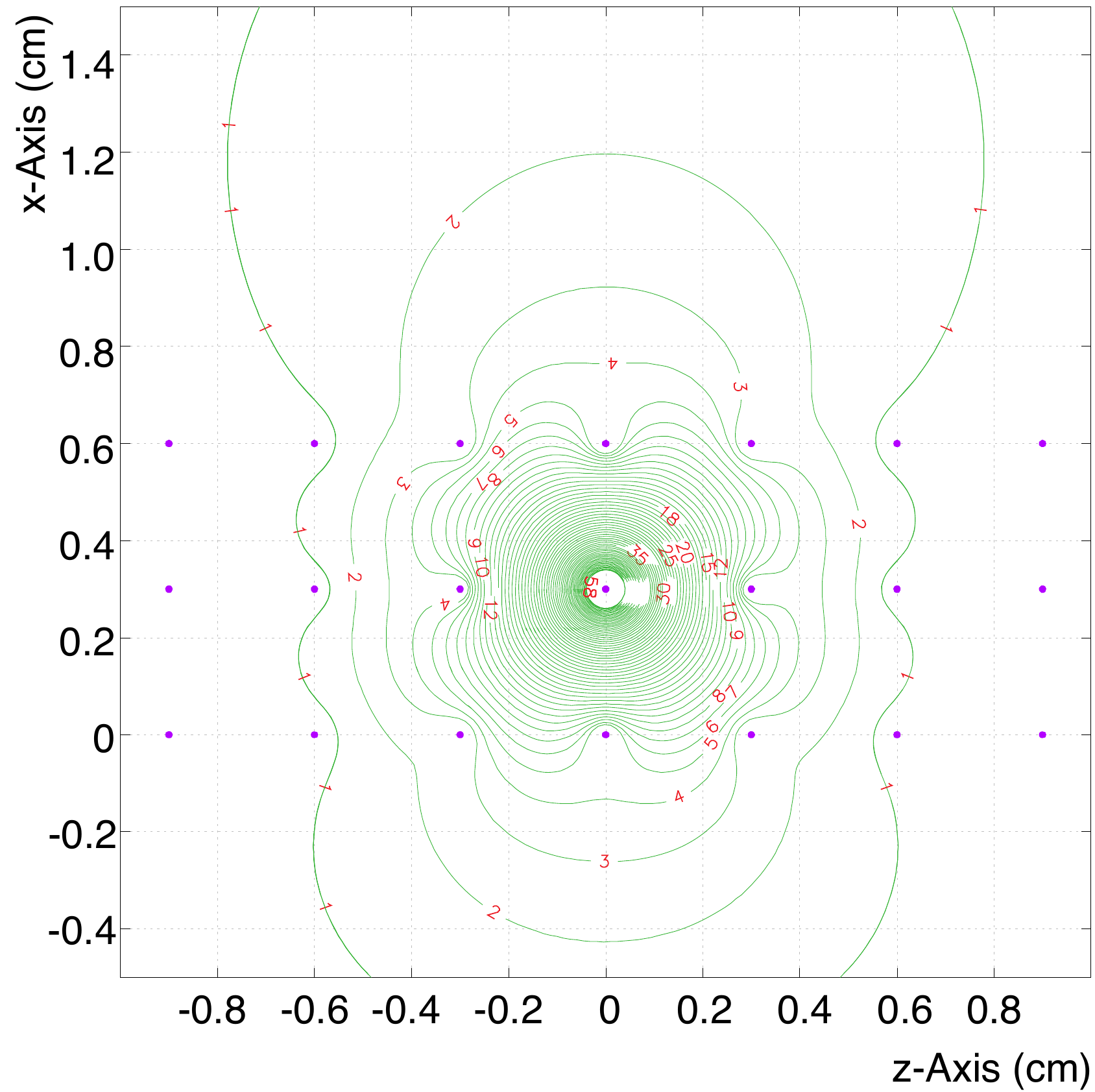}
    \caption[]%
    {{\small Weighting potential on a V wire.}}    
    \label{fig:v_weighting}
  \end{subfigure}
  \quad
  \begin{subfigure}[b]{0.475\textwidth}   
    \centering 
    \includegraphics[width=1.0\textwidth]{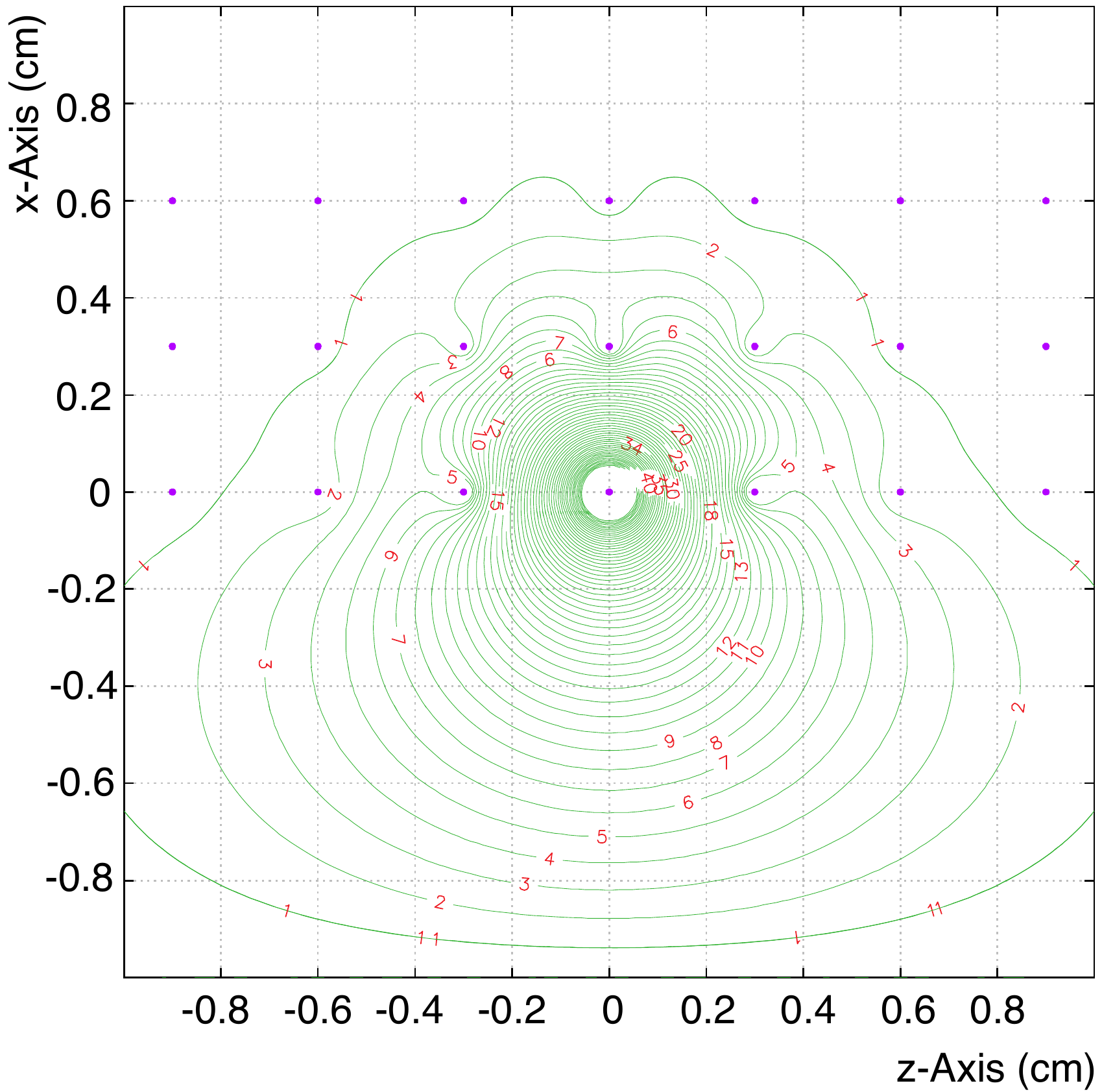}
    \caption[]%
    {{\small Weighting potential on a Y wire.}}    
    \label{fig:v_weigthing}
  \end{subfigure}
  \caption[]
  {\small A demonstration of electron drift paths in the applied
    electric field (panel a) and weighting potentials (panels b, c, d)
    on individual wires of the 2D MicroBooNE TPC model, using
    the Garfield program. The coordinates for each plane are defined in
    section~\ref{sec:topology_signal} as shown in
    figure~\ref{fig:coordinate}. The x-Axis is in the drifting field direction and
    the z-Axis is in the beam direction. The weighting potential is a dimensionless
    quantity, given as a value relative to the to the electric potential on the target wire.
    Values for the weighting potential are indicated in percentage on each equipotential
    line, ranging from 1\% for the farthest to 60\% for the closest illustrated.}
  \label{fig:fielddemo}
\end{figure}

When ionization electrons drift past the initial two induction wire
planes toward the final collection wire plane, current is induced on
nearby wires.  Henceforth, we refer to the induced current on one wire
due to a single electron charge as a field response function.  The
principle of current induction is described by Ramo's
theorem~\cite{ramo}.
An element of ionization charge $q$ in motion at a given location
induces a current $i$ on some electrode (wire),
\begin{equation}\label{eq:shockley_ramo}
  i = - q \vec{E}_w \cdot \vec{v}_q.
\end{equation}
This current is proportional to the inner product of a constructed
weighting field $\vec{E}_w$ for a given wire and the drift velocity
$\vec{v}_q$ of the charge at the given location.
The weighting field $\vec{E}_w$ only depends on the geometry of the electrodes and
can be calculated by removing the drifting charge, placing the targeted electrode
at unity potential, and setting all other conductors to ground. For a single medium,
the weighting potential is independent of the dielectric properties/constants.
For multiple media, the weighting potential must be calculated taking into account
each material's dielectric properties. This result is valid in the quasi-static
approximation and in arbitrary linear media where the permittivity is independent
of the potentials~\cite{ramo_extension, ramo_generalization}. A generalized form of
the weighting potential considering non-linear effects can also be found
in ~\cite{ramo_extension} and~\cite{ramo_generalization}. 
The charge's drifting velocity $\vec{v}_q$ is a function of the external
electric field, which depends on the geometry of the electrodes
as well as the applied drifting and bias voltages and liquid argon temperature.
Figure~\ref{fig:fielddemo} shows electron drift paths in an applied
electric field as well as lines of equal potentials for the weighting
field for the U, V, and Y wire planes in MicroBooNE.  
The electron drift paths and the weighting fields are calculated using
the Garfield~\cite{garfield} software. These calculations adopt a
2D model of a portion of MicroBooNE near a subset of wires.  Some
limitations in this model are discussed in section~\ref{sec:3D_field}. 

Although equation~\ref{eq:shockley_ramo} fully describes
a field response function for a given point charge at a given moment in time,
the easiest way to understand the
qualitative behavior of the field response function is through the integral
of the induced current and its connection to Green's reciprocity 
theorem. Let's consider a case where a point charge $q_m$ is moving in an
inter-electrode space. If we then assume that the charge $q_{m}$ is on an
infinitesimal electrode and the sensing electrode is labeled as electrode I,
then by Green's reciprocity,
\begin{equation}\label{eq:repo}
q_m \cdot V_m = Q_I \cdot V_I.
\end{equation}
Here $Q_I$ is the charge on the sensing electrode induced by $q_m$. $V_m$ is the potential at the location of $q_m$ introduced by the sensing electrode potential $V_I$. 
With equation~\ref{eq:repo}, we can derive the induced
current as
\begin{equation}\label{eq:repo1}
  i = \frac{dQ_I}{dt} = q_m \cdot \vec{\nabla} V_w \cdot \frac{d\vec{r}}{dt},
\end{equation}
where the weighting potential\footnote{This is equivalent to the definition
  from Ramo's theorem, where the voltage on the electrode under
  consideration is set to 1 V and all others are set to 0 V.}
is a dimensionless quantity defined as $V_w=V_m/V_I$.
It is easy to see that the above equation recovers 
equation~\ref{eq:shockley_ramo} where $\vec{\nabla} V_w$
corresponds to $-\vec{E}_w$ and
$\frac{d\vec{r}}{dt}$ is the velocity of the charge $q_m$.
Given equation~\ref{eq:repo1}, the integral of the induced current due to a charge $q_m$ moving along its drift path
\begin{equation}~\label{eq:voltage}
  \int i dt = q_m \cdot \left( V_w^{end} - V_w^{start}\right)
\end{equation}
is proportional to the difference of the weighting potential at the end and
start of the path. 

\begin{figure}[!htbp]
    \centering
    \includegraphics[width=0.8\textwidth]{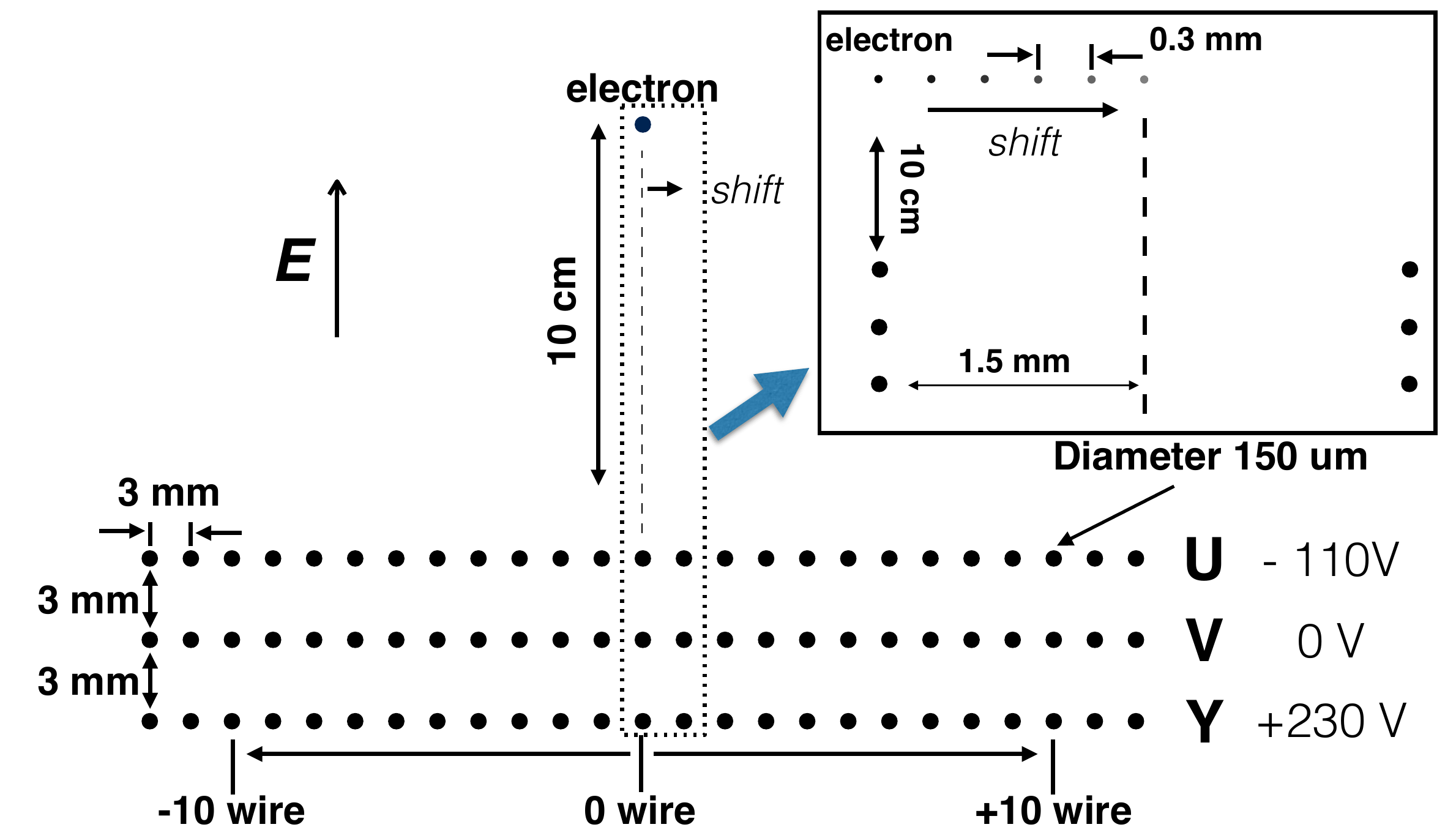} 
    \caption{Illustration of the 2D Garfield simulation scheme (dimensions
      not to scale), where black dots indicate individual wires. MicroBooNE's
      anode plane-to-plane spacing is \SI{3}{\mm}, with \SI{3}{\mm} wire pitch
      in each plane. The inset denotes the sub-pitch designation of
      electron drift paths whereupon the field response is calculated.}
    \label{fig:garfield_schem}
  \end{figure}

For signal processing and signal simulation, the field response functions for
a single ionization electron traveling over a number of possible discrete drift paths are
calculated with the Garfield program~\cite{garfield} using a 2D model for
MicroBooNE wires with a scheme illustrated in figure~\ref{fig:garfield_schem}.
The calculation
utilizes a region that spans \SI{22.4}{\cm} (the upper boundary at
\SI{20.4}{\cm} in front of the Y plane) along the nominal electron
  drift direction and \SI{30}{\cm} perpendicular to both the drift
  direction and the wire orientation. In the calculation, each wire
  plane contains 101 wires with \SI{150}{\micro\meter} diameter separated
  by a \SI{3}{\mm} wire pitch. The drifting field (273 V/cm) is achieved
  by setting the negative voltage at the upper boundary of the simulated
  area. The nominal MicroBooNE operating bias voltages for each wire plane are
  used in the calculation.

  There are two stages in calculating the field response functions.
  The first one is the calculation of the electron drift paths in
  the applied electric field as shown in figure~\ref{fig:efield}. 
  The second stage is the 
  calculation of the weighting electric potentials as shown in the remaining panels
  of figure~\ref{fig:fielddemo}. The induced current can be
  calculated following equation~\eqref{eq:shockley_ramo}.  The electron drift
  velocity as a function of electric field is taken from recent
  measurements~\cite{Li:2015rqa,lar_property}. For these single-electron
  simulations, diffusion is omitted.

  For a single drift path calculation, the electron starts from a point
  \SI{10}{\cm} away from the wire plane above the central wire (shown as
  ``0 wire'' in figure~\ref{fig:garfield_schem}).
  The field response functions for that central wire and $\pm$10 wires on both sides
  (21 wires in total) are calculated. The simulation is then repeated starting at
  different transverse locations, each shifted by \SI{0.3}{\mm} from the previous,
  spanning \SI{0}{\mm} to \SI{1.5}{\mm}.
  In total, 126 field responses (six electron positions $\times$
  21 readout wires) are calculated. In the described 2D scheme, the inter-plane wires are aligned. The shift in relative inter-plane 2D geometry is a 3D effect and has minimal impact on the calculated field response shape.

\begin{figure}[!htbp]
    \centering
    \includegraphics[width=0.7\textwidth]{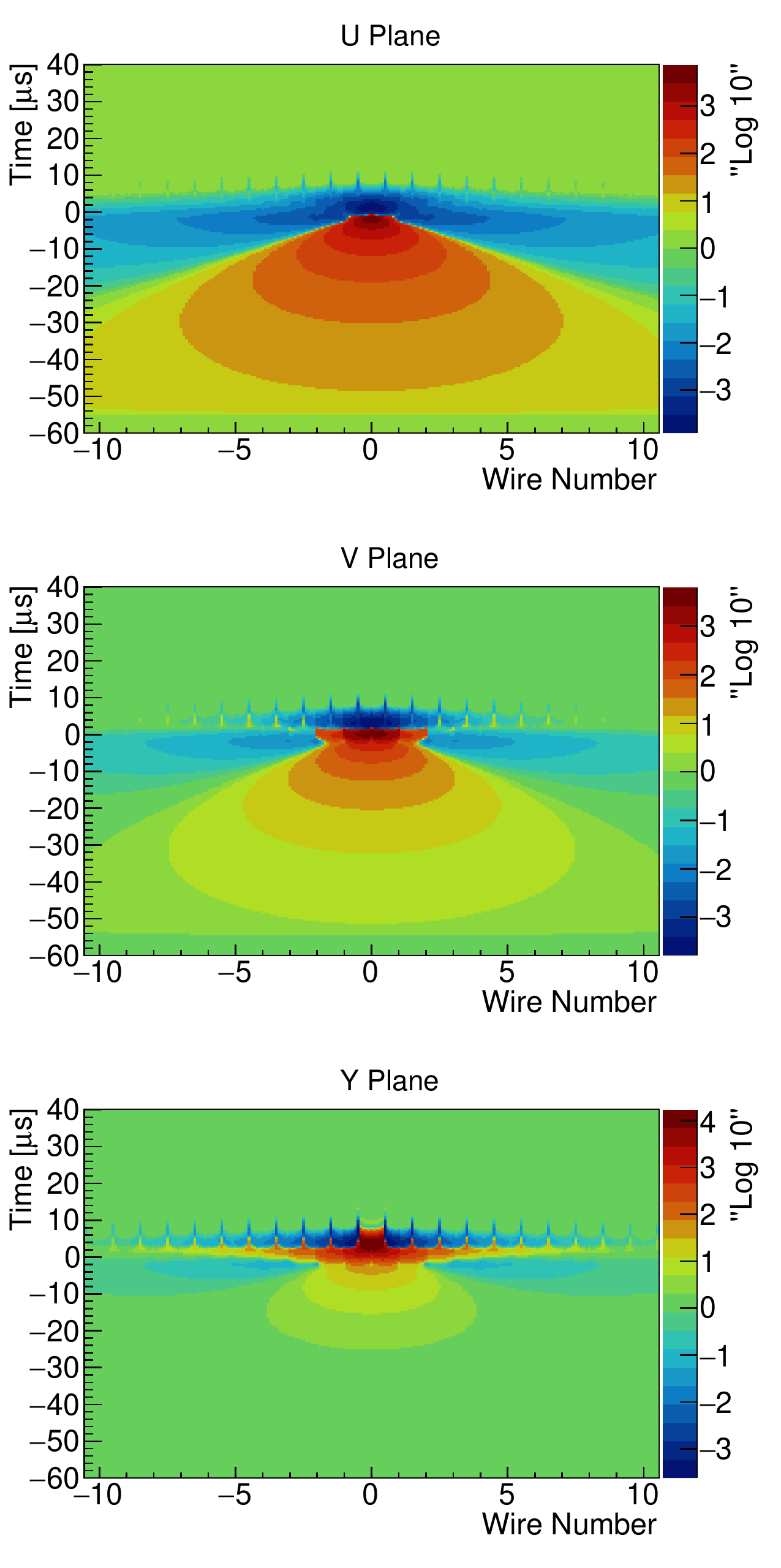}
    \caption{The overall response functions after convolving the field
      response function and an electronics response function with a \SI{2}{\us}
      peaking time are shown in two dimensions. All plots are shown
      in ``Log10'' scale.}
    \label{figs:overall_response}
  \end{figure}

  Figure~\ref{figs:overall_response} shows the
  overall response functions for each wire of interest for induction U (top panel),
  induction V (middle panel), and collection Y (bottom panel) wire planes, where the overall
  response is the field responses convolved with the electronics response (to be
  described in section~\ref{sec:electronics_resp}).
  The X-axis in figure~\ref{figs:overall_response}
  is the initial transverse position of the
  ionization electron relative to the central wire of each plane and
  expressed in units of wire number.
  Each wire region ($\pm$0.5 wire pitch) is sampled by 11 electron
  drift paths with starting points that are regularly separated by
  \SI{0.3}{\mm}.
  The field response on the central wire for a single path is thus
  represented by a slice of this plot at a corresponding location on the
  X-axis for the given path.  The Y-axis is the drift time
  relative to the electron arriving at the V plane. 
  Time is made discrete in units of \SI{0.5}{\us} (one ``tick'') which
  corresponds to the analog-to-digital converter (ADC) sampling period employed by 
  the MicroBooNE readout electronics.

  The
  normalization of the overall response function is chosen so that the integral of
  the response function of the closest wire from the collection Y plane is unity,
  which corresponds to a single electron.
  To
  emphasize the shape of the field response functions, a special scale labeling
  with ``Log10'' is used to set the color scale of the induced current $i$ (electrons
  per \SI{0.5}{\us}) in Figure~\ref{figs:overall_response}:

  \begin{equation}\label{eq:log10}
    i \text{ in ``Log 10'' } =
    \begin{cases}
      \text{log}_{10}(i\cdot10^{5}),& \text{if }i>1\times10^{-5},\\
      0,& \text{if }  -1 \times10^{-5} \leq i \leq 1 \times 10^{-5},\\
      -\text{log}_{10}(-1\cdot i\cdot10^{5})& \text{if }i<-1\times10^{-5}.\\
      \end{cases}
  \end{equation}


  The 2D nature of the model used for MicroBooNE wire planes in the
  Garfield calculation described above is an approximation of
  the actual detector.  No detector edge effects are considered and the
  2D nature of the model implies that wires are effectively infinite in
  length and that any effects due to the wires crossing near each other
  can not be included.  An initial set of field calculations has been
  performed with custom software that utilizes a 3D model of the MicroBooNE
  wire planes. More discussions can be found in section~\ref{sec:3D_field}.

\begin{figure}
    \centering
    \includegraphics[width=0.7\textwidth]{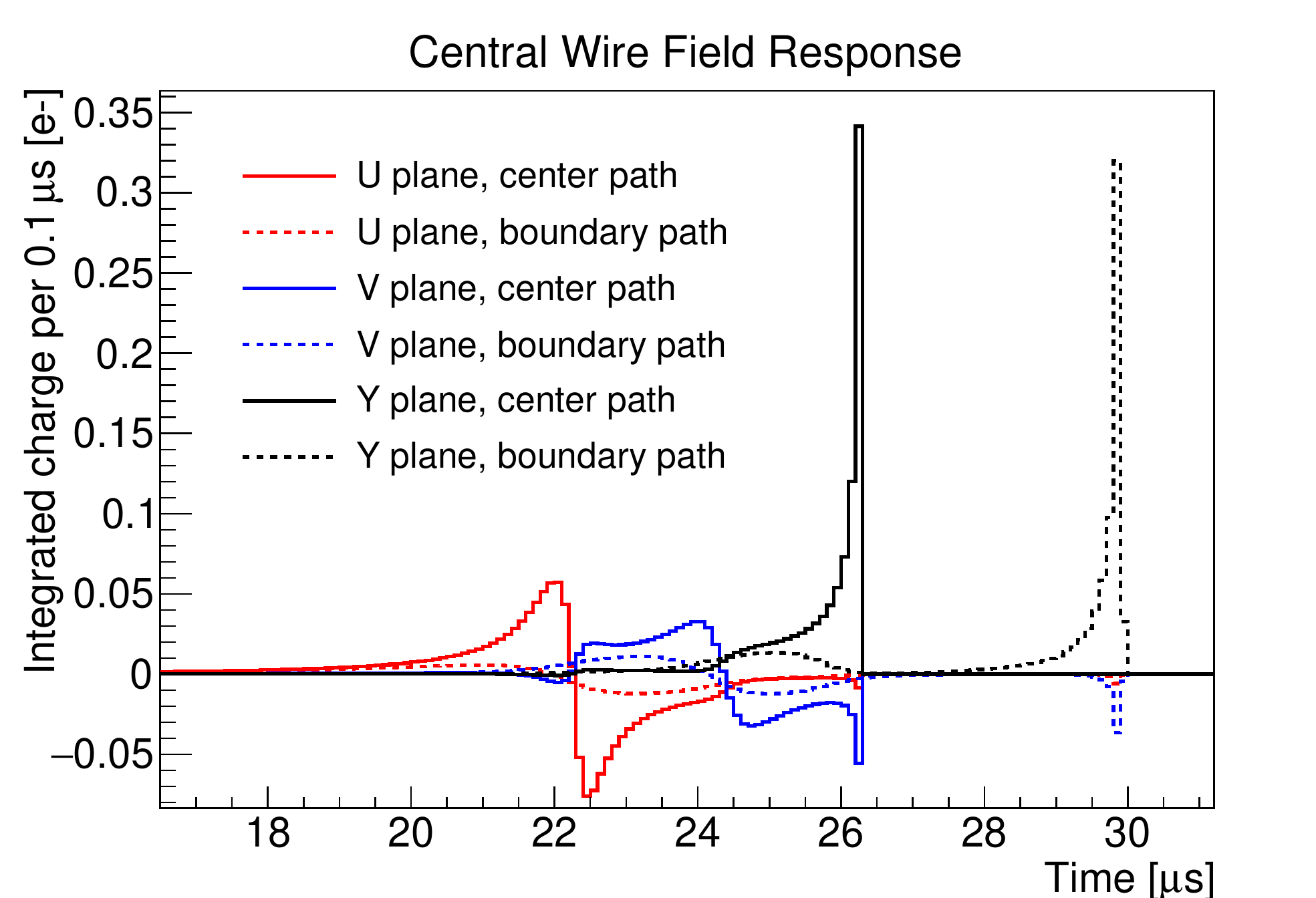}
    \caption{Field responses (induced-current) from various paths of one     
      drifting ionization electron for the three wire planes. Y-axis is the  
      integrated charge over \SI{0.1}{\micro\second}. Within the
      central wire pitch, a center path at \SI{0}{\milli\meter} (solid) and
      a boundary path at \SI{1.5}{\milli\meter} (dashed) are employed for
      this demonstration. See figure~\ref{fig:garfield_schem} for
      an illustration of the simulated geometry. The fine
      structures in the field responses are subject to the path of the drifting
      ionization electron and the weighting potential as shown in
      figure~\ref{fig:fielddemo}.}
    \label{fig:fieldstructure}
\end{figure}

Given the results shown above, we can conclude the following:
\begin{itemize}

\item As an ionization electron moves towards the closest induction
  plane wire, it climbs up the corresponding induction wire weighting
  potential; therefore, the induced current is negative and corresponds
  to the positive voltage waveform shown in this paper following the
  MicroBooNE electronics readout convention.
        
\item As an ionization electron passes the induction wire plane and
  moves toward the collection wire plane, the induction wire weighting
  potential decreases and the induced current changes sign. This results
  in the bipolar shape for the induction plane signals. See examples
        in figure~\ref{fig:fieldstructure}.

\item As an ionization electron drifts towards the wire on which it will
  eventually be collected (the central wire), the weighting function for
  that wire always increases. Consequently, collection wire signals are
  unipolar in shape.

\item For an ionization electron originating from the cathode plane
  (i.e. weighting potential zero), the integrated induction charge in
  an induction wire plane is nominally zero, since the electron ultimately
  ends up at a collection wire for which the corresponding induction
  wire weighting potential is also zero.  Similarly, the integrated
  induction charge in the collection wire should be equal to the
  charge of one electron, as the corresponding collection wire
  weighting potential is unity.

\item The above conclusion about the integrated induction charge
  is no longer true for an ionization electron generated inside the
  active LArTPC volume due to a non-zero weighting potential at the
  point of origin.  The deviation from zero integrated charge is
  largest for the first induction U plane, as V and Y wire planes are
  shielded by the U plane thus their weighting fields do not extend as
  far into the volume. However, even for the most extreme case, these
  deviations are generally limited to a few percent for
  points of origin 10 cm away from the U wire plane. We should also note that
  the induced current due to the sudden creation of the ionization
  electron is balanced by the creation of the positive argon ion.\footnote{At
    creation, $\int i dt = \sum_m q_m \cdot V_w^{end} =0$, since $\sum_m q_m = 0$. However, immediately
    following creation, the ion drift velocity is $\mathcal{O}(10^6)$ smaller than
    the electron drift velocity. Thus, drift of the Ar$^+$ contributes negligibly
  to the induced current.}

\item The strength of the induced current on an induction plane wire
  due to an ionization electron is related to the maximum weighting
  potential that the electron can reach. Therefore, we expect the
  induction signal to increase as a given electron is allowed to pass
  closer to the wire.  However, the bias voltage applied to each wire
  plane forces the electrons to divert from regions of maximum
  induction wire weighting potential.  This causes the induction signals
  to be generally smaller than those on collection wires.

\item Since the weighting field of the sense wire extends far beyond the sense
  wire region, an ionization electron drifting far away from a sense wire
  will still prompt induced current on the sense wire, although at a reduced
  strength. Therefore, the induced current depends on the local charge
  density, which is further determined by the event topology. 

\item The time duration of the induced current (field response) depends
  on the drift velocity of the ionization electron as well as the electrode
  geometry. For the U wire plane, the induced current becomes sizable when the
  ionization electron is about several cm away from the wire plane and ends
  when the electron is collected by the Y wire plane. For the V wire plane,
  the induced current becomes sizable when the electron is about to pass
  the U wire plane and also ends when the electron is collected by the Y
  wire plane. For the Y plane, the induced current becomes sizable as the
  electron is about to pass the U wire plane. These field response functions
  are further modified for a cloud of electrons due to different drift
  paths. In particular, broadening is expected for the collection wire
  response function for a cloud of electrons.

\item As illustrated in figure~\ref{fig:efield}, there exist
  variations in the drift paths of ionization electrons toward the collection
  wire. This results in fine structures
  (see figure~\ref{fig:fieldstructure}) in the field response which are dependent on
  the drifting path as well as the weighting potential shown in
  figure~\ref{fig:fielddemo}. In particular, an
  electron traveling along a path equidistant from two collection
  wires (at the boundary of one wire pitch) will be
  collected with a few \si{\micro\second} delay compared to one that
  arrives at its collection wire more directly.  The phase space for
  this delay is small so it affects a minority of drifting
  charge.  Integrated over a distribution of ionization electrons,
  this variation contributes to a broadening of the overall collection
  signal as shown in the bottom panel of figure~\ref{fig:parallelmip}. The second negative peak
  in the field response of the induction wire as shown in figure~\ref{fig:fieldstructure}
  originates from (and thus coincides with) the collection of the ionization electron
  in the collection wire. However, due to shielding from the V plane
  (see weighting potential in figure~\ref{fig:fielddemo}), this peak for the
  U plane is insignificant. 

\end{itemize}

\subsection{Electronics response}\label{sec:electronics_resp}

\begin{figure}[!h!tbp]
  \centering
    \begin{subfigure}[t]{0.45\textwidth}
        \centering
        \includegraphics[width=0.99\textwidth]{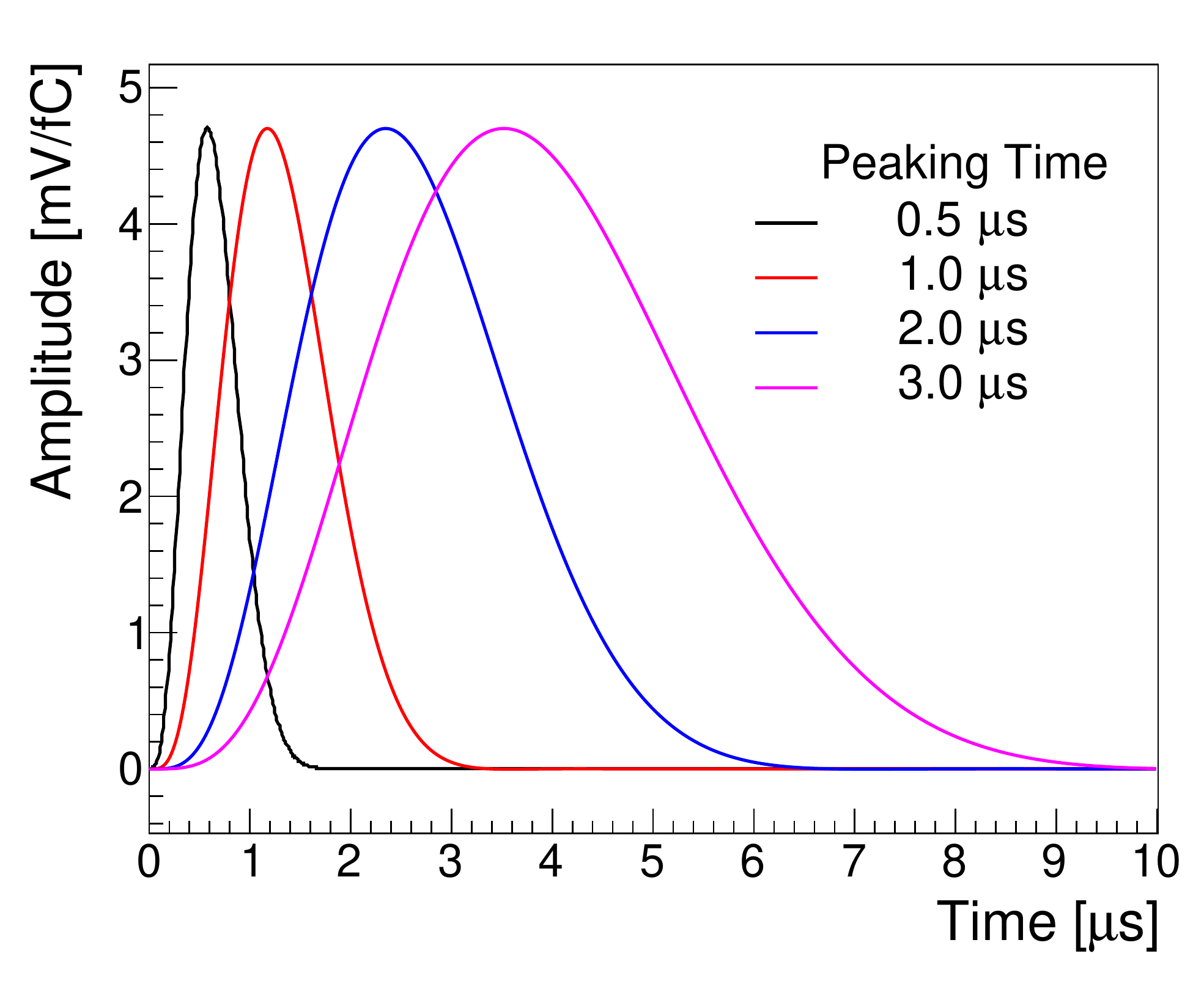}
        \caption{Pre-amplifier response function.}\label{fig:resp1_preamp}
    \end{subfigure}
    \begin{subfigure}[t]{0.45\textwidth}
        \centering
        \includegraphics[width=0.99\textwidth]{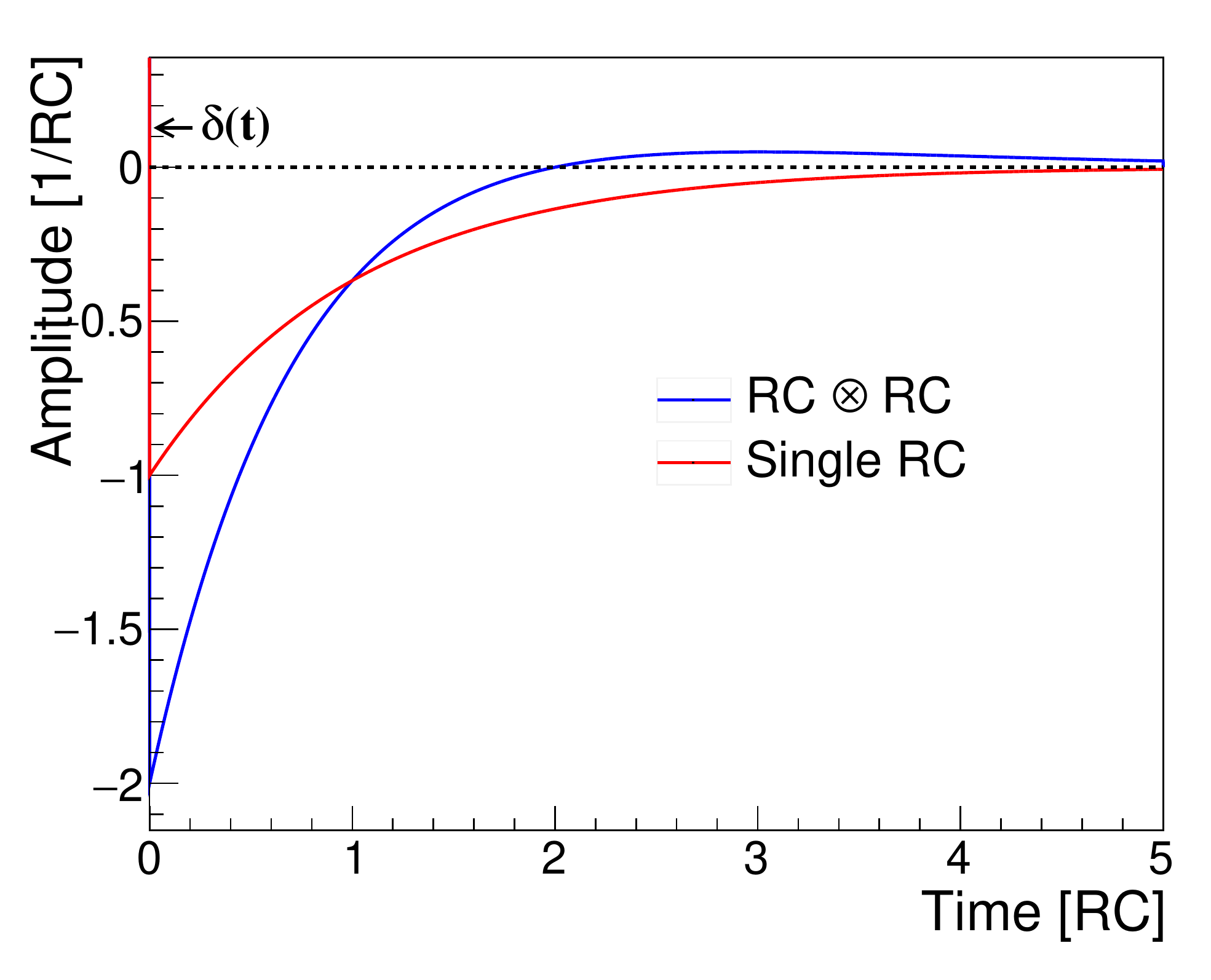}
        \caption{RC filter response function.}\label{fig:resp1_rc}
    \end{subfigure}
    \caption[elec-resp]{MicroBooNE pre-amplifier electronics impulse response
      functions are shown for (a) four peaking time settings at 4.7 mV/fC gain
      and (b) a single RC filter and two
      independent RC filters (RC $\otimes$ RC).}
  \label{fig:resp1}
\end{figure}

The induced current on the wire is received, amplified, and shaped by
a pre-amplifier. This process is described by the electronics response
function. The impulse response function in the time domain is shown in
figure~\ref{fig:resp1_preamp}.  The MicroBooNE front-end cold
electronics~\cite{cold} are designed to be programmable with four
different gain settings (\SIlist[per-mode=symbol]{4.7;7.8;14;25}{\mV\per\femto\coulomb})
and four peaking time settings (\SIlist{0.5;1.0;2.0;3.0}{\us}). In MicroBooNE,
 the gain is roughly 3.5\% lower than expected and the peaking time is 10\% higher than expected~\cite{SP2_paper}. For a fixed
gain setting, the peak of the impulse response is always at the same height
independent of the peaking time. The peaking time is defined as the time
difference between 5\% of the peak  at the rising edge and the peak. The
different gain settings allow for applications with differing ranges of
input signal strength. The four peaking time settings are provided to
satisfy the Nyquist criterion~\cite{nyquist} at different sampling rates. 
Two additional RC filters are exploited to remove the baseline from the
pre-amplifier and the intermediate amplifier. The intermediate amplifier provides an additional gain of 1.2 (dimensionless) to compensate for the loss without any shaping/filtering.
The time-domain impulse response is as follows (and is shown 
in figure~\ref{fig:resp1_rc}):
\begin{align}\label{eq:rcfilter}
    {\rm Single~RC:} &~ h(t) = \delta(t) - \frac{1}{\tau}\cdot e^{-t/\tau}u(t),\\
    {\rm RC \otimes RC:} &~ h(t) = \delta(t) + (\frac{t}{\tau}-2)\frac{1}{\tau}\cdot e^{-t/\tau}u(t),
\end{align}
where the time constant $\tau = RC$ and $\delta(t), u(t)$ are the delta
function and the step function, respectively. In general, the time constant
is \SI{1}{\milli\second} in MicroBooNE and the RC filter effect is visible
when the signal is large or long enough. 

The resulting signal waveform, as illustrated in figure~\ref{fig:parallelmip},
after electronics shaping is then digitally sampled at \SI{2}{\mega\hertz} by a
12-bit ADC with the input voltage ranging from 0 to \SI{2}{\volt}. More
details regarding the performance of MicroBooNE cold electronics can be found in~\cite{noise_filter_paper}.

\subsection{Topology-dependent TPC signals}\label{sec:topology_signal}
\begin{figure}[!h!tbp]
  \centering
  \includegraphics[width=0.625\textwidth]{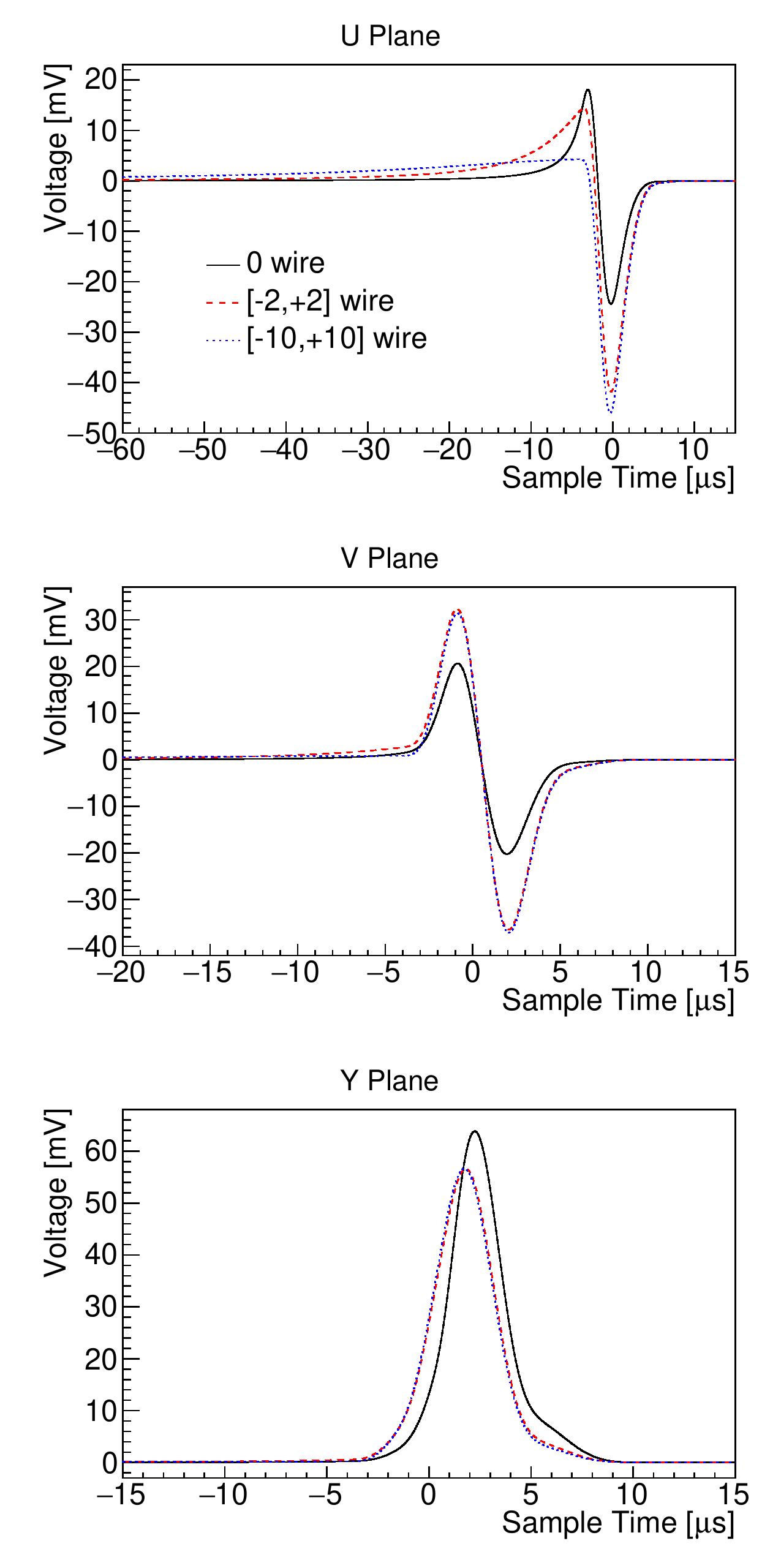}
  \caption[parallel-mip]{Simulated baseline-subtracted TPC signals for an
    ideal isochronous MIP
    track traveling perpendicular to each wire plane orientation
    ($\theta_{xz}=0^{\circ}$ and $\theta_y=90^{\circ}$, i.e. along the
    $z$-axis, wire pitch direction, for each wire plane) for MicroBooNE
    wire plane geometry. The track is an
    ideal line source that runs perpendicular to all wires, spans the
    transverse domain of the simulation, and is comprised of $\approx$4400
    ionization electrons per mm mimicking a MIP.
    Only the field response and pre-amplifier electronics response
    (\SI{2}{\us} peaking time and \SI{14}{mV} gain) are included;
    diffusion is neglected. ``0 wire'' depicts the signal for charge
    drifting within one-half pitch distance of a central wire.  The ``[-N,+N] wire''
    plots provide the contribution to the signal on the central wire
    from ionization electrons that drift in progressively more distant
    $N$ neighboring wire regions.}
  \label{fig:parallelmip}
\end{figure}

As shown in figure~\ref{fig:fielddemo}, the weighting potential is not
confined to just the region around one sense wire.  An element of
drifting ionization charge will induce signal on many wires in its
vicinity.  The signal of any one wire depends on the ensemble
distribution of charge in its neighborhood. To illustrate
this effect, we first consider the signal resulting from an
isochronous (parallel to the anode plane) minimum ionizing particle (MIP) track perpendicular to each anode plane wire orientation.
Figure~\ref{fig:parallelmip} shows the simulated central wire signal
when the contributions of ionization charge at different positions beyond the
central wire are considered. For all three wire planes, the signal contribution
due to long-range induced current for ionization charge is non-negligible.
For the induction V and collection Y wire planes, the
proportion of the wire signal is small for ionization charge beyond
a couple wires from the central wire. For the induction U wire plane, which is
the first wire plane facing the active TPC region, the
contribution of the wire signal from distant wires can be sizable for the ionization
charge as far as 10 wires away from the central wire. The modification
of the signal due to this long range induction effect is relatively
small for the collection wires, since the two induction planes
provide shielding.
The unipolar induction signal from any collected electrons on the collection
plane is large compared to the bipolar contributions from electrons which
collect on neighboring wires. On the other hand, signal distortion due to
long range induction can be sizable
for the induction wires due to their smaller field response functions and the
potential cancellation of multiple bipolar signals. This is also true
for adjacent collection plane wires which did not actually collect the ionization charge. They
effectively behave as induction wires.

To describe the signal dependence on track topology and later for evaluation
of the signal processing technique (section~\ref{sec:quan_eva}), the detector
Cartesian coordinate as well as each wire plane's coordinate are defined.

\begin{figure}[htbp]
    \centering
    \begin{subfigure}[t]{0.8\textwidth}
    \includegraphics[width=1.0\textwidth]{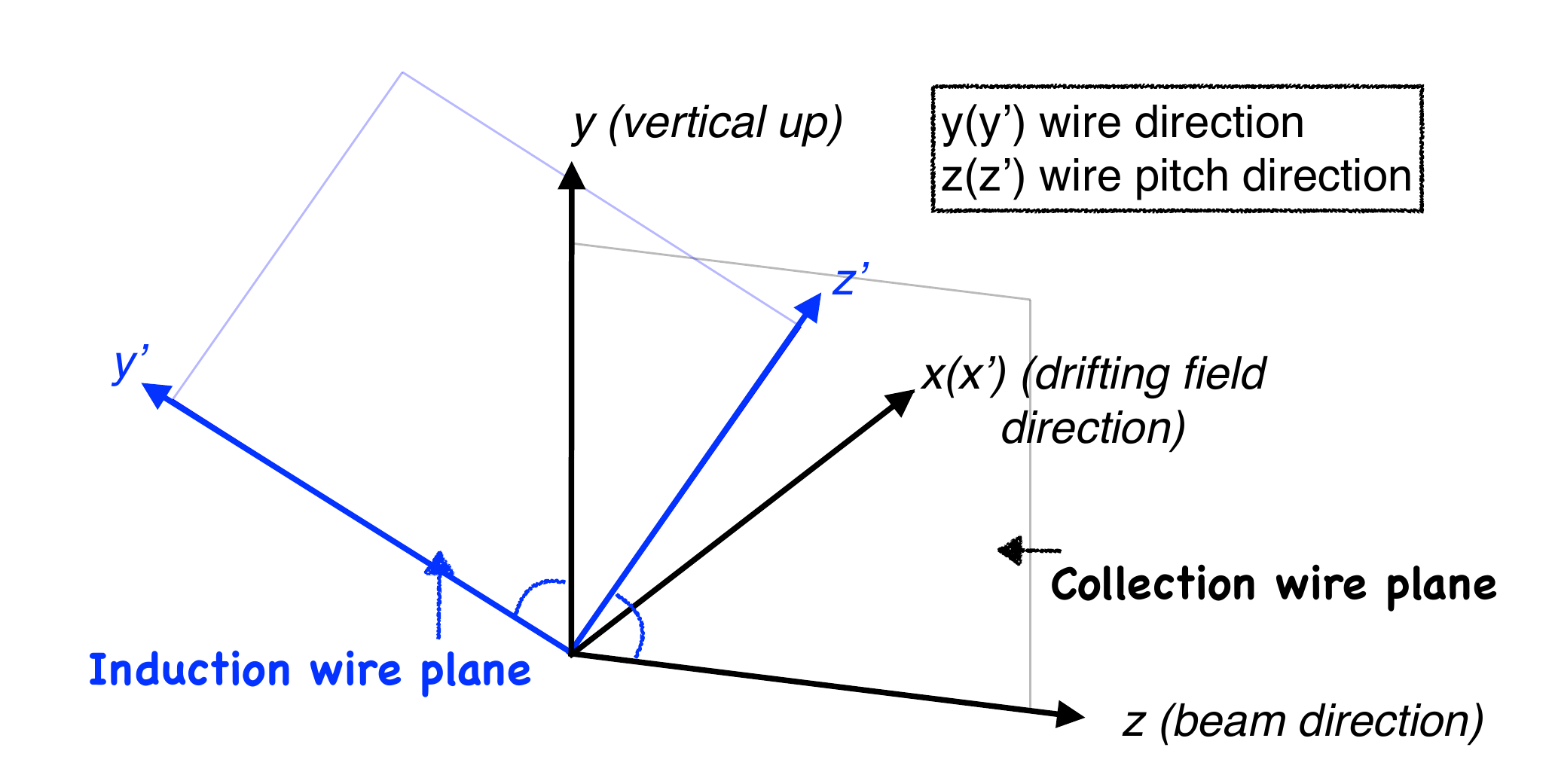}
    \caption{Coordinates for collection and induction planes. The $y'$ ($z'$)
      axis is rotated by 60$^{\circ}$ around the $x$ axis from the $y$ ($z$) axis. 
      The illustrated induction plane's coordinate is consistent with the V plane's and the U plane's is with a rotation in the opposite direction.}
    \label{fig:coordinate}
    \end{subfigure}
    \begin{subfigure}[t]{0.8\textwidth}
    \includegraphics[width=1.0\textwidth]{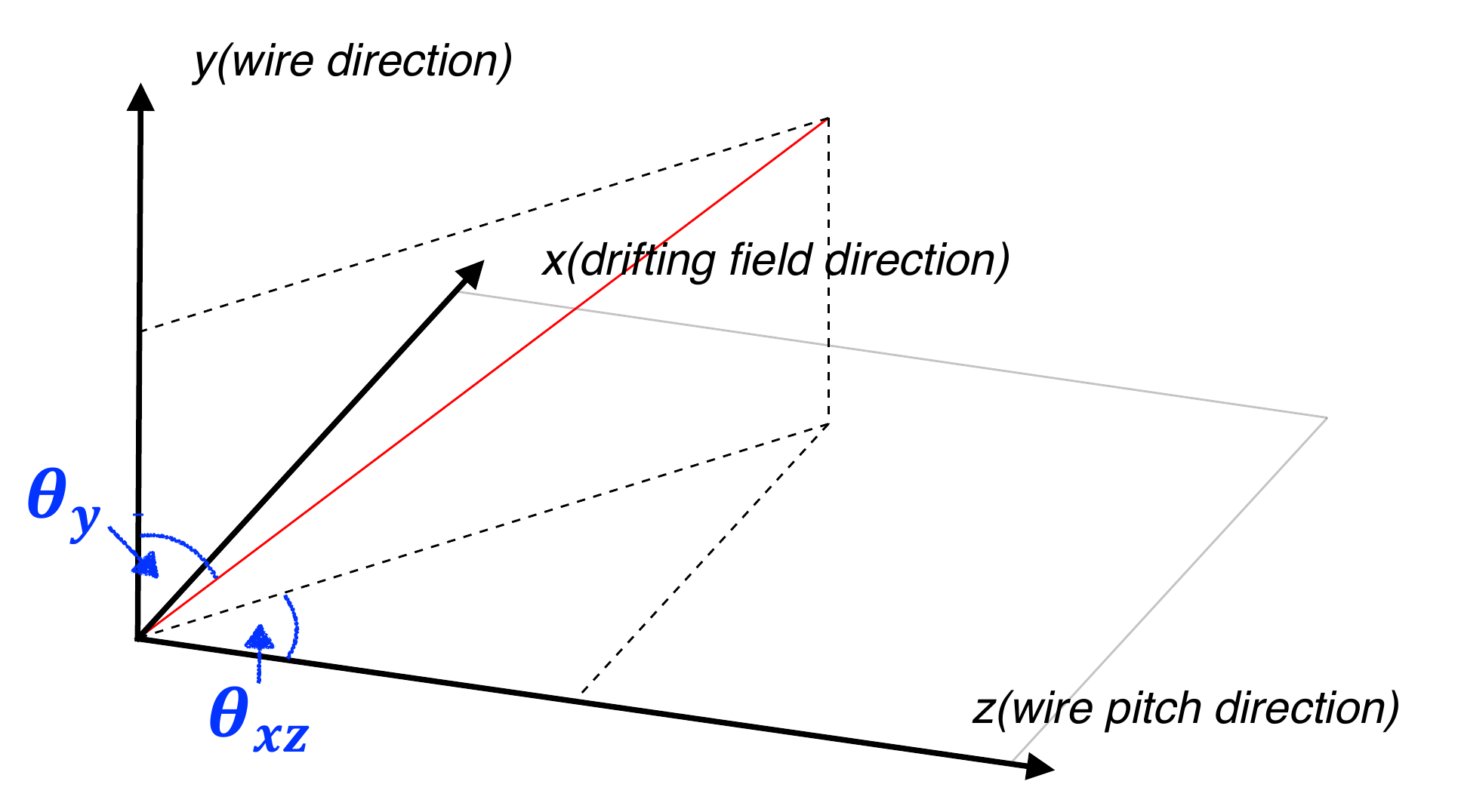}
    \caption{Definition of two angles, $\theta_{xz}$ and $\theta_y$.}
    \label{fig:angle}
    \end{subfigure}
    \caption{Geometric coordinates and angles for topology description.}
    \label{fig:geometry}
\end{figure}

\begin{description}
\item{\textit{Coordinate system for each wire plane}} -- As shown in
  figure~\ref{fig:coordinate}, for each wire plane, the $x$-axis is along the drifting
  field direction, the $y$-axis is along the wire orientation, and the $z$-axis is
  along the wire pitch direction. The nominal (default) detector coordinate system
  is identical to the collection plane's coordinate system for which the $y$-axis is
  vertical in the upwards direction and the $z$-axis is along the wire pitch direction.
\item{\textit{Angles for topology description}} -- Based on the predefined coordinate for each
  wire plane, two angles are defined as well to describe the topology of the track.
  As shown in figure~\ref{fig:angle}, $\theta_y$ is the angle between the track and
  the $y$-axis, and $\theta_{xz}$ is the angle between the projection onto the $x-z$
  plane and the $z$-axis. 
\end{description}
      
\begin{figure}[!h!tbp]
  \centering
  \includegraphics[width=0.7\textwidth]{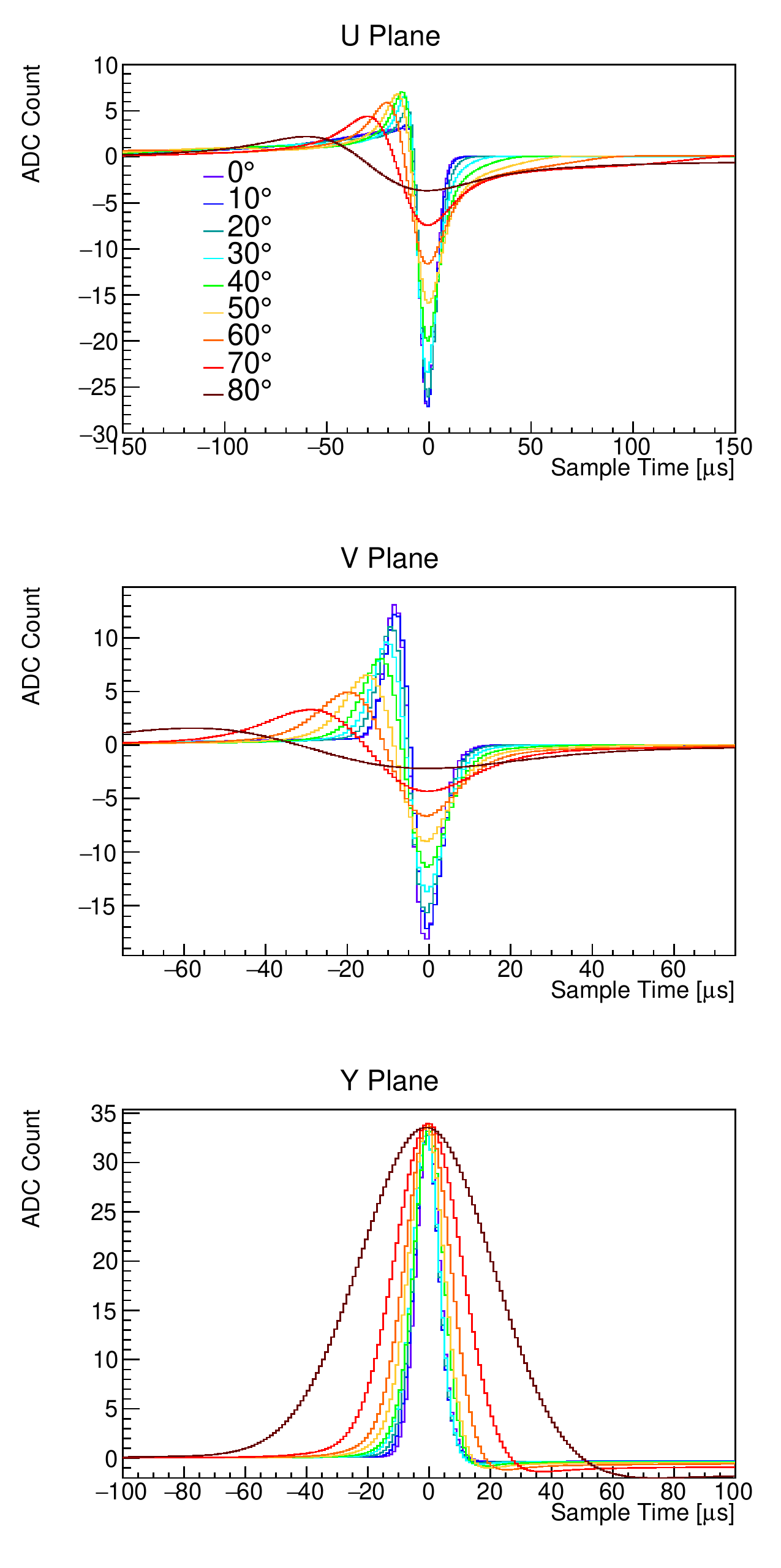}
  \caption[sim-perpendicular-mip]{Simulated baseline-subtracted MicroBooNE
    TPC signals for a 1 meter long MIP track ($\approx$4400 ionization
    electrons per mm) traveling perpendicular to each wire
    plane orientation ($\theta_y=90^{\circ}$) with $\theta_{xz}$
    varying in the $x-z$ plane with respect to the $z$-axis. Detector physics
    effects and the nominal MicroBooNE electronics response~\cite{SP2_paper}
    were included. The signal shape is solely determined by $\theta_{xz}$,
    independent of $\theta_y$.}
  \label{fig:sim_perpendicularmip}
  \end{figure}

\begin{figure}[!h!tbp]
  \centering
  \includegraphics[width=0.7\textwidth]{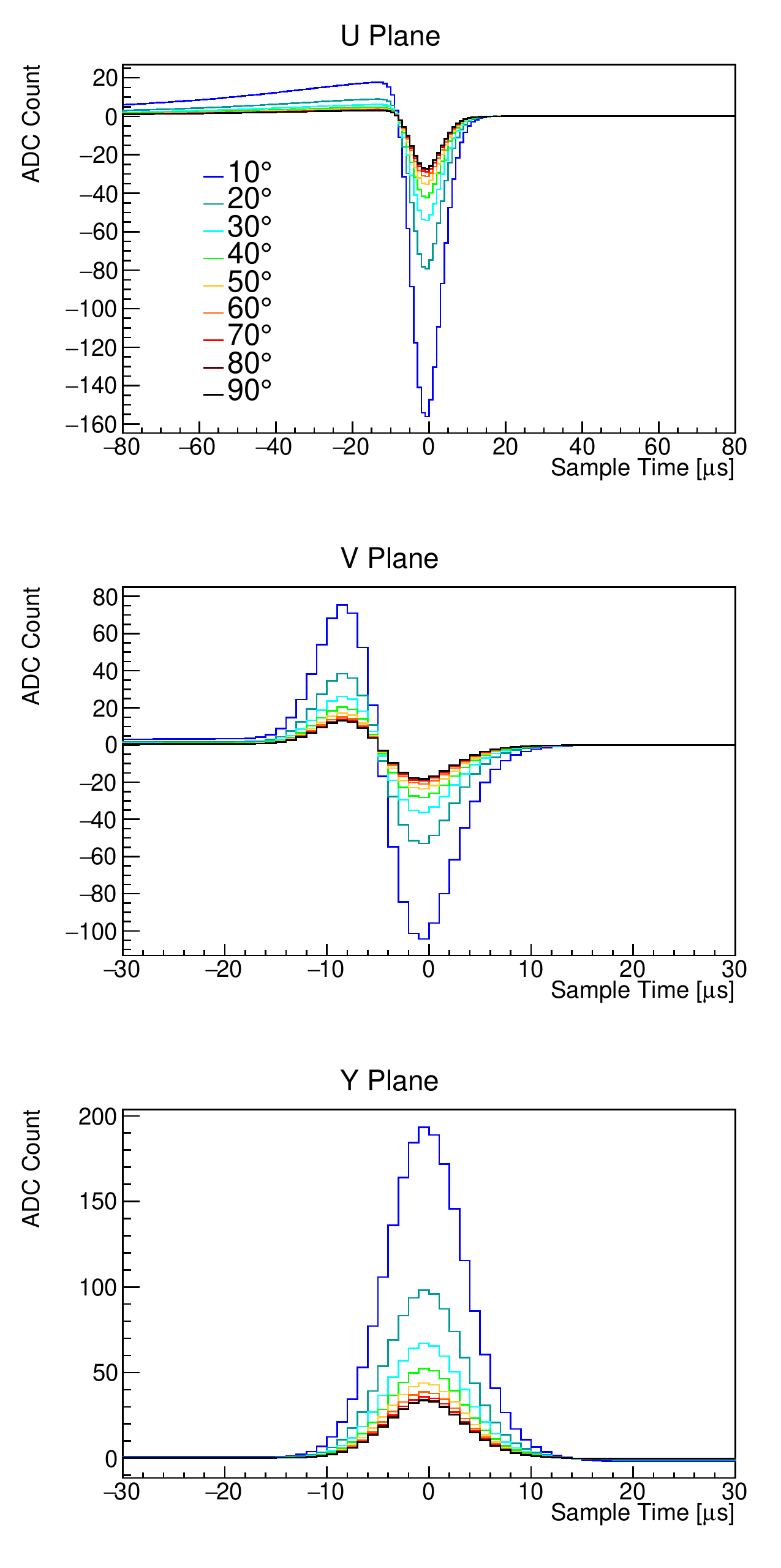}
  \caption[sim-parallel-mip]{Simulated baseline-subtracted MicroBooNE TPC
    signals for a 1 meter long MIP track ($\approx$4400 ionization
    electrons per mm), isochronous ($\theta_{xz}=0^{\circ}$) with
    $\theta_y$ varying with respect to the wire orientation. Detector physics
    effects and the nominal MicroBooNE electronics response~\cite{SP2_paper}
    were included. Given a $\theta_{xz}$, $\theta_y$ just changes the
    signal amplitude.}
  \label{fig:sim_parallelmip}
  \end{figure}

Since the $x$ component determines the time extent of the track and the $z$
component determines the range of sense wires (channels), $\theta_{xz}$ alone
determines the shape of the TPC signal, assuming the field response is identical
along the $y$-axis (wire orientation). This assumption exactly holds in the current
2D field response calculation and corresponds to roughly the average of the
3D field response. 
The $y$ component is proportional to the length of the charge
deposition projection on the wire direction. It simply scales the
charge deposition within one wire pitch by
$1/(\cos\theta_{xz}\cdot \sin\theta_y)$. As an example,
figure~\ref{fig:sim_perpendicularmip} and figure~\ref{fig:sim_parallelmip} demonstrate
the TPC signal topology-dependency on $\theta_{xz}$ and $\theta_y$. Note that the
discussions above are related to the individual coordinates and angles for each
wire plane. One track in the detector coordinate has different angles with
respect to each wire plane's coordinate.  

Compared to the collection plane signals, the induction plane signals
can be much smaller due to the cancellation of the various bipolar
signals from all nearby elements of an extended event topology. In
particular, for a track traveling in a direction close to normal to
the wire planes (commonly referred to as a \textit{prolonged track}),
its induction plane signals will have low amplitude and a long duration in time
(figure~\ref{fig:sim_perpendicularmip}). This amplitude can be comparable to
noise levels. Having the lowest achievable inherent electronics noise~\cite{noise_filter_paper},
avoiding excess noise sources, and applying proper signal processing are
crucial to resolve the induction plane signals. Recovering these signals enables new opportunities to take full advantage of the LArTPC's capability and reduce its residual ambiguities in later reconstruction. In order to
minimize the inherent electronics noise, MicroBooNE uses a custom designed
complementary metal-oxide-semiconductor (CMOS) analog front-end
application-specific integrated circuit (ASIC)~\cite{asic}
operating at cryogenic temperatures inside the liquid argon. The close proximity of the preamplifier
to the sense wire minimizes the input capacitance. 
The low temperature of LAr further reduces the electronics noise
of the ASIC. The residual equivalent noise charge (ENC) after the
software noise filtering varies with wire length and is found to be around 400
electrons for the longest wires (4.7 m) in MicroBooNE. This noise level is
significantly lower than previous experiments utilizing warm front-end
electronics. More details can be found in~\cite{noise_filter_paper}.

\section{Reconstruction of drifted electron distribution}~\label{sec:charge_extract}
In this section, we describe the principle for reconstructing the
charge distribution of drifted ionization electrons (section~\ref{sec:principle})
from the measured TPC signal waveform and the actual implementation of algorithms
applied in analyzing MicroBooNE data (section~\ref{sec:method}).

\subsection{ Signal extraction principles}\label{sec:principle}

As described in section~\ref{sec:signal_form}, the raw digitized TPC signal is a
convolution of the arriving ionization electron distribution, the field response
describing the induced current on wires due to moving charge, and the overall
electronics response. The goal of TPC charge extraction is to unfold the field
and electronics responses from the raw TPC signal and to recover the number of
ionization electrons passing by each wire at each sampled time. 
  
\subsubsection{Deconvolution and software filters}\label{sec:deconvolution}

The principal method to extract charge is deconvolution.
This procedure in its one-dimensional (1D) form has been used
in the data analysis of previous liquid argon
experiments~\cite{icarus_sulej, Baller:2017ugz}. This technique
has the advantages of being robust and fast. It is an essential step in the
overall drifted-charge profiling process.

Deconvolution is a mathematical technique to extract the \textit{original signal}
$S(t)$ from the \textit{measured signal} $M(t')$.  The measured signal is modeled
as a convolution integral over the original signal $S(t)$ and a given detector
\textit{response function} $R(t,t')$, which gives the instantaneous portion of
the measured signal at some time $t'$ due to an element of original signal at
time $t$:
\begin{equation}\label{eq:decon_1}
  M(t') = \int_{-\infty}^{\infty}  R(t,t') \cdot S(t) dt.
\end{equation}
If the detector response function is time-invariant, then $R(t,t') \equiv R(t'-t)$
and we can solve the above equation by performing a Fourier transformation, yielding
$M(\omega) = R(\omega) \cdot S(\omega)$, where $\omega$ is in units of
angular frequency. We can derive the signal in the frequency domain by taking the
ratio of the measured signal and the given response function:
\begin{equation}\label{eq:decon_2}
  S(\omega) = \frac{M(\omega)}{R(\omega)}.
\end{equation}

In principle, the original signal in the time domain $S(t)$ can then be
obtained by applying the inverse Fourier transformation from the frequency
domain $S(\omega)$. However in practice, there are two intermixed complicating
effects.  First, the measured signal $M(t')$ contains an additional contribution
from various electronics noise sources~\cite{noise_filter_paper}.
Second, in realistic detectors, the response function $R$ decreases
substantially at high frequency (large $\omega$).  These two factors
lead to high-frequency components of the noise spectrum being artificially
amplified through equation~\ref{eq:decon_2}. If left unchecked, the derived
signal $S(\omega)$ would be completely overwhelmed by noise.  To address
this issue, a \textit{filter function} $F(\omega)$ is introduced, yielding
\begin{equation}\label{eq:decon_filt}
  S(\omega) = \frac{M(\omega)}{R(\omega)} \cdot F(\omega).
\end{equation}
Its purpose is to attenuate the problematic high frequency noise. 
The addition of the filter function $F$ can be considered as a replacement
of the response function $R$. In essence, the deconvolution replaces the real
field and electronics response function with an effective software filter
response function. The advantage of this procedure is most pronounced on
the induction plane where the irregular bipolar field response function is
replaced by a regular unipolar response function through the inclusion of
the software filter.

A common choice of software filter is the Wiener filter\footnote{A discussion of the application of the Wiener filter in the data
  unfolding problem can be found in~\cite{Tang:2017rob}.}~\cite{wiener},
which is based on the expected signal $\overline{S^2(\omega)}$ and noise
$\overline{N^2(\omega)}$ frequency spectra:
\begin{equation}\label{eq:wiener}
  F(\omega) = \frac{\overline{R^2(\omega)S^2(\omega)}}{\overline{R^2(\omega)S^2(\omega)} + \overline{N^2(\omega)}}.
\end{equation}
With this construction, the Wiener filter is expected to achieve the best signal to
noise ratio with minimal mean square error (the sum of the variance and the
squared bias) of the deconvolved distribution. However, naively applying the
Wiener filter to TPC signal processing is problematic for three reasons. Firstly,
as described in section~\ref{sec:topology_signal}, the TPC signal $S(\omega)$ varies
substantially depending on the exact nature of the event topology.  The
electronics noise spectrum is also a function of the duration of the time window
over which it is observed. A longer time window allows for observation of more low
frequency noise components. Therefore, it is impractical to achieve a universal
Wiener filter yielding the best signal-to-noise ratio  for all signals of varying time
windows.  Secondly, given the definition of the Wiener filter in
equation~\ref{eq:wiener}, we have $F(\omega = 0)<1$. Considering the addition
of the filter as a replacement of the response function $R(t'-t)$, we can see
that the Wiener filter does not conserve the total number of ionization electrons. Thirdly, as shown in
equation~\ref{eq:decon_filt}, the filter acts as a
replaced response function and smears the extracted ionization electron
distribution along the drift time dimension. Since the drift time is equivalent to
the drift distance, a filter that can alter the charge distribution in an extended (non-local) time range instead of in a short (local) one is undesirable. For induction wire planes, none of the
ionization electrons are expected to be collected, which leads to a bipolar
signal in the time domain and a low-frequency suppressed signal in the frequency
domain. A direct construction of the Wiener filter with this low-frequency
suppression would lead to a non-local charge smearing. To overcome these
shortcomings associated with the Wiener filter, we use a Wiener-inspired filter.
Details are elaborated upon in section~\ref{sec:method_filter}.
  
\subsubsection{2D deconvolution}\label{sec:2D_deconvolution}

As described in section~\ref{sec:signal_form}, the induced current on the sense wire
receives contributions not only from ionization charge passing by the sense wire,
but also from ionization charge drifting in nearby wire regions. Naturally, the
contribution of charge farther from the target sense wire is
smaller. Ignoring the variation of the strength of the field response within one
wire region, equation~\eqref{eq:decon_1} can be expanded to
\begin{equation}\label{eq:decon_2d_1}
    M_i(t_0) = \int_{-\infty}^{\infty} \left( ... + R_1(t_0-t)\cdot S_{i-1}(t) + R_0(t_0-t) \cdot S_i(t) + 
    R_1(t_0-t) \cdot S_{i+1} (t) + ...\right) \cdot dt,
\end{equation}
where $M_i$ represents the measured signal from wire $i$.  $S_{i-1}$, $S_i$, and
$S_{i+1}$ represent the real signal inside the boundaries of wire $i$
and its adjacent neighbors. $R_0$ represents an average
response function for an ionization charge passing through the wire
region of interest.  The average is taken over all possible drift paths
through the wire region. Similarly, $R_1$ represents
the average response function for an ionization charge drifting past
in the next adjacent wire region.  One can expand this
definition to $n$ number of neighbors by introducing terms up to $R_n$.
Equation~\eqref{eq:decon_2d_1} assumes translational invariance in
the response function (i.e. the $R$ does not depend on the actual
location of the wire).  In section~\ref{sec:MCtruth}, we will discuss
the impact of ignoring position dependence of the field response at small
scales within the wire region of interest. 

If we then apply a Fourier transform on both sides of equation~\eqref{eq:decon_2d_1},
we have:
\begin{equation}\label{eq:decon_2d_2}
    M_i(\omega) = ... + R_1(\omega) \cdot S_{i-1}(\omega) + R_0(\omega) \cdot S_i(\omega) + R_1(\omega) \cdot S_{i+1} (\omega) + ...,
\end{equation} 
which can be written in matrix notation as:
\begin{equation}
  \begin{pmatrix}
    M_1(\omega)\\
    M_2(\omega)\\
    \vdots\\
    M_{n-1}(\omega)\\
    M_{n}(\omega)
  \end{pmatrix}
  =
  \begin{pmatrix}
    R_0(\omega) & R_1(\omega) & \ldots & R_{n-2}(\omega) & R_{n-1}(\omega) \\
    R_1(\omega) & R_0(\omega) & \ldots & R_{n-3}(\omega) & R_{n-2}(\omega) \\
    \vdots  & \vdots      & \ddots & \vdots          & \vdots \\
    R_{n-2}(\omega) & R_{n-3}(\omega) & \ldots & R_0(\omega) & R_1(\omega) \\
    R_{n-1}(\omega) & R_{n-2}(\omega) & \ldots & R_1(\omega) & R_0(\omega) \\
  \end{pmatrix}
  \cdot
  \begin{pmatrix}
    S_1(\omega)\\
    S_2(\omega)\\
    \vdots\\
    S_{n-1}(\omega)\\
    S_{n}(\omega)
  \end{pmatrix}
  \label{eq:matrix_expansion}
\end{equation}
Now, if we assume that we know all response functions (i.e. the matrix $R$), the 
problem converts into deducing the vector $S$ from the measured signal $M$ by an inversion of matrix $R$, 
provided that the wires of interest are distant from the wire plane edges, the matrix $R$ is symmetric and Toeplitz\footnote{A Toeplitz matrix is one for which
  each diagonal descending from left to right has all elements equal.
  Multiplication by a Toeplitz matrix is equivalent to an operation of
  discrete convolution.}, and the inverse problem can be solved using discrete-space Fourier
transformation techniques as discussed in section~\ref{sec:deconvolution}, on
$M_i(\omega)$, $S_i(\omega)$, and $R_i(\omega)$. Therefore, instead of 1D a
deconvolution involving only the time dimension, a two-dimensional (2D) deconvolution involving both
time and wire dimensions is performed to recover the ionization electron
distribution. An additional Wiener-inspired filter is applied to the wire
dimension deconvolution in analogy to that of the time domain deconvolution.
These software filters will be discussed in section~\ref{sec:method_filter}.

\begin{figure}[htb]
  \centering
  \includegraphics[width=0.99\textwidth]{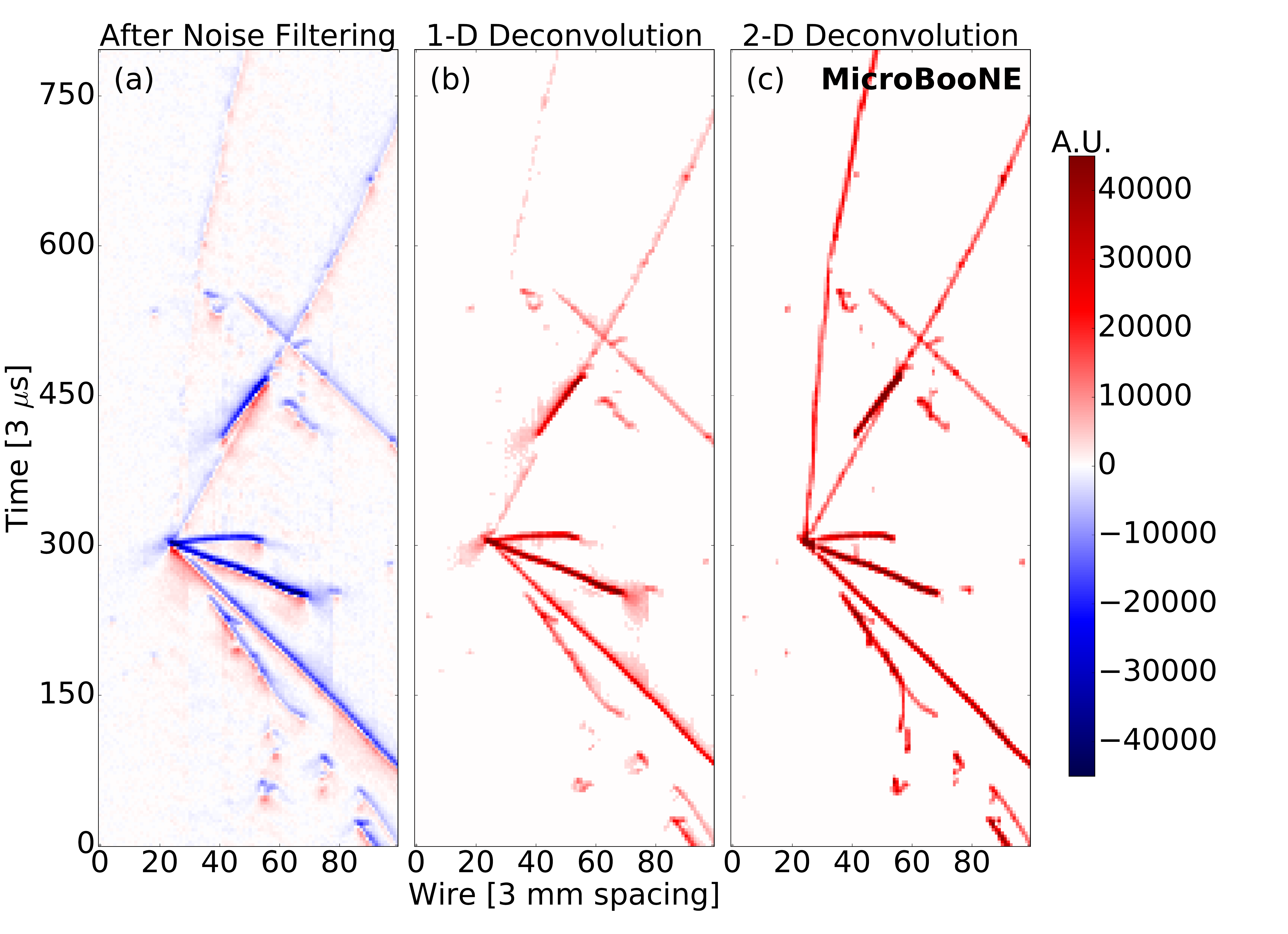}
  \caption{ A neutrino candidate from MicroBooNE data
    (event 41075, run 3493) measured on the U plane. (a) Raw waveform after
    noise filtering in units of average baseline subtracted ADC scaled by 250
    per \SI{3}{\us}. (b) Charge spectrum in units of electrons per
    \SI{3}{\us} after signal processing with 1D deconvolution. (c) Charge
    spectrum in units of electrons per \SI{3}{\us} after signal processing
    with 2D deconvolution.}
  \label{fig:1dvs2d}
\end{figure}

An example comparison of the 1D and 2D deconvolution results in a data event can be seen in figure~\ref{fig:1dvs2d}, demonstrating the charge recovery
using the 2D deconvolution approach in contrast to the 1D method. Figure~\ref{fig:1dvs2d}
highlights the signal dependence on topology and long range
induction. An evaluation of 2D deconvolution signal processing performance
given the topological dependence of the signal and the long range
induction inherent to the signal is considered in section~\ref{sec:evaluation}.

\subsubsection{Region of interest (ROI)}

\begin{figure}[htb]
  \centering
    \begin{subfigure}[t]{0.48\textwidth}
        \includegraphics[width=1.0\textwidth]{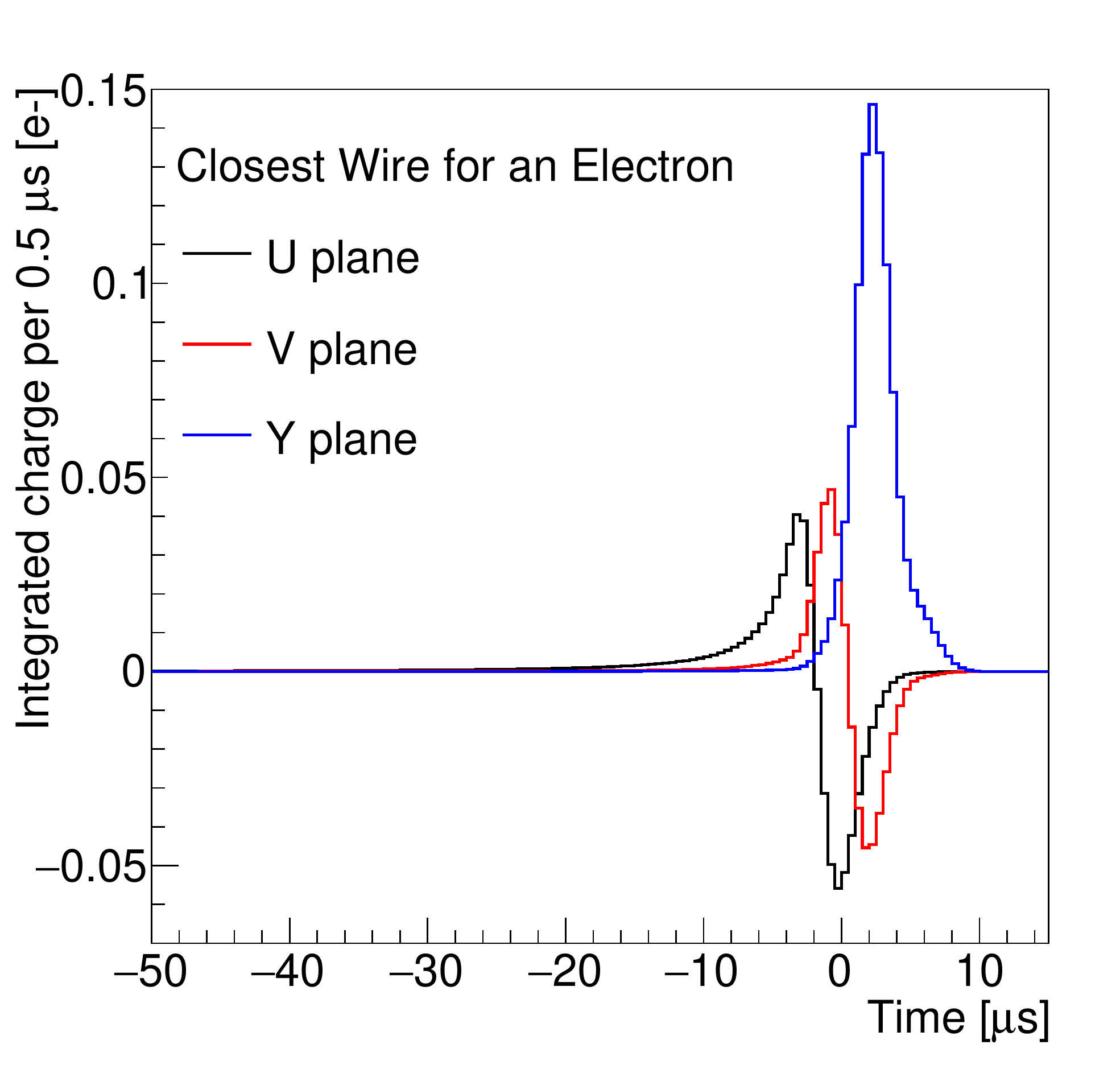}
        \caption{Average response in time domain.}
        \label{fig:induction_field_a}
    \end{subfigure}
    \begin{subfigure}[t]{0.48\textwidth}
        \includegraphics[width=1.0\textwidth]{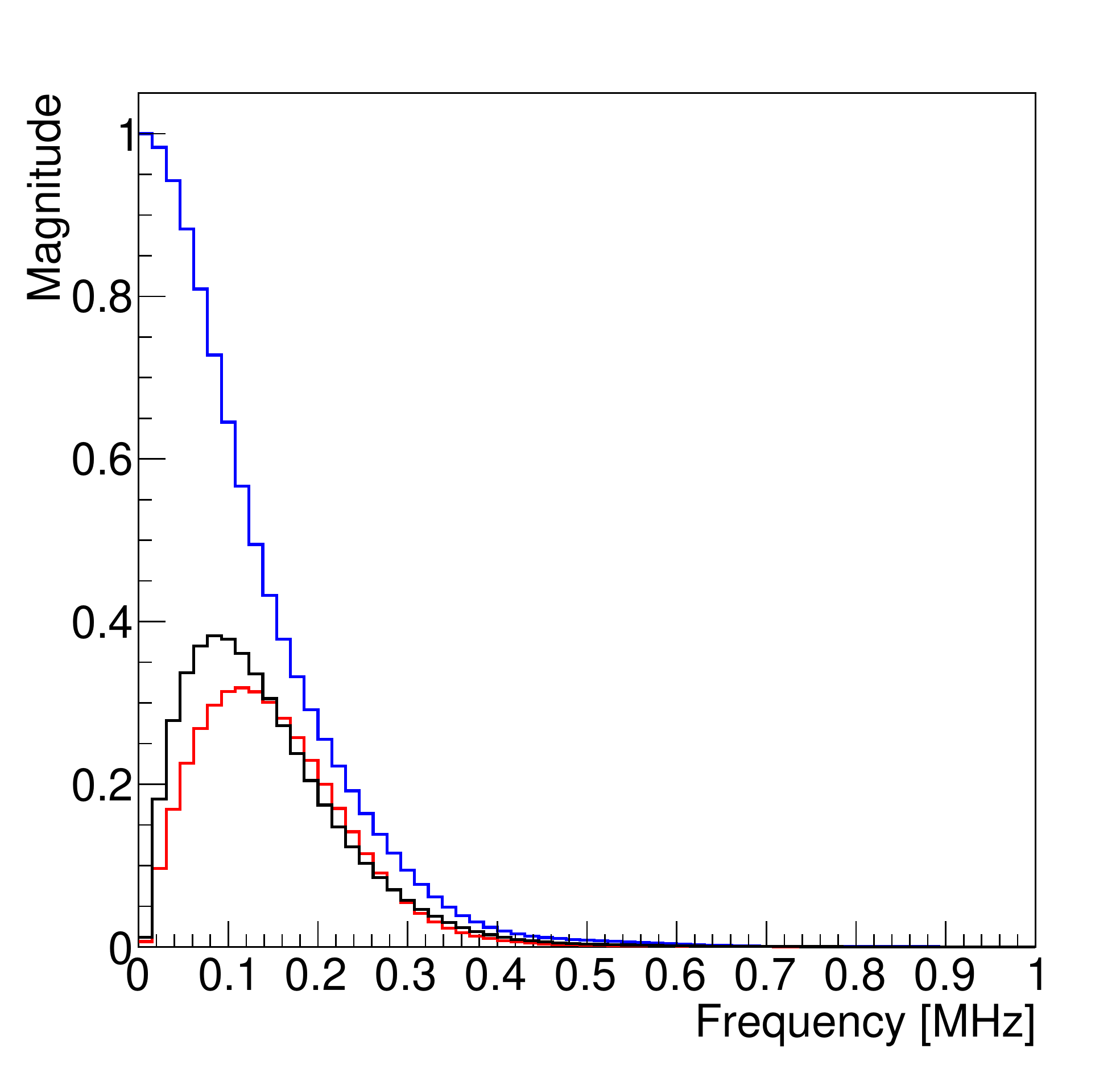}
        \caption{Average response in frequency domain.}
        \label{fig:induction_field_b}
    \end{subfigure}
    \caption{Examples of simulated average response functions for
      induction (black and red) and collection (blue) wires in the time (a)
      and frequency (b) domains.}
  \label{fig:induction_field}
\end{figure}

The 2D deconvolution procedure described in section~\ref{sec:2D_deconvolution}
provides a robust and computationally-efficient method to extract the distribution of
ionization electrons. While successful for the collection plane the procedure is
still not optimal for the induction planes due to the bipolar nature of the measured
induction plane signals. In order to illustrate this point, the average response functions on the
closest wire for a point-like ionization charge is shown in figure~\ref{fig:induction_field}.
The average response function includes both the average field response (averaged over all possible
electron drift paths within the wire region as simulated by Garfield without diffusion) and
the electronics response (\SI{2}{\us} peaking time). The normalization of the overall
response function is chosen so that the integral of the collection plane response function is unity, corresponding to a single electron. Figure~\ref{fig:induction_field_b}
shows the frequency components of the average response functions for the three wire planes. All responses have suppressions at
high frequency, where the filter is required to stabilize the deconvolved results
(e.g. equation~\ref{eq:decon_filt}). Compared to the collection wire response, the induction
wire responses exhibit suppressions at low frequency. In particular, at zero frequency,
the frequency components are equivalent to the integral of the response function over
time and should be close to zero as indicated by equation~\ref{eq:voltage}.

Similar to the situation at high frequencies, the suppression of the induction responses
at low frequencies is problematic for the proposed deconvolution procedure. As shown in~\cite{noise_filter_paper}, the  measured signal contains electronics noise, which is not
necessarily as suppressed at low frequencies. Therefore, following
equation~\eqref{eq:decon_filt}, the low frequency noise will be amplified in the deconvolution
process. The amplification of low frequency noise can be seen clearly in figure~\ref{figs:ROI_example_a}. 
Left unmitigated, the amplification of low frequency noise would lead to  an unacceptable uncertainty in the charge estimation. 


In principle, the amplification of the low-frequency noise through the deconvolution process
can be suppressed through the application of low-frequency (high-pass) filters
similar to  the filters suppressing high-frequency (low-pass) noise. However, as explained
in section~\ref{sec:deconvolution}, applying such a low-frequency filter would lead
to an alteration of the charge distribution in extended (non-local) time ranges, which is
not desirable. Instead we turn to the technique of selecting a signal region
of interest (ROI) in the time domain.  


The region of interest (ROI) technique was proposed~\cite{Baller:2017ugz} to reduce the
data size and to speed up the deconvolution process. The idea is to limit the deconvolution
to a small time window that is slightly bigger than the extent of the signal it contains.
The entire event readout window (4.8 ms for MicroBooNE) is replaced by a set of
ROIs. For induction wire signals, the ROI technique also limits the low frequency noise.
To illustrate this point, we consider a time window with $N$ samples. MicroBooNE samples at 
 intervals of \SI{0.5}{\us} and the highest frequency that can be resolved is the
Nyquist frequency of \SI{1}{\MHz}.  After a discrete Fourier transform, the bin above the
first bin (zero frequency) starts at $\frac{\SI{1}{\MHz}}{N/2}$. The noise in the zero-frequency bin represents a baseline shift after the ROI is transformed back into
the time domain.  Therefore, once we identify the signal region and create a ROI
just big enough to cover the signal, we can naturally suppress the low-frequency noise 
at the cost of having to correct for the baseline shift. This correction is performed
through a linear interpolation of the two baselines, determined by samples at the start and 
end of the ROI window in the time domain. 

\subsection{Method}\label{sec:method}

In this section, we describe the inputs and the detailed algorithms to extract
the ionization charge spectrum from the digitized TPC wire plane signals,
based on the principles described above. The algorithm is implemented
and available at~\cite{WCT_SP}.

In general, the full chain contains four major steps:
\begin{itemize}
\item {\bf Noise filtering: } \\
  Apply specific noise filters to remove possible excess
  noise apart from the inherent electronics noise. The results of this
  step have been previously reported in~\cite{noise_filter_paper}.
\item {\bf 2D deconvolution:} \\
  Apply a 2D deconvolution to the digitized TPC wire signals, resulting
  in a deconvolved charge distribution. An average field response involving
  multiple sense wires is calculated and utilized as shown in
  section~\ref{sec:response_function}.
  Two types of deconvolved charge spectra are obtained, corresponding to
  different software filters for the time domain. A Wiener-inspired filter is applied to
  maximize the signal-to-noise ratio with better time resolution
  and a Gaussian filter is applied to achieve a non-distorted charge
  spectrum except for a Gaussian smearing. Details will be explained in section~\ref{sec:method_filter}.
\item {\bf ROI finding and refining:} \\
  Perform ROI finding with the deconvolved charge distribution after
  the Wiener-inspired filter. The principle of these algorithms will be
  explained in section~\ref{sec:finding_ROIs} and ~\ref{sec:refining_ROIs}.
  In short, loose and tight high-pass
  filters are combined to optimize the purity and efficiency of ROI finding.
\item {\bf ROI application:} \\
  Apply the identified ROI window to the deconvolved charge distribution after
  the Gaussian filter and extract the ionization charge, with a linear baseline
  subtraction for the induction planes based on the start/end bins of the ROI
  window.
\end{itemize}

\begin{figure}
    \centering
    \includegraphics[width=0.95\textwidth]{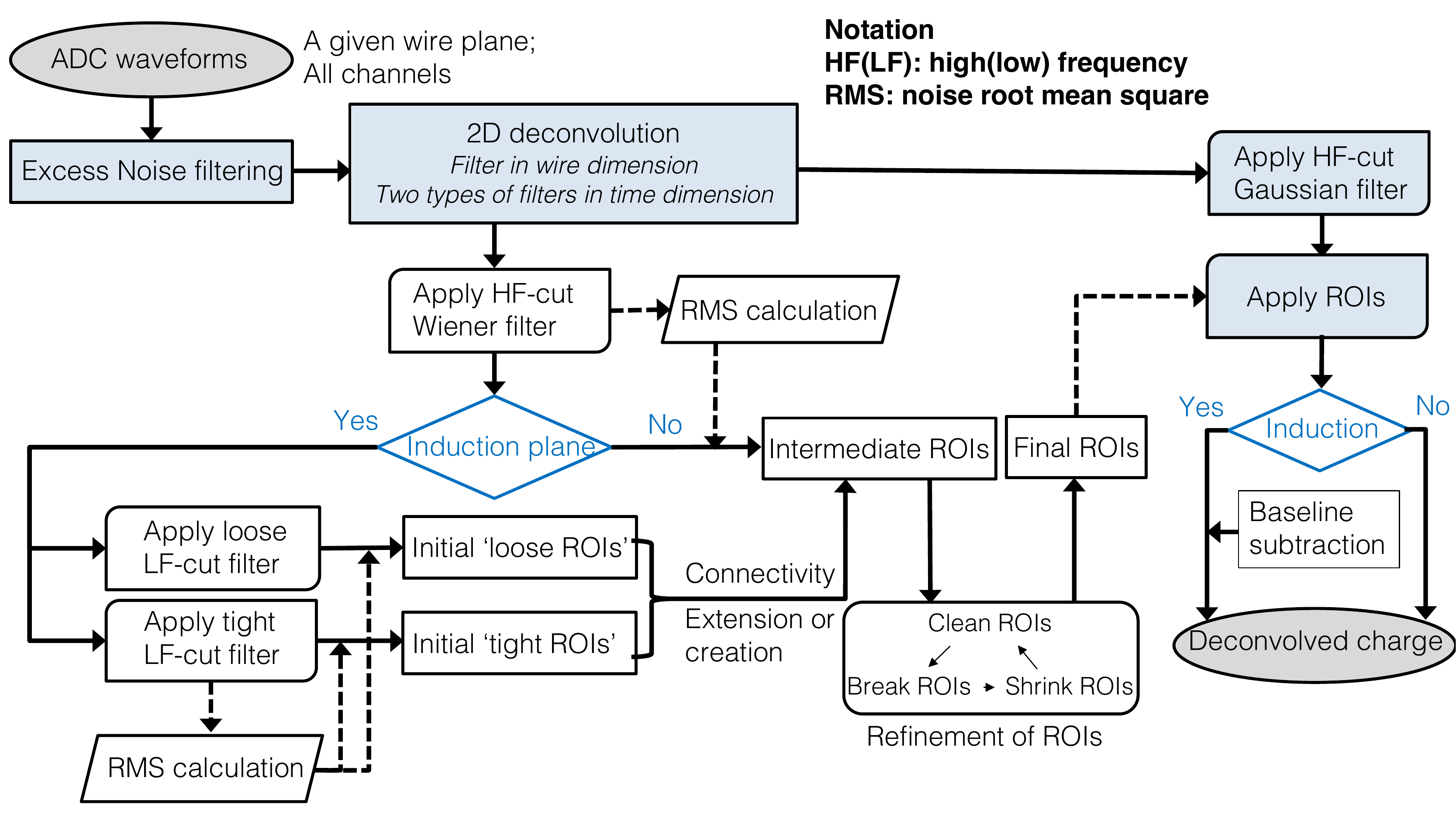}
    \caption{A flow chart of the full chain of signal processing. See text for explanation.}
    \label{fig:SPdiagram}
\end{figure}

A flow chart of the full chain of the aforementioned signal processing can be seen in figure~\ref{fig:SPdiagram}.



\subsubsection{Position-averaged response functions}\label{sec:response_function}
\begin{figure}[!htbp]
  \begin{subfigure}[t]{0.48\textwidth}
    \includegraphics[width=1.0\textwidth]{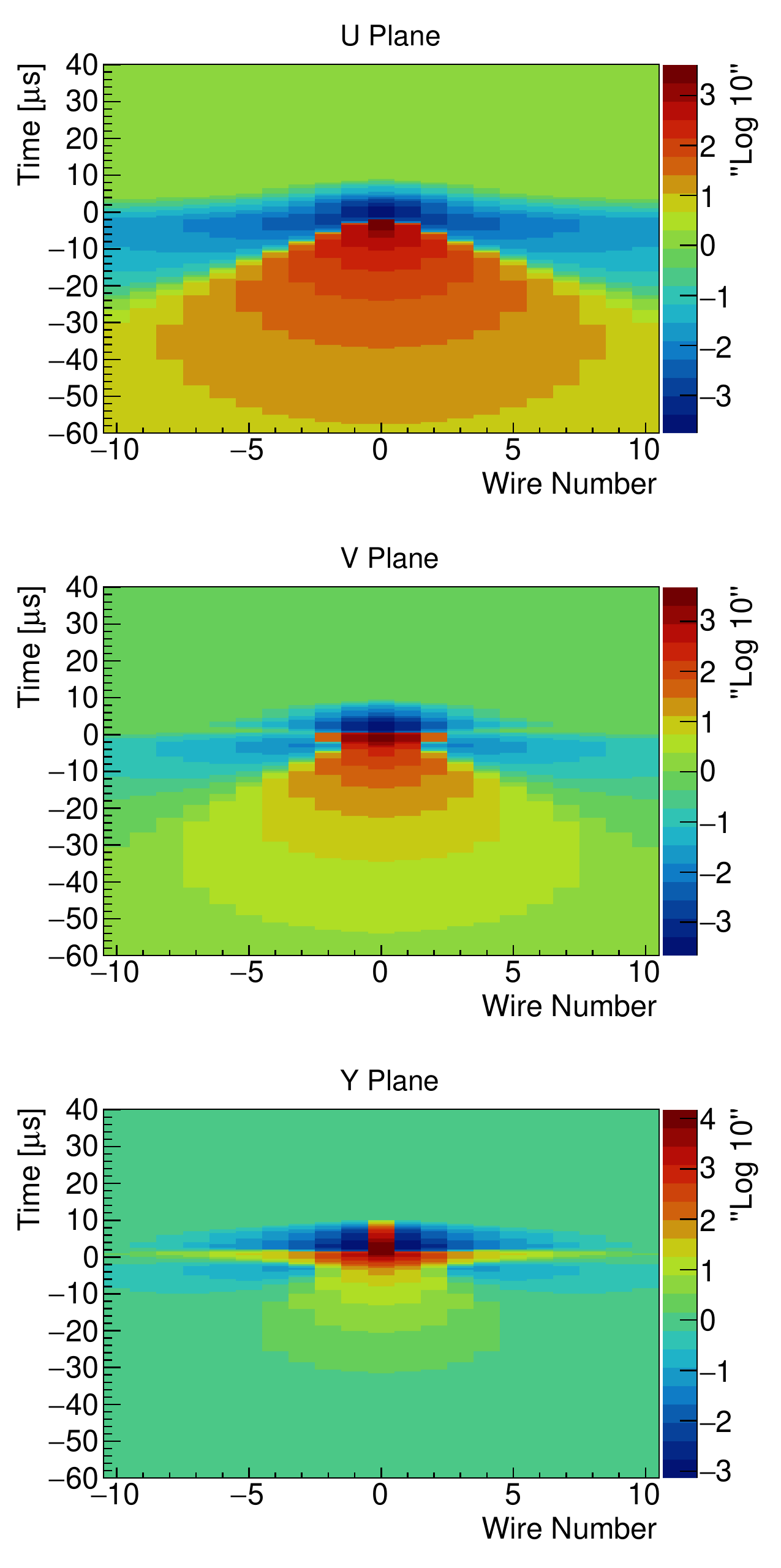}
    \caption{Wire number versus time in ``Log10'' scale.}
    \label{figs:decon_overall_response_a}
  \end{subfigure}
  \begin{subfigure}[t]{0.48\textwidth}
    \includegraphics[width=1.0\textwidth]{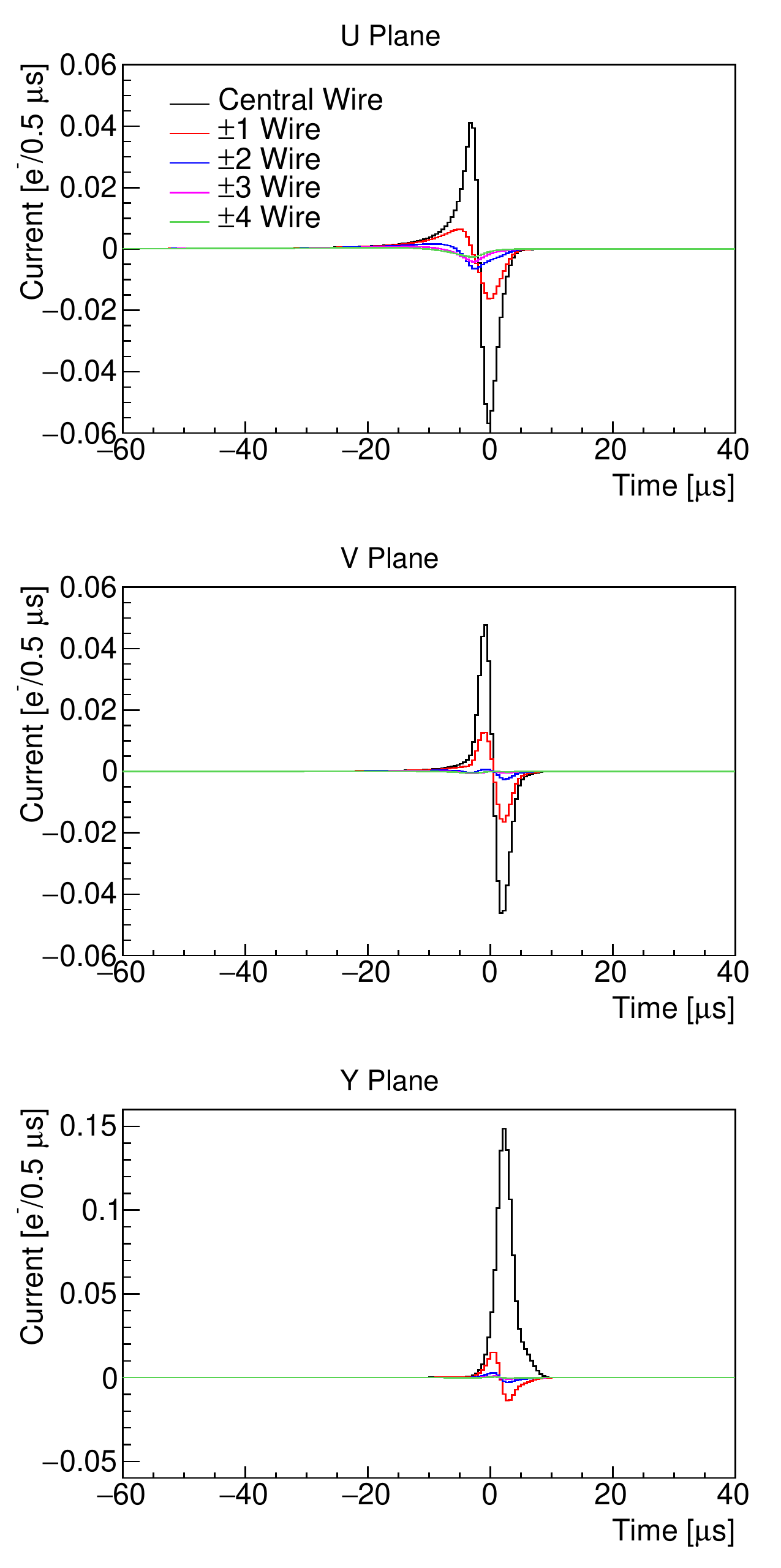}
    \caption{Response functions for each of the first five wires.}
    \label{figs:decon_overall_response_b}
  \end{subfigure}
  \caption{The position-averaged response functions after convolving the field response
    function and an electronics response function at \SI{2}{\us} peaking time in ``Log 10'' scale (a) and linear scale for the first 5 wires (b).
  }
  \label{figs:decon_overall_response}
  \end{figure}

Since the signal in a readout wire is recorded without any prior knowledge of the transverse
position distribution of electrons within a wire region, the average response functions
are used in the 2D deconvolution as discussed in section~\ref{sec:2D_deconvolution}.
The average response function for a single electron within a particular wire region is
obtained through the following summation:
\begin{equation}
  R_i=\frac{0.5\times R^{0.0~mm}_i + R^{0.3~mm}_i + R^{0.6~mm}_i + R^{0.9~mm}_i + R^{1.2~mm}_i + 0.5\times R^{1.5~mm}_i}{5},
\end{equation}
where $R_i^{z}$ represents the response function at the $i$th wire for an electron
starting at $z$ transverse position. The impact of applying the average response function in
the 2D deconvolution will be explained in section~\ref{sec:MCtruth}.
These average responses for 21 wires (the central wire $\pm$ 10 wires on both sides) are shown in figure~\ref{figs:decon_overall_response}.
The left panel shows the response function in 2D with the ``Log10'' scale. The X-axis
represents the wire number. The Y-axis represents the drift time with 1 tick of
\SI{0.5}{\us} in each bin. The normalization is the same as that in
figure~\ref{figs:overall_response}.
The right panel shows the average response function in a linear scale for the first
five wires. For the V and Y wire planes, the strength of the response function drops quickly
for wires further away from the central wire (i.e. negligible beyond \num{\pm4} wires).
For the U induction wire plane, the strength of the response function is still sizable at
\num{\pm4} wires. This is due to the fact that the U induction wire plane is
the first wire plane facing the active TPC volume without any shielding.

\subsubsection{Software filters}\label{sec:method_filter}

\begin{figure}[!htbp]
    \begin{subfigure}[t]{0.48\textwidth}
        \includegraphics[width=1.0\textwidth]{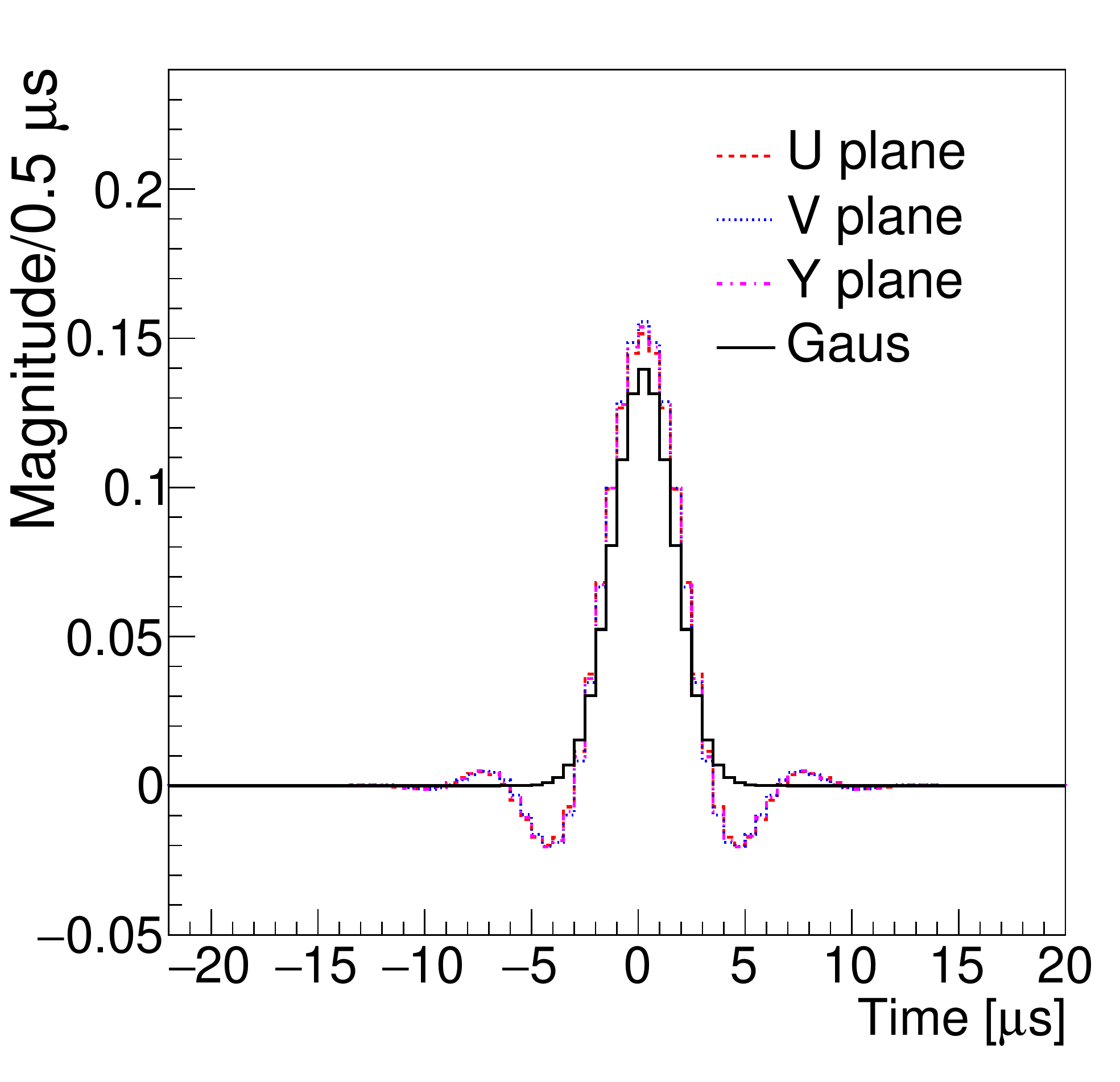}
        \caption{Filters in the time domain.}
        \label{figs:filter_hf_a}
    \end{subfigure}
    \begin{subfigure}[t]{0.48\textwidth}
        \includegraphics[width=1.0\textwidth]{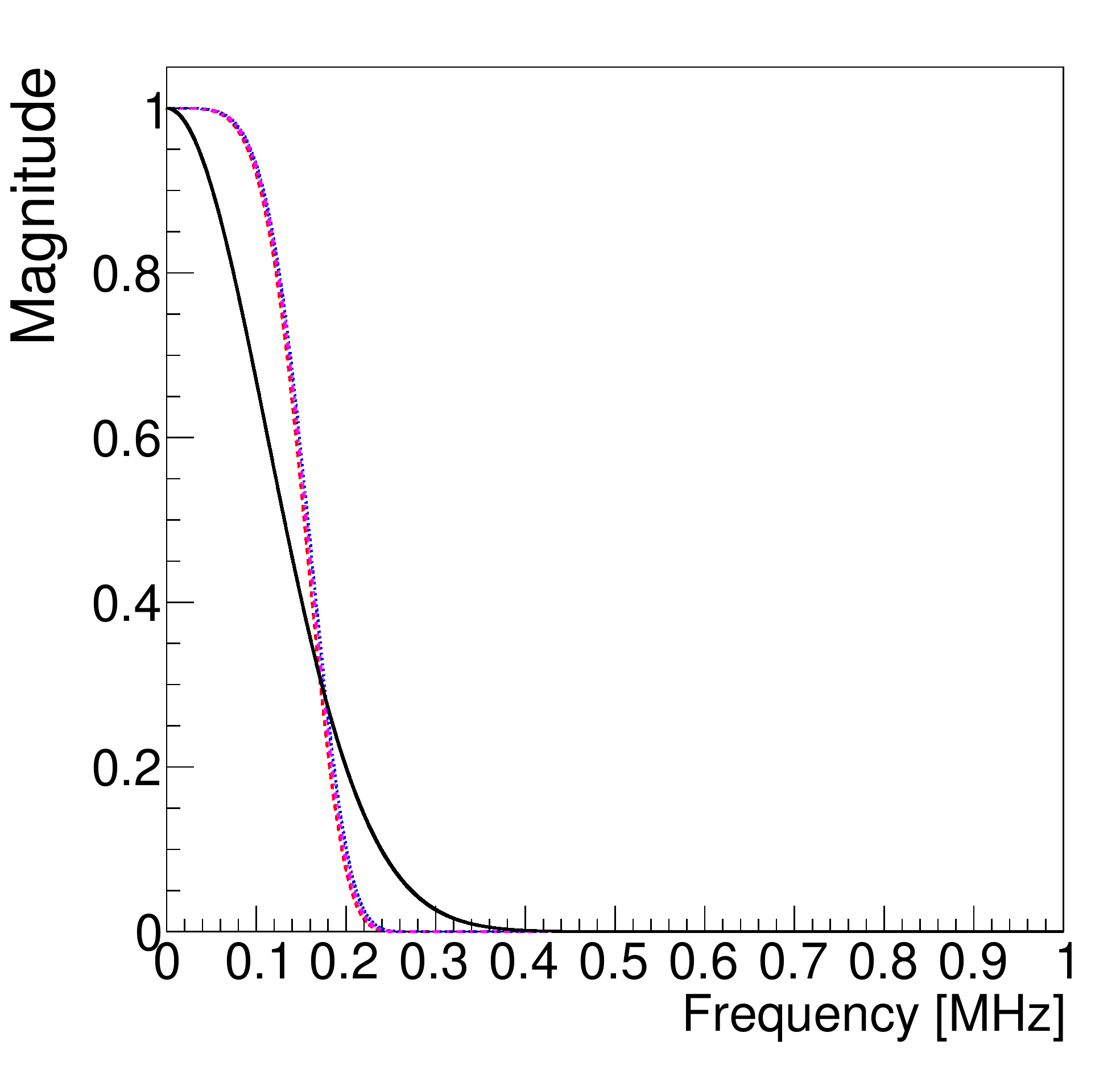}
        \caption{Filters in the frequency domain.}
        \label{figs:filter_hf_b}
    \end{subfigure}
    \caption{Wiener-inspired (dashed line) and Gaussian (solid line) software filters 
      are shown in (a) the time domain and (b) the frequency domain for each
      of the three wire planes. In the time domain, we commonly
      refer to the filter function as the smearing function. With the chosen
      functional forms, the local behavior of the filters is apparent from
      their shapes.
  }
  \label{figs:filter_hf}
\end{figure}

In this section, we describe the derivation of the software filters
used in the signal processing.  As mentioned in section~\ref{sec:deconvolution},
the chosen Wiener-inspired filter is based on the Wiener filter in
equation~\ref{eq:wiener} constructed from simulation. The time window is
chosen to be 100 $\mu$s which generally performs well in a variety of cases
with proper additional smearing of the signal. 
The signal is chosen to be an isochronous MIP track traveling
perpendicular to the wire orientation. The total number of ionization electrons
is assumed to be $1.6\times10^4$ per wire pitch. The RMS of the spread in  the drift time
due to diffusion is taken to be 1 mm, corresponding to the average drift
distance in the MicroBooNE detector. The gain and peaking time are 14 mV/fC
and 2 $\mu$s, which is the same as the nominal running condition~\cite{noise_filter_paper}. The electronic
noise is simulated in an analytic way, as will be described in
section~\ref{sec:noise_model}. The results following equation~\eqref{eq:wiener}
were then fitted with the following functional form in order to exclude
low-frequency suppressions presented in the original Wiener filters:
\begin{equation}\label{eq:fit}
  F(\omega) = c\cdot e^{- \frac{1}{2} \cdot \left( \frac{\omega}{a} \right)^b},       
\end{equation}
where $a$, $b$, $c$ are free parameters to be determined by the fit.
This functional form of the filter guarantees that the corresponding
smearing function (filter) in the time domain is local. Given the fit,
the filter is chosen to be
\begin{equation}
  F(\omega) = 
  \begin{cases}
    e^{- \frac{1}{2} \cdot \left( \frac{\omega}{a} \right)^b} &  \omega >0 \\
    0 &  \omega = 0, \\
  \end{cases}
\end{equation}
with $a$ and $b$ being the same parameters as in equation~\ref{eq:fit} with
$c$ removed. The modification of the filter takes into account the following
considerations:
\begin{itemize}
\item The filter is explicitly zero at $\omega = 0$ in order to remove any
  DC component in the deconvolved signal. This removes information about the
  baseline. A new baseline is calculated and restored for the waveform after
  deconvolution.
\item The above functional form of the filter leads to
  \begin{equation}
    \lim_{\omega \rightarrow 0} F(\omega) = 1.
  \end{equation}
  This means that the integral of the corresponding smearing function in the time
  domain is unity, which does not introduce any extra factor in the overall
  normalization.
\end{itemize}
Figure~\ref{figs:filter_hf} shows the three Wiener-inspired filters and a
Gaussian filter,
\begin{equation}
  F(\omega) = 
  \begin{cases}
    e^{- \frac{1}{2} \cdot \left( \frac{\omega}{a} \right)^2} &  \omega >0 \\
    0 &  \omega = 0, \\
  \end{cases}
\end{equation}
for charge extraction. Compared to the Wiener-inspired filters, the
Gaussian filter is expected to have a slightly worse signal-to-noise ratio. However,
such a filter has advantages in calculating the charge and is better matched with a
Gaussian hit finder that may be used in later stages of event reconstruction.

\begin{figure}[!htbp]
    \begin{subfigure}[t]{0.48\textwidth}
        \includegraphics[width=1.0\textwidth]{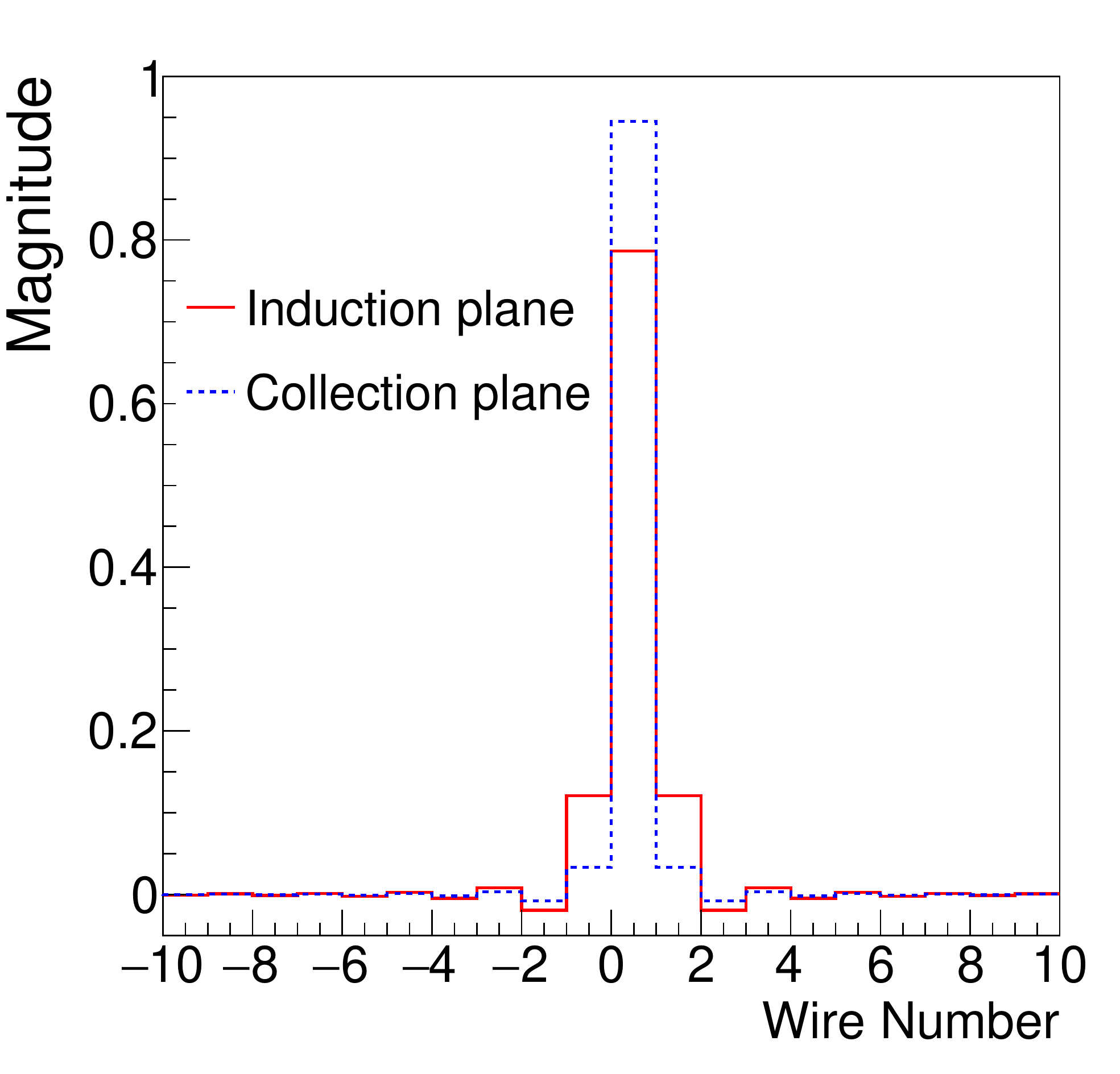}
        \caption{Wire filters in the wire number domain.}
        \label{figs:filter_wire_filter_a}
    \end{subfigure}
    \begin{subfigure}[t]{0.48\textwidth}
        \includegraphics[width=1.0\textwidth]{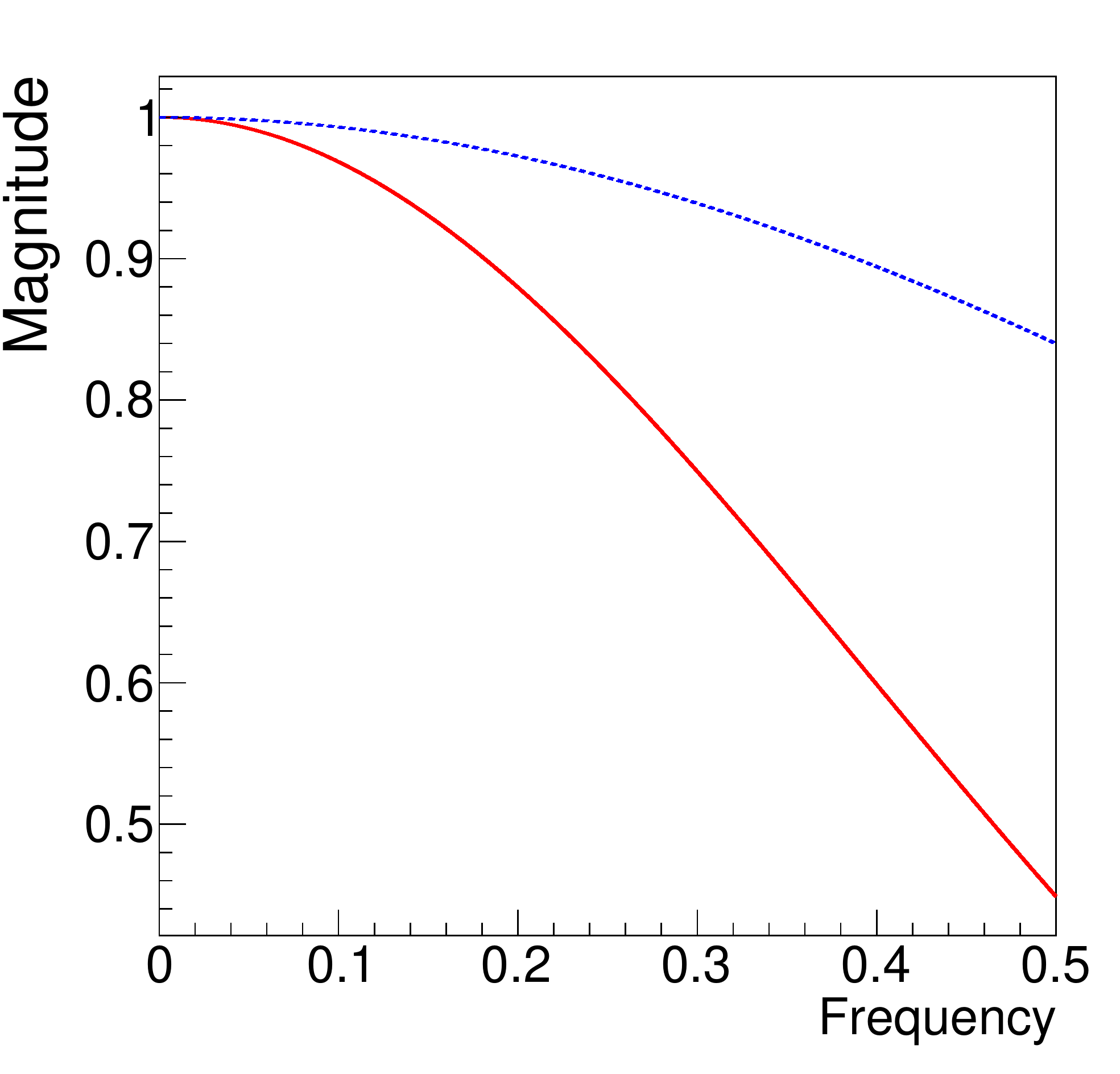}
        \caption{Wire filters in the ``frequency'' domain.}
        \label{figs:filter_wire_filter_b}
    \end{subfigure}
  \caption{ Filters using a Gaussian form are applied for the wire
    dimension shown in both the wire number (a) and ``frequency'' (b) domains.
    Different parameters are chosen for induction and collection
    wire planes. }
  \label{figs:filter_wire_filter}
\end{figure}

In 2D deconvolution, a similar filter in the wire dimension is constructed using
the Gaussian form
\begin{equation}
  F(\omega_w) = 
    e^{- \frac{1}{2} \cdot \left( \frac{\omega_w}{a} \right)^2}
\end{equation}
where the ``frequency'', $\omega_w$, is the Fourier transform over the wire number instead of time.
Figure~\ref{figs:filter_wire_filter} shows the filters in the
wire domain. Different parameters are chosen for
induction and collection wire planes. 

\subsubsection{Identification of signal ROIs}\label{sec:finding_ROIs}

\begin{figure}[!htbp]
    \centering
    \begin{subfigure}[]{0.49\textwidth}
    \centering
        \includegraphics[width=0.8\textwidth]{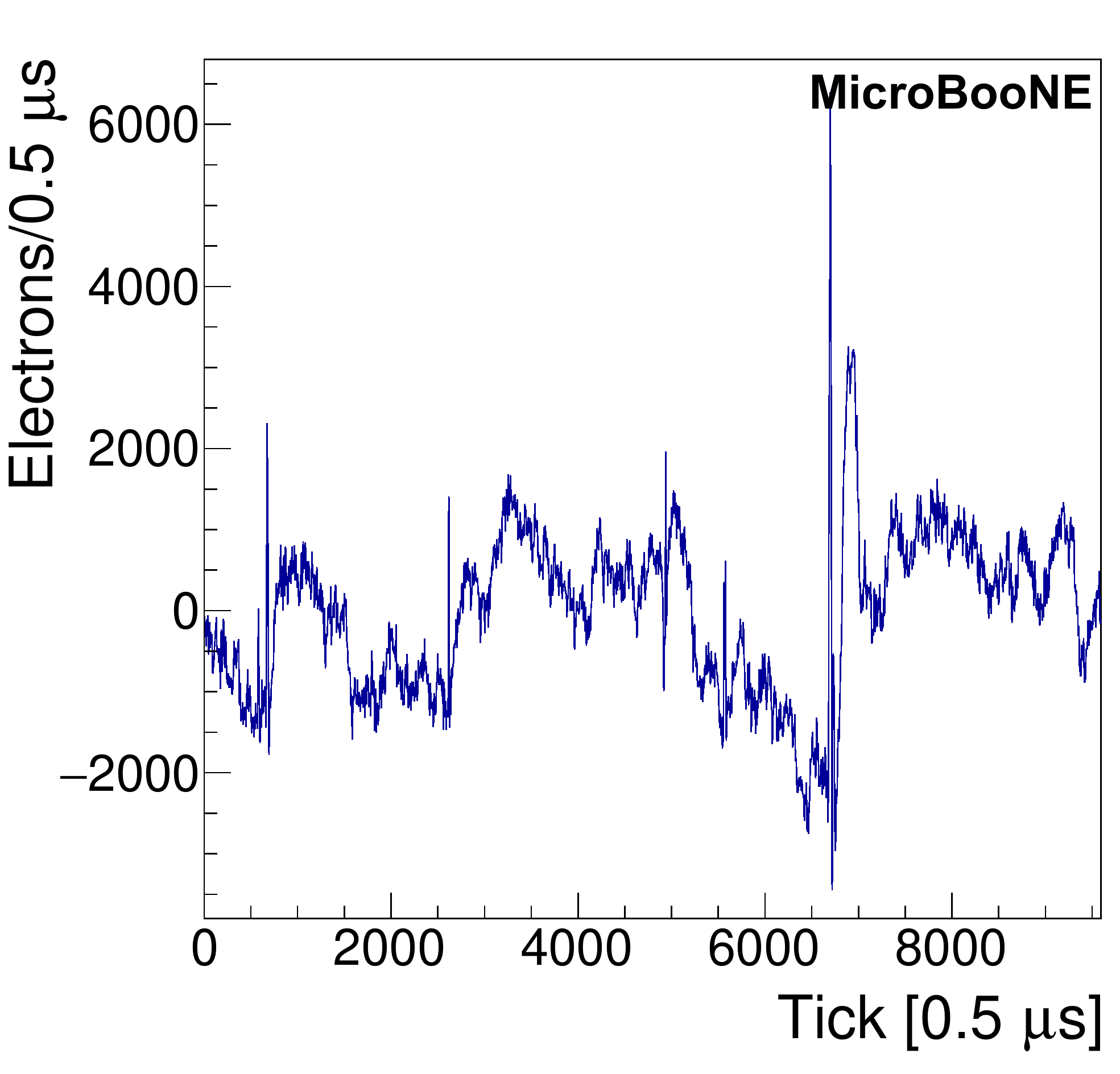}
        \caption{Without low-frequency filter.}
        \label{figs:ROI_example_a}
    \end{subfigure}
    \begin{subfigure}[]{0.49\textwidth}
    \centering
        \includegraphics[width=0.8\textwidth]{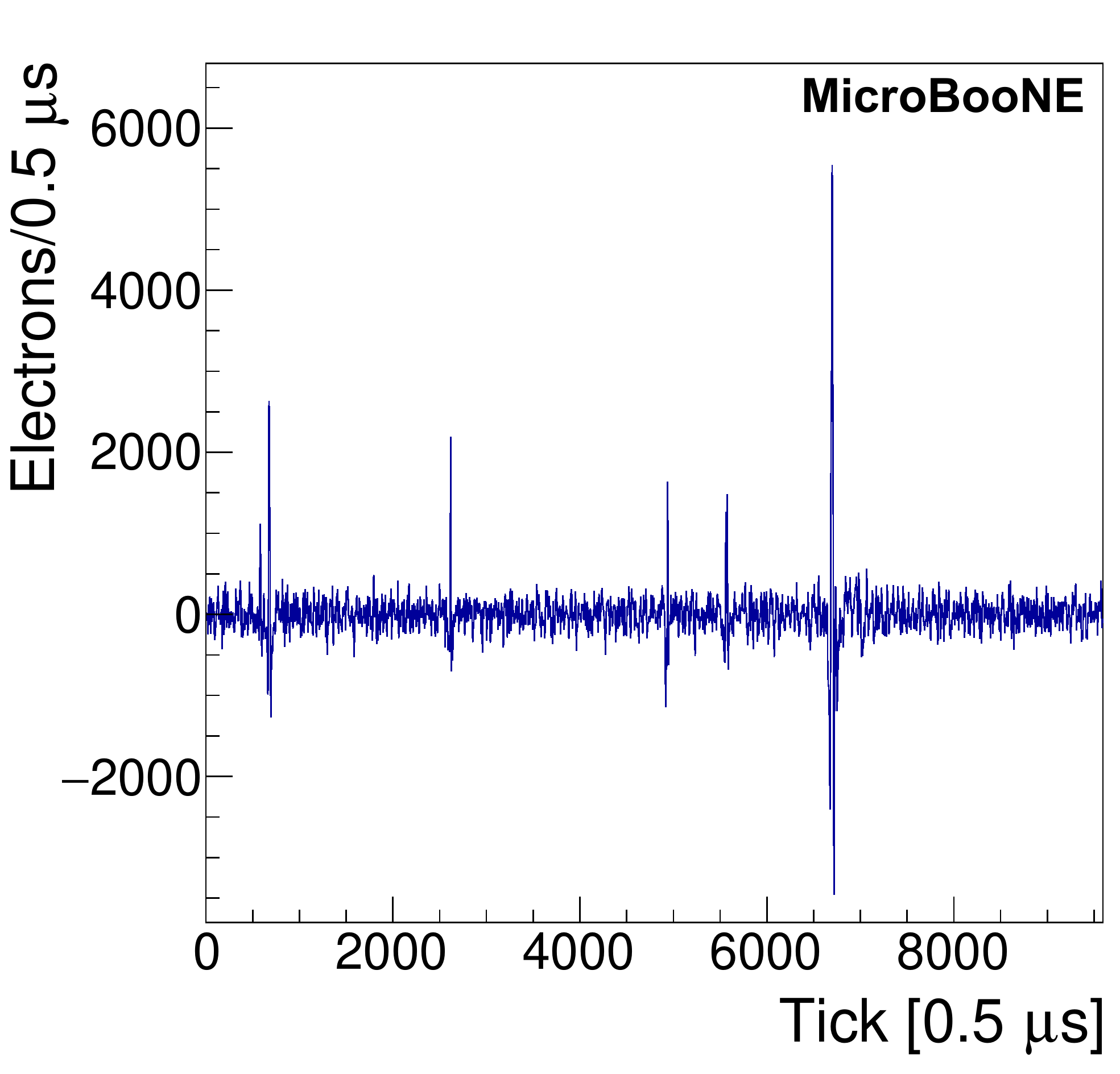}
        \caption{With tight low-frequency filter.}
        \label{figs:ROI_example_b}
    \end{subfigure}
    \begin{subfigure}[]{0.49\textwidth}
    \centering
        \includegraphics[width=0.8\textwidth]{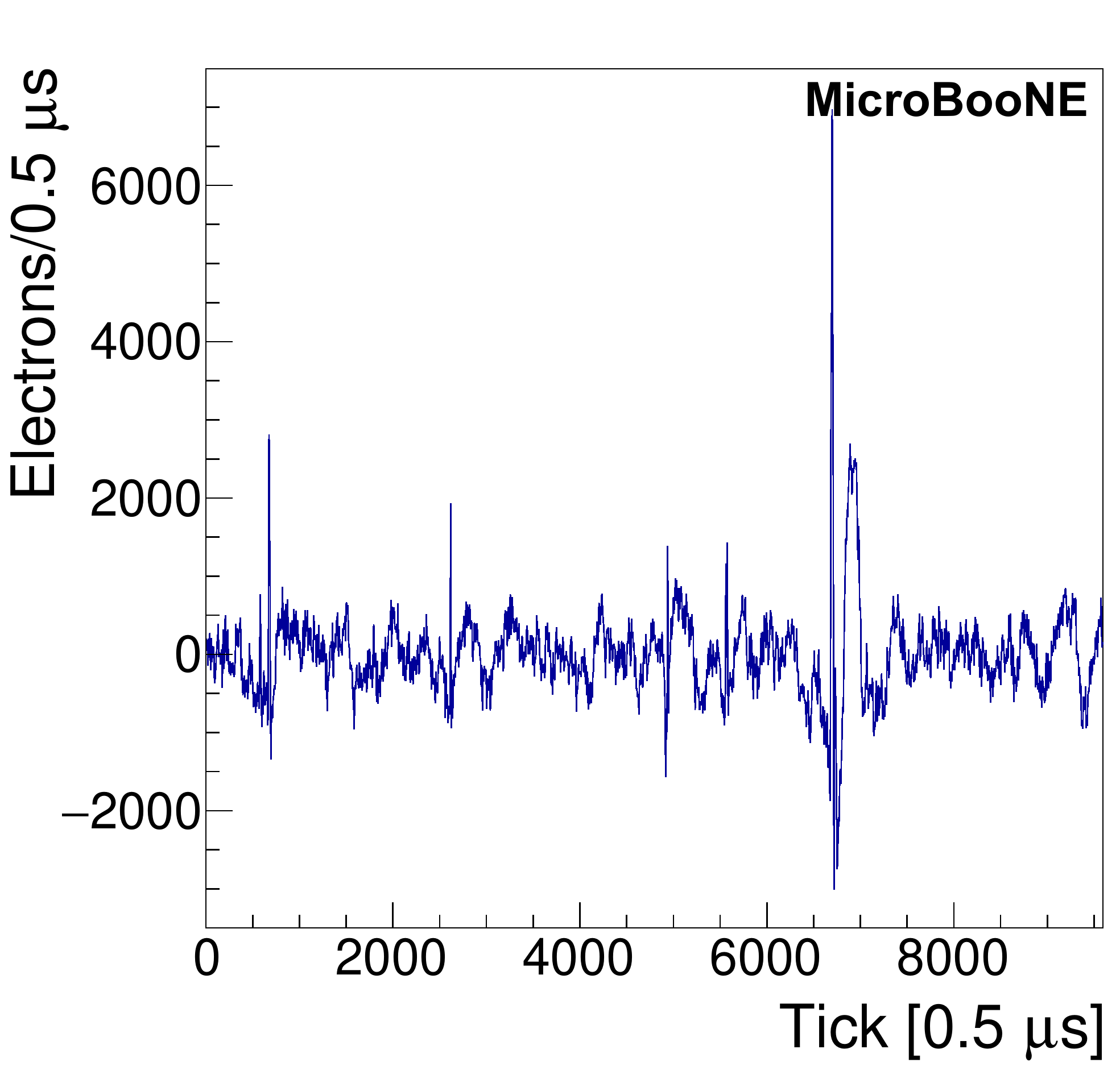}
        \caption{With loose low-frequency filter.}
        \label{figs:ROI_example_c}
    \end{subfigure}
  \caption{ Comparison of the deconvolved signal on the U induction plane 
    (a) without the low-frequency filter,
    (b) with the tight low-frequency filter,
    and (c) with the loose low-frequency filter. }
  \label{figs:ROI_example}
\end{figure}

As shown in figure~\ref{figs:ROI_example_a}, the direct
application of the deconvolution procedure significantly amplifies the low-frequency
noise for the induction wire planes. In order to identify the signal regions of interest
(signal ROIs or ROIs for short), additional low-frequency filters with a
functional form
\begin{equation}
  F_{\rm LF}(\omega) = 1 - e^{\left(\frac{\omega}{a}\right)^2},
\end{equation}
are applied to the deconvolved charge distribution for the induction wire planes to search
for ROIs. These low-frequency filters are not used for the collection wire plane signal.

\begin{figure}[!htbp]
    \begin{subfigure}[t]{0.48\textwidth}
        \includegraphics[width=1.0\textwidth]{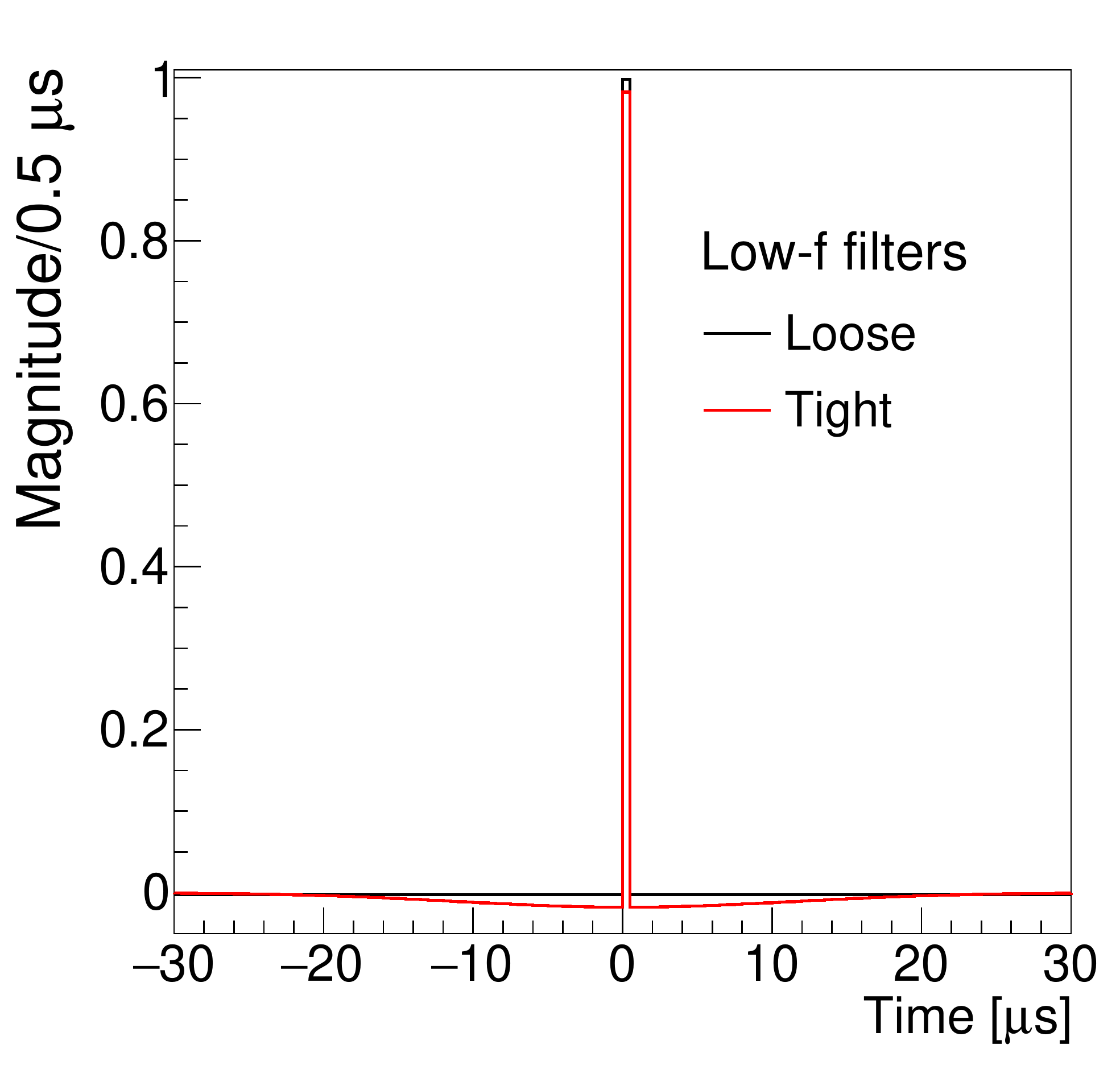}
        \caption{Low-frequency filters in the time domain.}
        \label{figs:filter_lf__a}
    \end{subfigure}
    \begin{subfigure}[t]{0.48\textwidth}
        \includegraphics[width=1.0\textwidth]{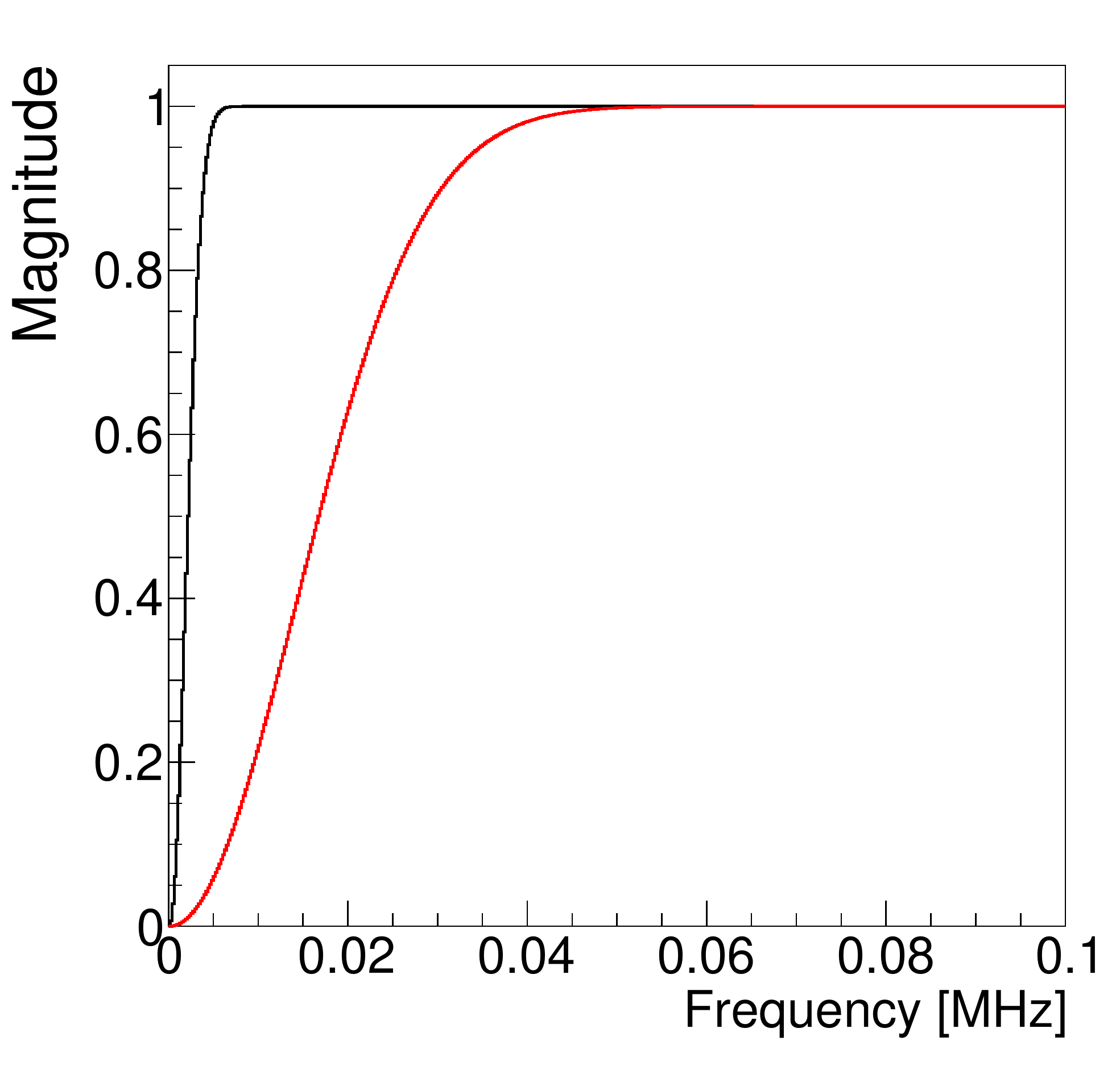}
        \caption{Low-frequency filters in the frequency domain.}
        \label{figs:filter_lf_a}
    \end{subfigure}
  \caption{ Low-frequency filters used in identifying ROIs for induction
    wire planes shown in both the time (a) and frequency (b) domains.}
  \label{figs:filter_lf}
\end{figure}

Figure~\ref{figs:filter_lf} shows the two low-frequency filters used to identify ROIs for
induction wire planes. Due to suppressions at low frequencies, the corresponding smearing
functions in the time domain exhibit a long negative tail. The magnitude
of the negative tail is larger for the tight low-frequency filter, whereas the
tail extends to longer times for the loose low-frequency filter. Since
such a long-range behavior is not desired for the filters used to obtain
the charge signal, these filters are exclusively used to identify ROIs.

Figure~\ref{figs:ROI_example} shows the impact of these low-frequency filters. Without the 
filter, the low-frequency noise totally overwhelms the signal (figure~\ref{figs:ROI_example_a}).
After applying the tight low-frequency filter (figure~\ref{figs:ROI_example_b}), the signal-to-noise ratio
improves for short (in time) signals. However, the long signal at around 7000 time ticks is
removed by the tight low-frequency filter. It is recovered by the loose low-frequency
filter (figure~\ref{figs:ROI_example_c}). 

The deconvolved charge distribution for each channel within the entire readout window
($\sim$9600 ticks) is used to calculate the RMS (root mean square) noise,
to set the threshold of the ROI identification. The RMS is calculated using a 68\%
quantile range relative to the ADC count distribution mean value. 
This RMS calculation is insensitive to the true signals in the deconvolved charge distribution.
For the collection plane, the threshold is set at 5 times the RMS noise which is about 300 electrons/tick on average.
For the induction planes, the deconvolved charge distribution with the tight low-frequency
filter is employed to calculate the RMS noise and the threshold is set at 3.5 times the RMS noise, which is about 350 electrons/tick and 500 electrons/tick on average for the U plane and V plane, respectively. 

\begin{figure}[!htbp]
     \centering
     \begin{subfigure}[]{0.49\textwidth}
        \centering
        \includegraphics[width=0.9\textwidth]{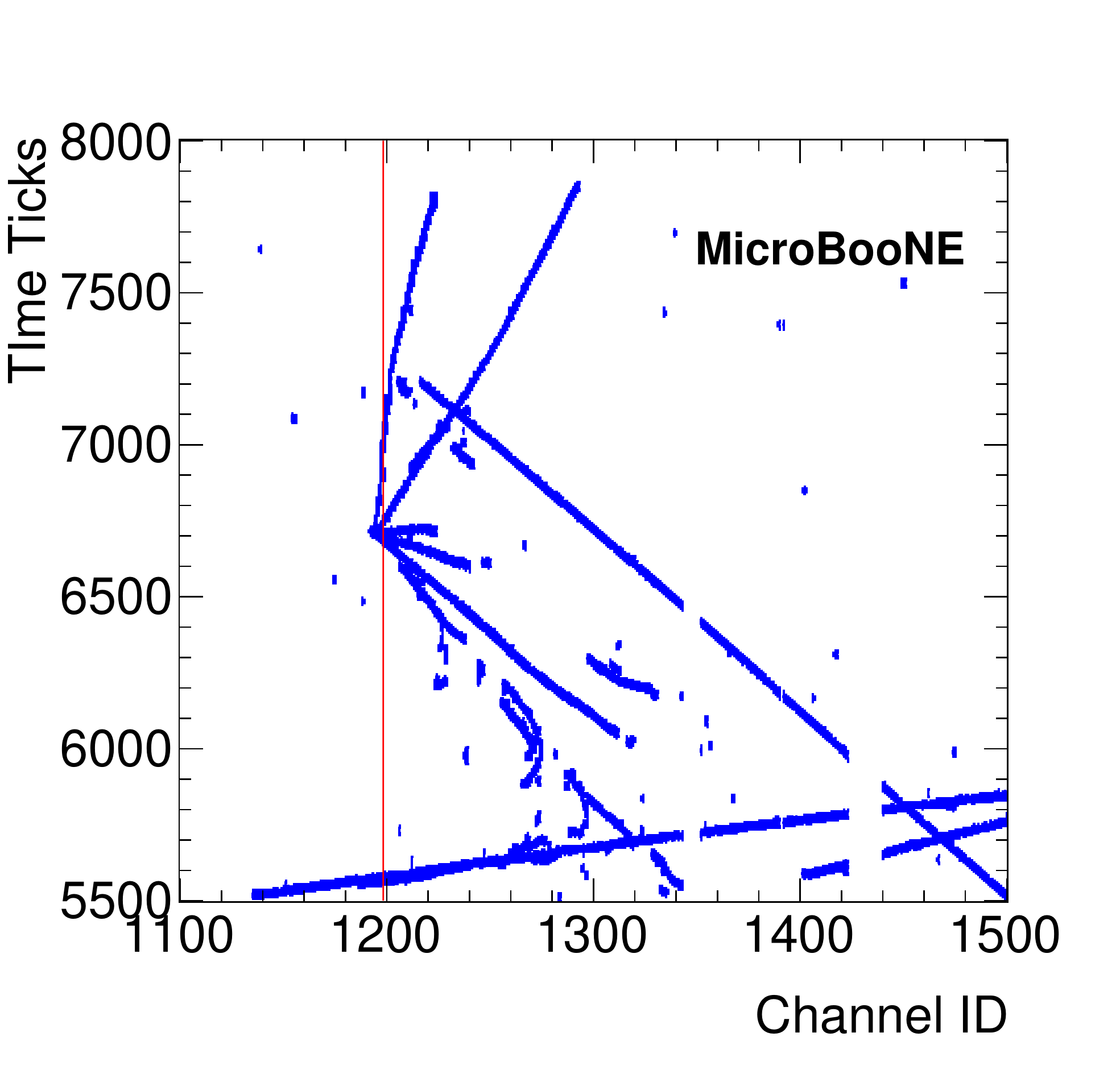}
         \label{figs:loosetight_ROI_example_b}
         \caption{Deconvolved signal with ``loose and tight ROIs''.}
     \end{subfigure}
     \begin{subfigure}[]{0.49\textwidth}
        \centering
        \includegraphics[width=0.9\textwidth]{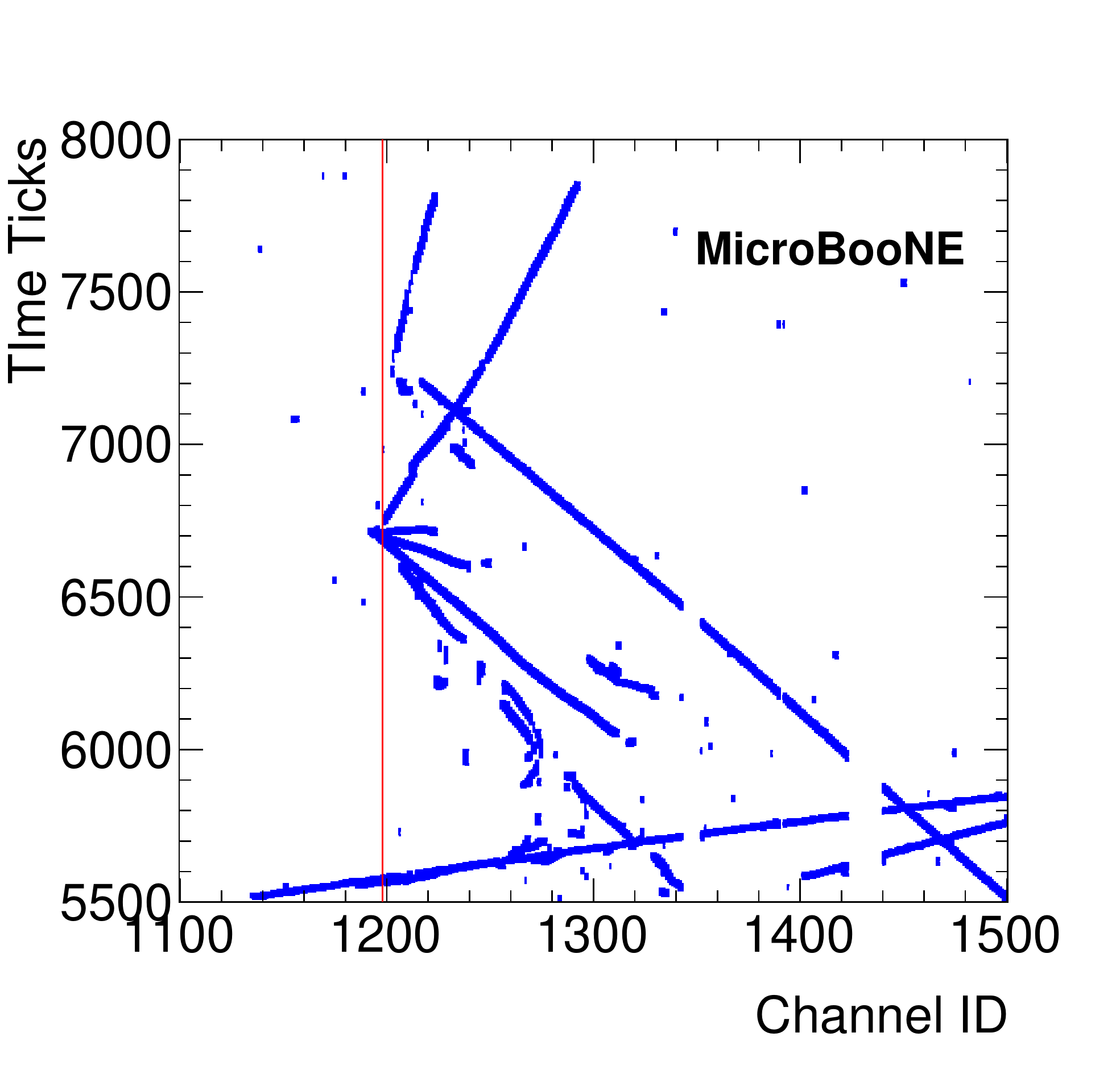}
         \label{figs:loosetight_ROI_example_a}
         \caption{Deconvolved signal with ``tight ROIs''.}
     \end{subfigure}
     \begin{subfigure}[]{0.49\textwidth}
        \centering
        \includegraphics[width=0.95\textwidth]{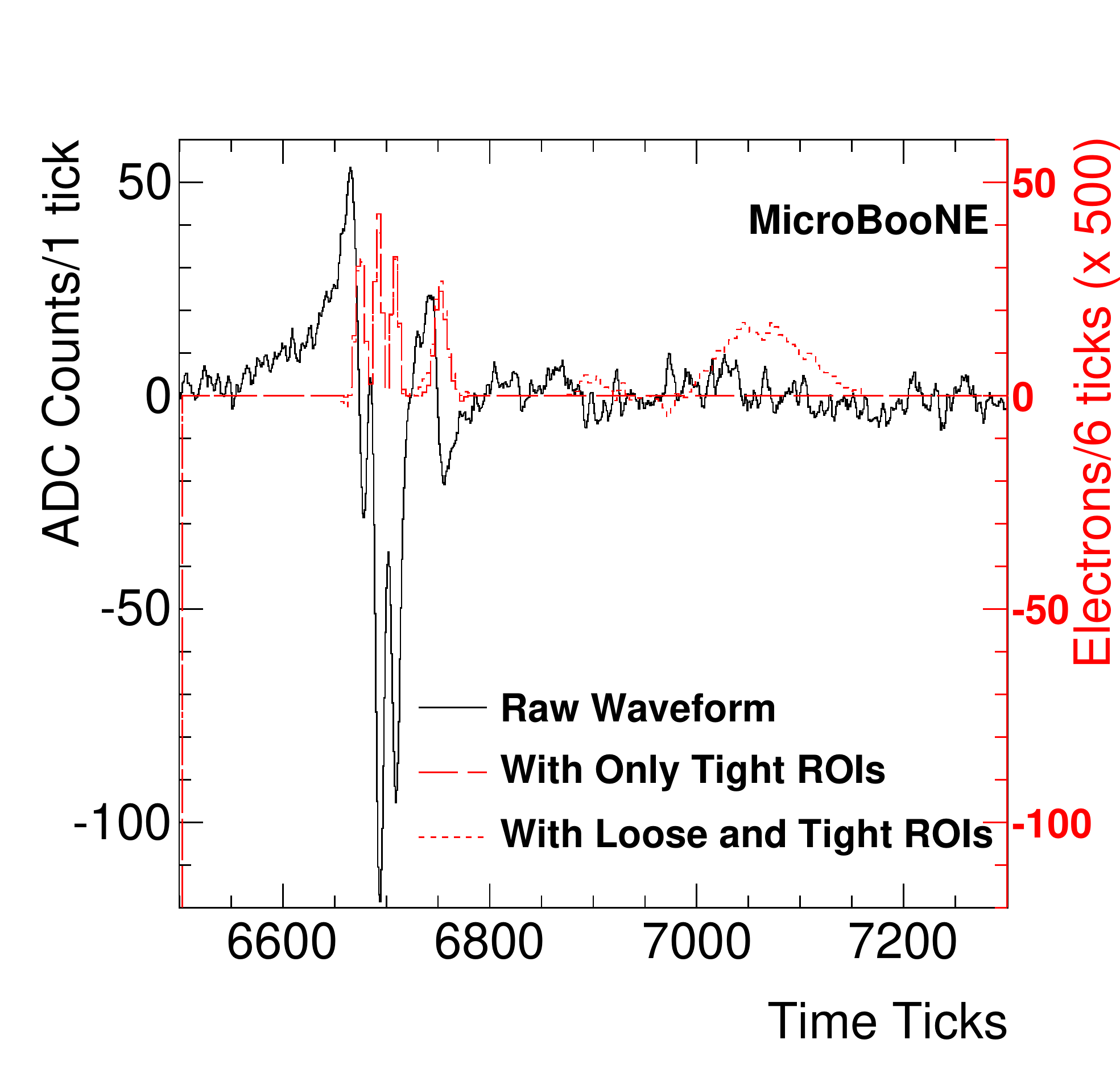}
         \label{figs:loosetight_ROI_example_c}
         \caption{Waveform comparison for the channel marked with a
           red line in (a) and (b).}
     \end{subfigure}
     \caption{A neutrino candidate from MicroBooNE data (event 41075, run 3493)
       on the U plane is shown in an event display with (a) and
       without (b) ``loose ROIs''.
       In (c), the raw baseline-subtracted waveform (black) and the
       deconvolved signal
       with (dotted red) and without (dashed red) ``loose ROIs'' are
       presented. }
  \label{figs:loosetight_ROI_example}
  \end{figure}

ROIs are then extracted from these deconvolved signals. For the induction wire planes,
there are two types of ROIs: tight and loose ROIs,
which are extracted from the deconvolved signal after applying tight and loose low-frequency
filters, respectively.  The goal of the tight ROIs is to achieve high purity in terms of
containing real signal. However, it is expected that tight ROIs have a low efficiency,
in particular for long (in time) signals. On the other hand, the goal of the loose ROIs
is to achieve high efficiency in terms of containing real signal. The trade-off is that we
expect the purity of the loose ROIs to be lower. ROIs are then extracted by searching for
signal above noise. Each ROI is then extended in time to cover the signal tails. For (ROIs) tight ROIs in the (collection) induction plane, additional ROIs are created by examining
the connectivity of the existing ones.
Each of the loose ROIs is then compared with the tight ROIs on the same wire. If one
loose ROI overlaps with a tight ROI, the loose ROI is extended to ensure the tight
ROI is contained. If a tight ROI is not contained by a loose ROI, a new loose ROI with
the exact range of the tight ROI is created. The operations above ensure that each tight ROI
is contained by a loose ROI. Figure~\ref{figs:loosetight_ROI_example} shows the impact
of including tight and loose ROIs for the induction plane signal processing. 

\subsubsection{Refinement of ROIs}\label{sec:refining_ROIs}


As explained previously, the loose ROIs for the induction wire plane are expected to have a
high efficiency in containing the real signal, but low purity. Therefore, an additional
refining procedure using connectivity information is applied to exclude fake ROIs. The applicability of this methodology to a variety of event topologies demonstrates its robustness.

The basic components of the ROI refinement include:
\begin{itemize}
\item {\bf Clean ROIs: } \\
  The motivation of this step is to remove fake ROIs---ROIs containing no signal. In particular, 
  each loose and tight ROI is examined to ensure that part of the bin content inside the ROI
  is above a predefined threshold. ROIs failing this examination are removed. Loose ROIs
  are clustered according to connectivity information.  For each loose ROI cluster, if
  none of its loose ROIs contain one or more tight ROIs, the cluster is removed entirely.

\item {\bf Break ROIs:} \\
  The motivation of this step is to separate a loose ROI into a few small ROIs.
  Sometimes a few separated tracks (e.g. near the neutrino interaction vertex)
  can be quite close to each other along the drift time direction. Often a single loose
  ROI would be created to contain these tracks given the presence of low-frequency noise.
  Therefore, a special algorithm is needed to identify this scenario and separate
  the ROIs. In particular, each loose ROI is examined to search for multiple independent
  peaks. If found, the loose ROI is separated into several loose ROIs.
  Figure~\ref{figs:break_ROI_example} shows the impact of the ``break ROIs'' step.

  \begin{figure}[!htbp]
  \centering
     \begin{subfigure}[]{0.49\textwidth}
        \centering
        \includegraphics[width=0.9\textwidth]{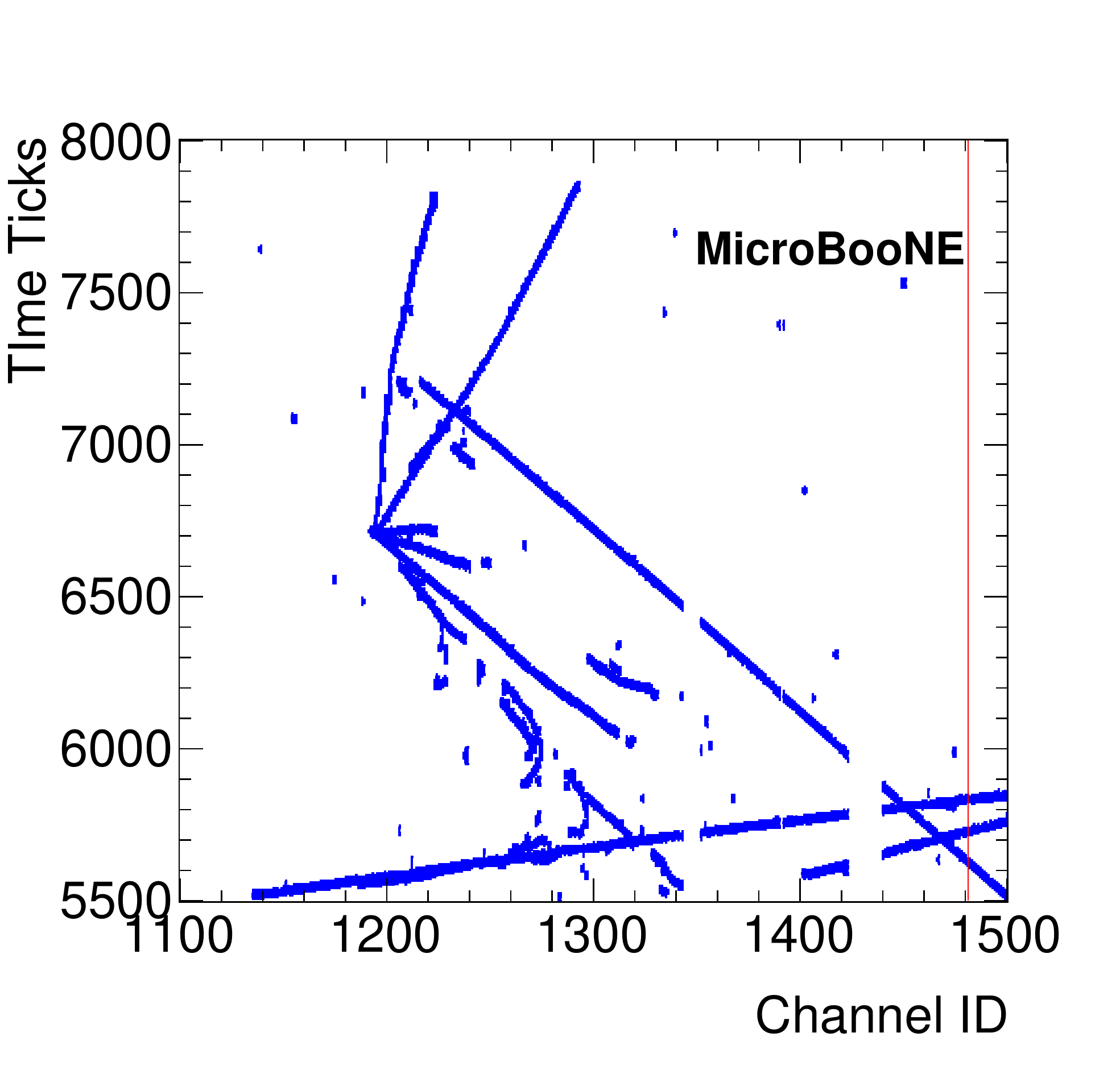}
         \label{figs:break_ROI_example_b}
         \caption{Deconvolved signal with ``break ROIs''.}
     \end{subfigure}
     \begin{subfigure}[]{0.49\textwidth}
        \centering
        \includegraphics[width=0.9\textwidth]{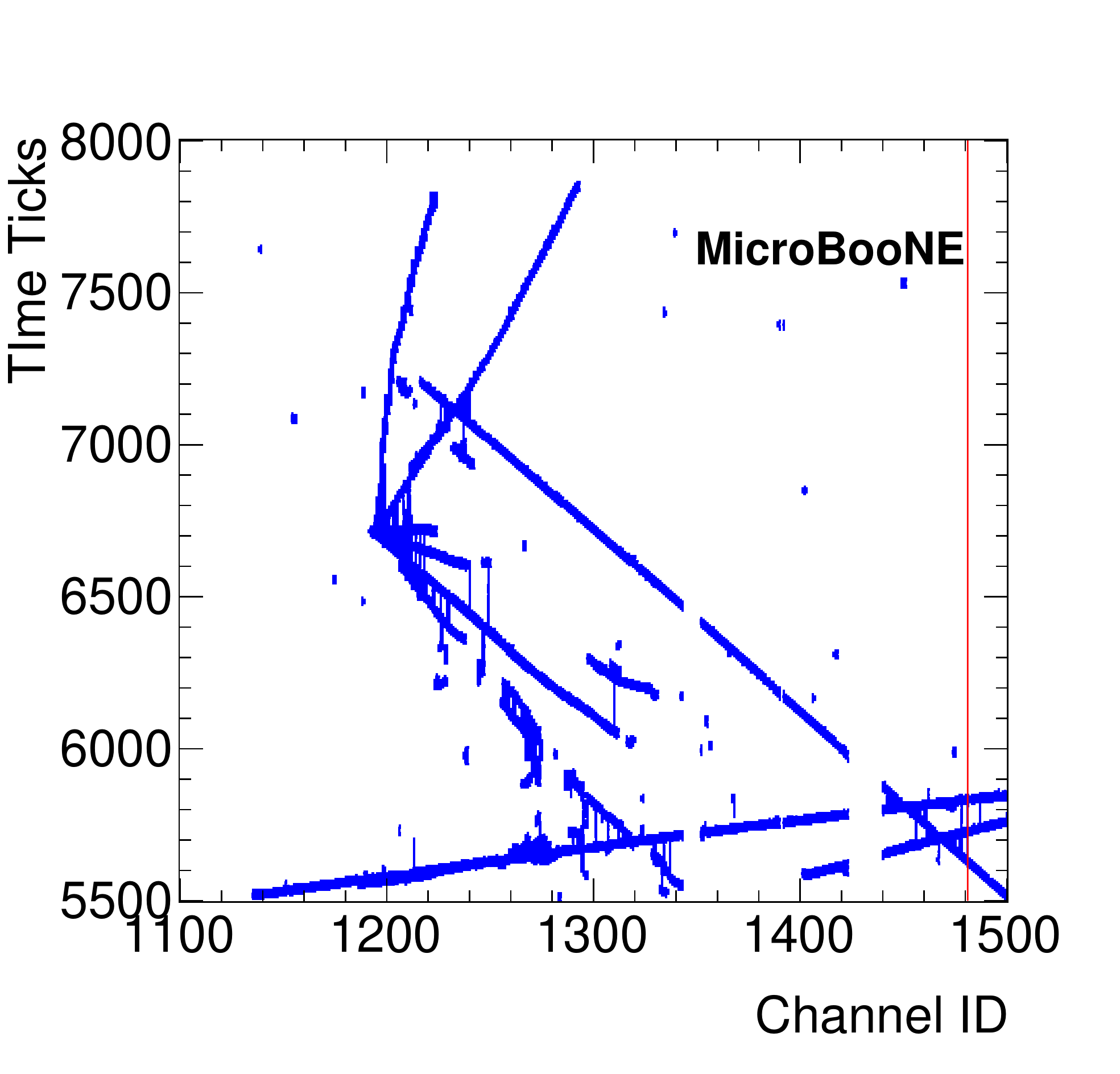}
         \label{figs:break_ROI_example_a}
         \caption{Deconvolved signal without ``break ROIs''.}
     \end{subfigure}
     \begin{subfigure}[]{0.49\textwidth}
        \centering
        \includegraphics[width=0.9\textwidth]{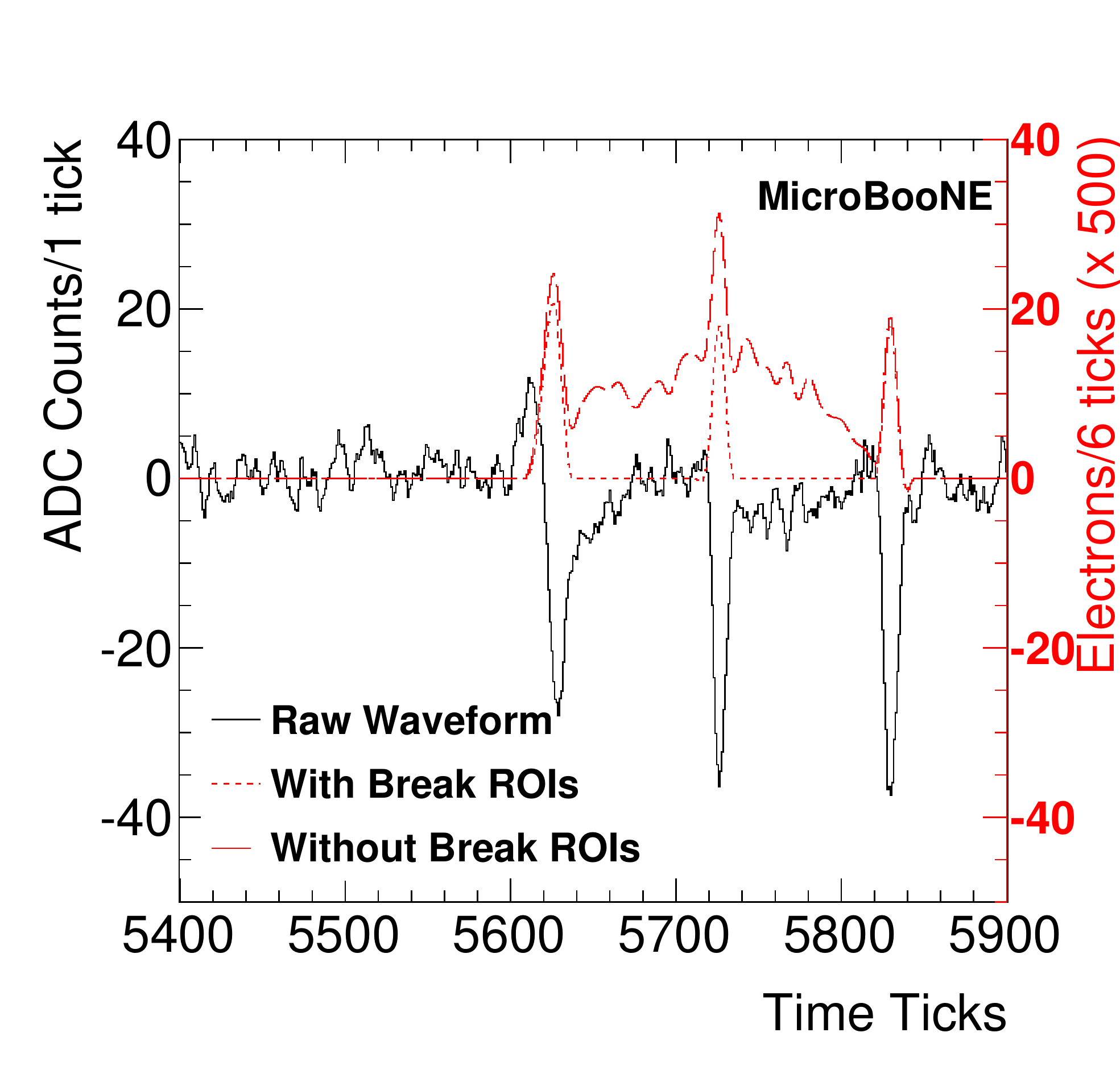}
         \label{figs:break_ROI_example_c}
         \caption{Waveform comparison for the channel marked with a
           red line in (a) and (b).}
     \end{subfigure}
     \caption{The same neutrino candidate from
       figure~\ref{figs:loosetight_ROI_example}
       is shown in an event display with (a) and without (b) ``break ROIs''.
       In (c), the raw baseline-subtracted waveform (black) and the
       deconvolved signal with (dotted red) and without (dashed red)
       ``break ROIs'' are presented. }
  \label{figs:break_ROI_example}
  \end{figure}

\item {\bf Shrink ROIs:} \\
  The motivation of this step is to reduce the length of a ROI that contains real
  signal that had otherwise been accidentally extended into a much broader time range due to the presence of
  low-frequency noise. In particular, the range of each loose ROI is reduced according
  to the tight ROIs they contain as well as those in the adjacent loose ROIs.
  Figure~\ref{figs:shrink_ROI_example} shows the effect of the
  ``shrink ROIs'' step.


 
 \begin{figure}[h!]
  \center
     \begin{subfigure}[]{0.49\textwidth}
        \centering
        \includegraphics[width=0.9\textwidth]{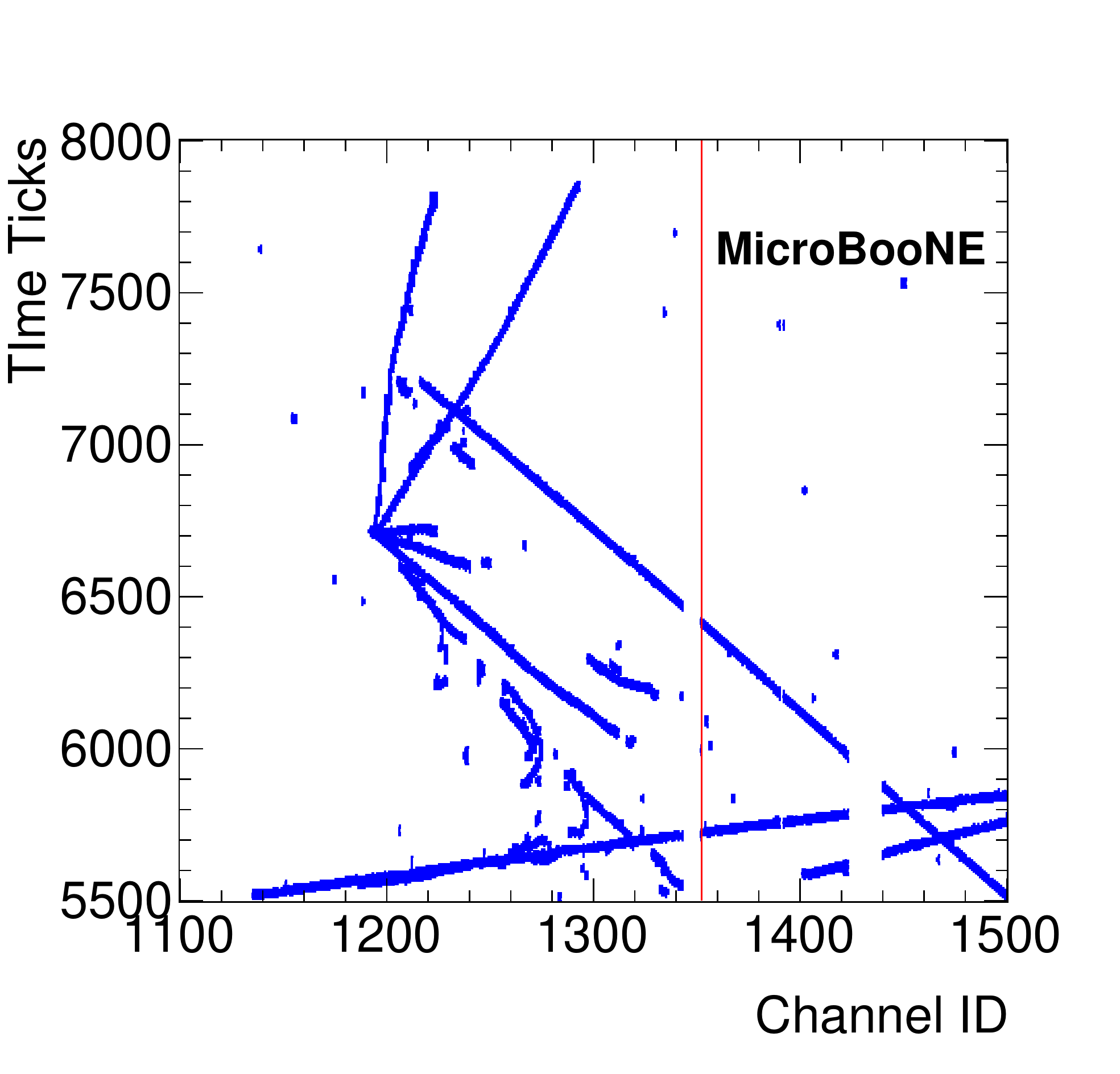}
         \label{figs:shrink_ROI_example_b}
         \caption{Deconvolved signal with ``shrink ROIs''.}
     \end{subfigure}
     \begin{subfigure}[]{0.49\textwidth}
        \centering
        \includegraphics[width=0.9\textwidth]{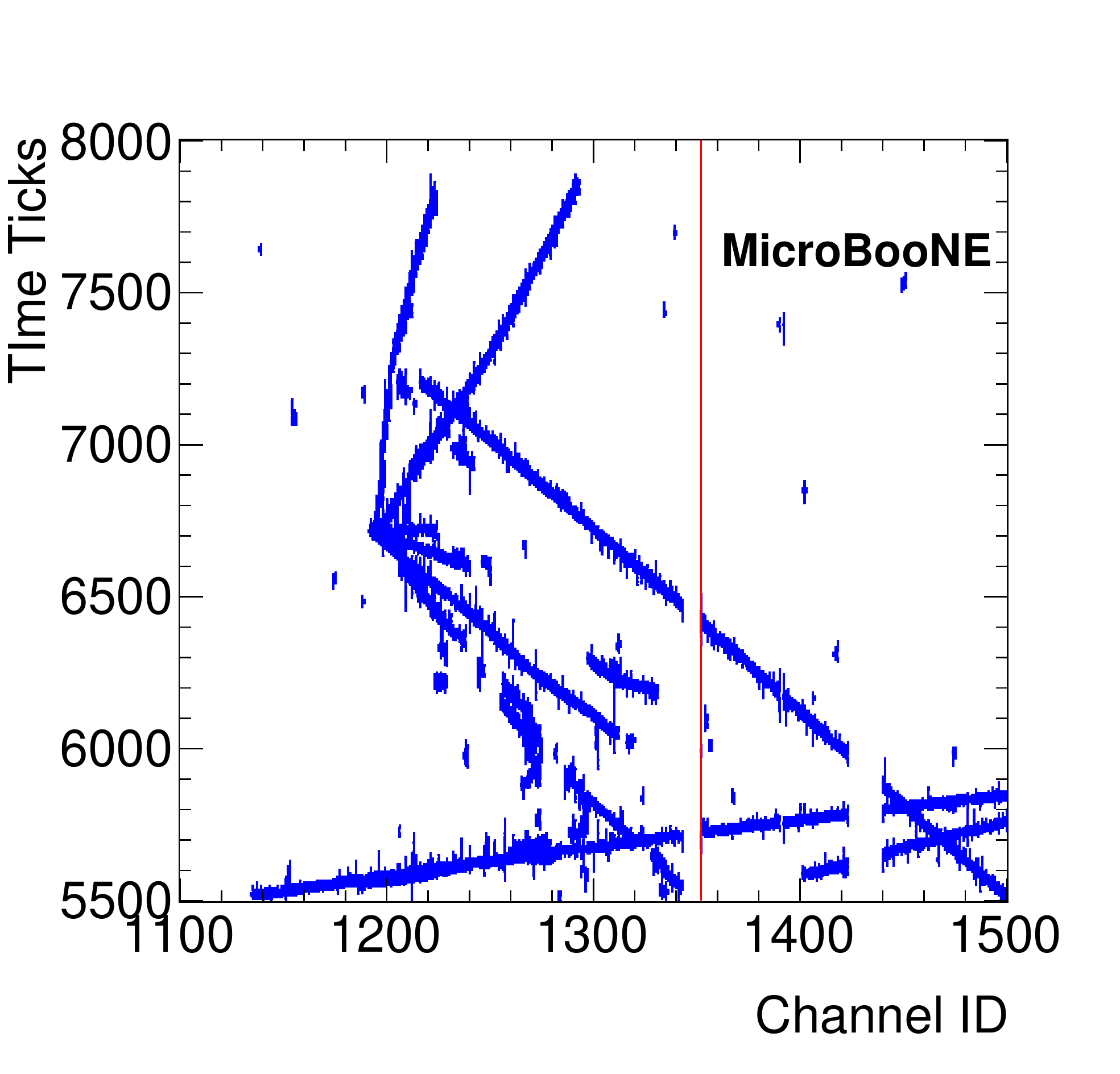}
         \label{figs:shrink_ROI_example_a}
         \caption{Deconvolved signal without ``shrink ROIs''.}
     \end{subfigure}
     \begin{subfigure}[]{0.49\textwidth}
        \centering
        \includegraphics[width=0.9\textwidth]{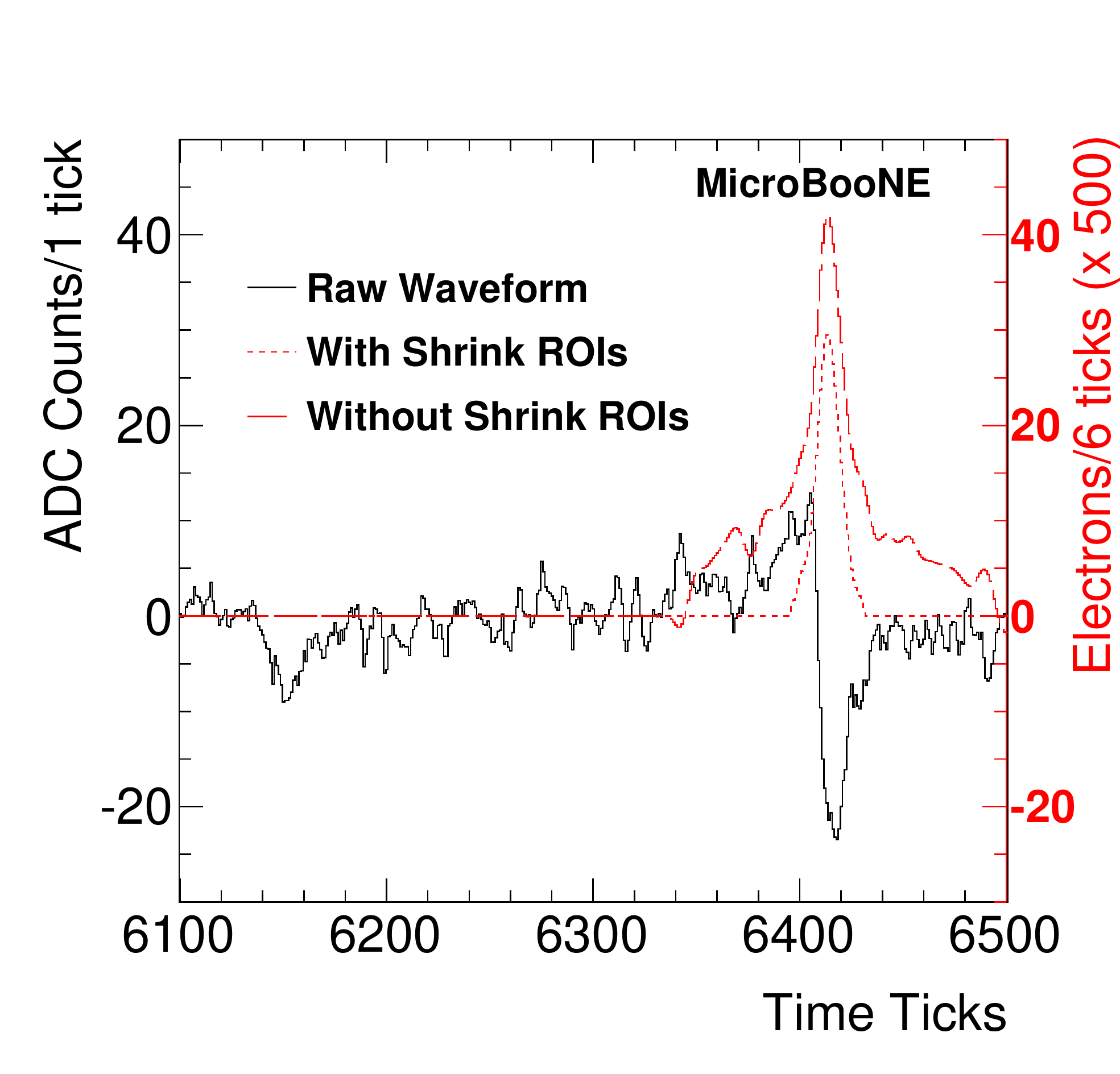}
         \label{figs:shrink_ROI_example_c}
         \caption{Waveform comparison for channel marked with red line in (a) and (b).}
     \end{subfigure}
     \caption{ The same neutrino candidate from
       figure~\ref{figs:loosetight_ROI_example}
       is shown in an event display with (a) and
       without (b) ``shrink ROIs''.
       In (c), the raw baseline-subtracteed waveform (black) and the
       deconvolved signal with (dotted red) and without (dashed red)
       ``shrink ROIs'' are presented. }
  \label{figs:shrink_ROI_example}
  \end{figure}
\end{itemize}

The overall ROI refinement takes an iterative approach by applying the above steps in
sequence. The final remaining ROIs are then applied to the deconvolved signal without
the low-frequency filter. A linear correction to the baseline is applied so that the
bin content of the ROI boundaries are exactly zero.

\section {Evaluating TPC signal processing with simulation} \label{sec:evaluation}

This section describes the quantitative evaluation of the TPC signal
processing using simulation. First, we describe some key improvements
in the simulation. Next, the details of the simulation algorithms are shown.
Finally, a quantitative evaluation of the signal processing using this
simulation is demonstrated including measures of reconstructed charge
resolution, bias, and efficiency. The simulation itself is validated
against MicroBooNE detector data in~\cite{SP2_paper}.

\subsection{Key simulation improvements}

Evaluation of the signal processing relies, in part, on an improved
signal and noise simulation. The signal simulation has two key
improvements compared to prior implementations.

Previous implementations~\cite{larsoft} adopted a simplified 
model to consider the induction charge from all drifting electrons, in which case the field response was extracted from an entire track at a single angle.
An additional simplification was made to use a common, average field response
independent of the transverse distance between the wire and the drifting charge.

The improved simulation addresses these two issues by employing the field
response calculations described in section~\ref{sec:field_resp}. 
It supports a long-range induction model that more correctly incorporates angle-dependent effects and records the
current induced not only on the closest wire but also current which is
induced from the ten wires to either side.  This allows for the accounting of
contributions to the total induced current down to the sub-percent
level.  The simulation also takes into consideration the fine-grained
variability that exists for different possible paths within a single
wire region.

The second improvement relates to the treatment of the complexity
inherent in the initial distribution of energy depositions by particles
interacting in the LAr and its resulting distribution of ionization
electrons.  In past implementations, the ionization electrons were
grouped into spatial bins and the contribution to induced current
was extracted from each bin. To improve on this, the new simulation retains
the identity of each
energy deposition produced by a given step of the particle tracking
simulation.  The effects of diffusion and absorption are applied to
each ionization point by associating a 2D Gaussian charge
distribution.  The distributions for all ionization points are kept
distinct during drifting until they reach the wire planes.  There, they
are sampled and interpolated onto the regular 2D grid defined in the
transverse and longitudinal directions by the field response functions
described in section~\ref{sec:field_resp}.  At this point, a fine binning
(\SI{0.5}{\us} $\times$ \SI{0.3}{\mm} for the MicroBooNE
implementation) is applied.

The simulation improvements described above highlight an important
connection between the choice of signal simulation and the signal
processing models which may be employed.  The former primarily applies
a convolution of the ionization charge distribution and the field (and
electronics) response functions.  The latter primarily performs their
deconvolution.  Prior simulation implementations used the same
kernel in both the
convolution and deconvolution processes.  In the absence of noise and up to the
spatial binning simulated, the prior simulation approach produces an
exact recovery of the initial ionization charge distribution, which is
not representative of real detector effects.  The new
approach supports a better approximation of reality by allowing the
variation of field response across a wire region to be accounted for
in the convolution.  However, this variation cannot be accounted for
in the deconvolution as there exists no \apriori knowledge of the
fine-grained charge distribution.
Quantification of this effect, i.e. the realistic performance of signal processing, is one of the goals of the evaluation shown here.



\subsection{Simulation overview}
\label{sec:sim}

\begin{figure}[htbp]
  \centering
    \begin{subfigure}[t]{0.49\textwidth}
        \centering
        \includegraphics[width=0.99\textwidth]{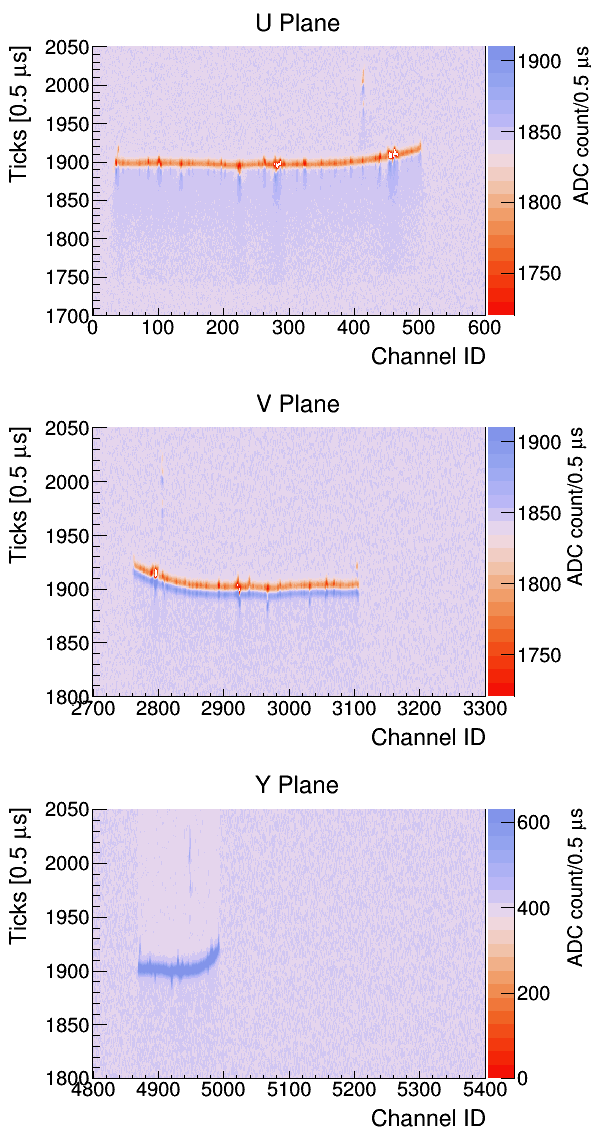}
        \caption{Simulated waveforms.}\label{fig:simevent_wf}
    \end{subfigure}
    \begin{subfigure}[t]{0.49\textwidth}
        \centering
        \includegraphics[width=0.99\textwidth]{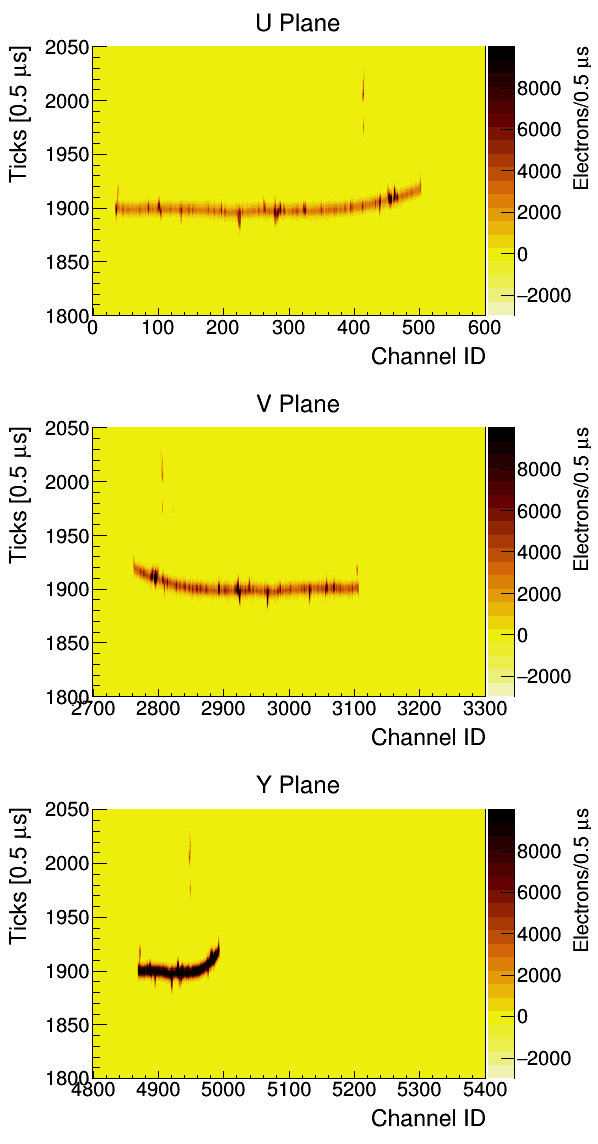}
        \caption{Extracted charge distributions.}\label{fig:simevent_charge}
    \end{subfigure}
    \caption[Simulated isochronous event.]{Simulated waveforms and reconstructed
      charge from a nearly isochronous track of a GEANT4-based 5-GeV muon
      in liquid argon. The X-axis represents the channel (wire) from the
      three wire planes of MicroBooNE. The Y-axis represents the sampled time ticks
      (per \SI{0.5}{\micro\second}). (a) Simulated waveforms including inherent
      electronics noise in units of ADC counts. (b) Reconstructed charge
      in units of $e^-$ corresponding to the simulated waveforms in (a).}
  \label{fig:simevent}
\end{figure}

The TPC detector simulation spans detector physics ranging from
measures of energy loss by particles traversing the detector to the digitized
waveforms produced from a model of the front-end electronics. A data-driven,
analytical simulation of the inherent electronics noise is also performed. 
Figure~\ref{fig:simevent} shows one example event produced by this simulation.

The signal simulation, i.e. the ADC waveform on a given channel, 
\begin{equation}
  \label{eq:sim-convolution}
    M = (Depo \otimes Drift \otimes Duct + Noise) \otimes Digit, 
\end{equation}
is conceptually a convolution of five functions:
\begin{description}
\item[$Depo$] represents the initial distribution of the 
  ionization electrons created by energy depositions in space and time. 
\item[$Drift$] represents a function that transforms an initial charge
  cloud to a distribution of electrons arriving at the wires.
  Physical processes related to drifting, including attenuation due to
  impurities, diffusion and possible distortions to the nominal
  applied electric field, are applied in this function.
\item[$Duct$] is a family of functions, each is a
  convolution $F \otimes E$ of the field response functions $F$
  associated with the sense wire and the element of drifting charge
  and the electronics response function $E$ corresponding to the
  shaping and amplification of the front end electronics.
\item[$Digit$] models the digitization electronics according to a given
  sampling rate, resolution, and dynamic voltage range and baseline 
  offset resulting in an ADC waveform. 
\item[$Noise$] simulates the inherent electronics noise by producing a
  voltage level waveform from a random sampling of a Rayleigh distributed 
  amplitude spectrum and uniformly distributed phase spectrum.  The noise
  spectra used are from measurements with the MicroBooNE detector
  after software noise filters, which have excess (non-inherent)
  noise effects removed.
\end{description}
These functions are defined over broad ranges and with fine-grained resolution. The
characteristic scales are set by the variability and extent of the
field response functions and the sampling rate of the digitizer.
Their range is set by the size of the detector, which is related to the
length of time over which one exposure is digitized.  With
sub-millimeter field variability, \SI{0.5}{\micro\second}
sampling, a detector of several meters in extent, and a readout over
several milliseconds, the size of each dimension of these
functions is $10^3 - 10^5$.  A naive implementation of the required
convolution is not possible with commercial computing hardware.
The remainder of this section describes each term in equation~\ref{eq:sim-convolution} including the methods
employed to reduce the dimensionality.

%

\subsubsection{Initial distribution of charge depositions}
\label{sec:sim-depo}

The simulation takes as input a distribution of initial charge depositions - free ionization electrons from charged particles traversing the liquid argon medium.
In order to reduce the size of the input, the simulation requires charge depositions to
be defined as a set of discrete, localized depositions 
\begin{equation}
  \label{eq:sim-depo-def}
    Depo_i = (q_i, t_i, \vec{r}_i), 
\end{equation}
where $q_i$ is the number of ionization electrons at the given
point $(\vec{r}_i,t_i)$ in space and time.  

The scale at which charge deposition is localized must be chosen to balance
computation time and to ensure that discreteness is smoothed out by
subsequent convolution terms.  In practice, this limitation is well
matched to the form of results produced by tracking simulations such
as those based on GEANT4~\cite{geant4} implemented in LArSoft (Liquid Argon Software)~\cite{larsoft}.

The user may directly supply $q_i$ in terms of ionization electrons or
in terms of the amount of energy lost by particles over a given step
of the tracking simulation (i.e. $dE/dx$ over distance $dx$).  If the
latter is provided, the simulation will apply the appropriate Fano
factor~\cite{fano_factor}, recombination~\cite{recombination1, recombination2, recombination3}
and their associated statistical fluctuations in generating the ionization
charge.

\subsubsection{Drift transport and physics}
\label{sec:sim-drift}

The simulation supplies a full set of physical
processes related to the transport of ionization electrons through 
liquid argon under the influence of an applied electric field. These
processes include electron attenuation (corresponding to electron attachment
on impurities in liquid argon), longitudinal and transverse
diffusion~\cite{Li:2015rqa}, and transport.  The default transport is performed
neglecting distortions due to the build up of space charge~\cite{space-charge}.
As a result, initial point-like depositions are typically provided as points but they may
have characteristic widths ($\sigmaT,\sigmaL$) in the transverse direction
and along the nominal drift, respectively, as a 3D Gaussian cloud
\begin{equation}
  \label{eq:sim-drift}
    Depo_i \otimes Drift \to Depo_i(q_i,t_i+t_{drift}, \vec{r}_i\rvert_{x=x_{rp}}, \sigmaTi{i}, \sigmaLi{i}),
\end{equation}
where $\sigma \cong \sqrt{2\cdot D\cdot t_{drift}}$, $t_{drift}$ is the drift
time, and $D$ is transverse ($D_T$) or longitudinal ($D_L$) diffusion
coefficients, respectively~\cite{Li:2015rqa}.
The transport term of the convolution transforms each deposition as
in expression~\ref{eq:sim-drift} independently until the center of the resulting
distribution of ionization electrons reaches a \textit{response plane}.
The response plane is normal to the nominal drift direction which is
along the $x$ coordinate axis and is located near the wire planes.
Its location $x_{rp}$ exactly coincides with the plane containing the
starting locations (\SI{10}{\cm}) of the drift paths on which the field response functions are defined/simulated.

Diffusion gives the initial deposition a finite extent in the
transverse and longitudinal directions.  Each deposition is then
trifurcated for each wire plane.  The transverse location and extent
are transformed to measures $\rho$ along the pitch direction for the
given plane.  The longitudinal extent is transformed to a time-based
measure using the nominal drift speed relative to the time at
which the center of the deposition reached the response plane.  This
trifurcation transforms a single 3+1 space-time dimensional problem
into three smaller 1+1 problems.  The single remaining space
coordinate is the transverse measure along the pitch direction of each
wire plane.

\subsubsection{Detector field and electronics response}
\label{sec:sim-duct}

The $Duct$ term is itself a convolution $F_j \otimes E$ of the field
and electronics responses, respectively.  The first function $F_j$
describes the current induced on a sense wire due to the passage of a
nearby unit charge along a path $j$ starting a distance $\rho_j$ in the wire pitch direction. Given a wire at $\rho = 0$,
these paths are shown in figure~\ref{fig:fielddemo}. The field response
employed in the simulation is a 2D Garfield calculation, meaning that it is independent of the position along the wire direction for
each wire plane. However, this 2D calculation is meant to serve as a
good approximation of the realistic 3D field response averaging along the wire direction.

The second function $E$ includes two parts as shown in figure~\ref{fig:resp1}
corresponding to the pre-amplifier and two RC filters. 
In addition, an intermediate gain of 1.2 is included. 
$E$ determines the voltage response of the front-end
amplifiers to the instantaneous application of a unit charge as their input.
In simulation, the electronics response is taken to be
constant for all channels.

The convolution $F_j \otimes E$ is independent of the
distribution of drifted charge and can be calculated separately.
Results of such a calculation are shown in figure~\ref{figs:overall_response}.
Instead of performing the convolution through a direct integration in the
time domain, a discrete Fourier transform (DFT) is applied on each term,
followed by a multiplication in frequency-space and a
final inverse DFT.  However, in the simulation, subsequent
convolution is required in order for the intermediate frequency-space
result to be cached for later reuse.

The sampling period is fixed according to the electronics digitization in $E$, as the instantaneous current in $F_j$ must be integrated
over each sample. In the
discussion below, this convolution of field and electronics responses
is referred to generically as a response function.

\begin{figure}[htbp]
  \centering
  \includegraphics[width=0.85\textwidth]{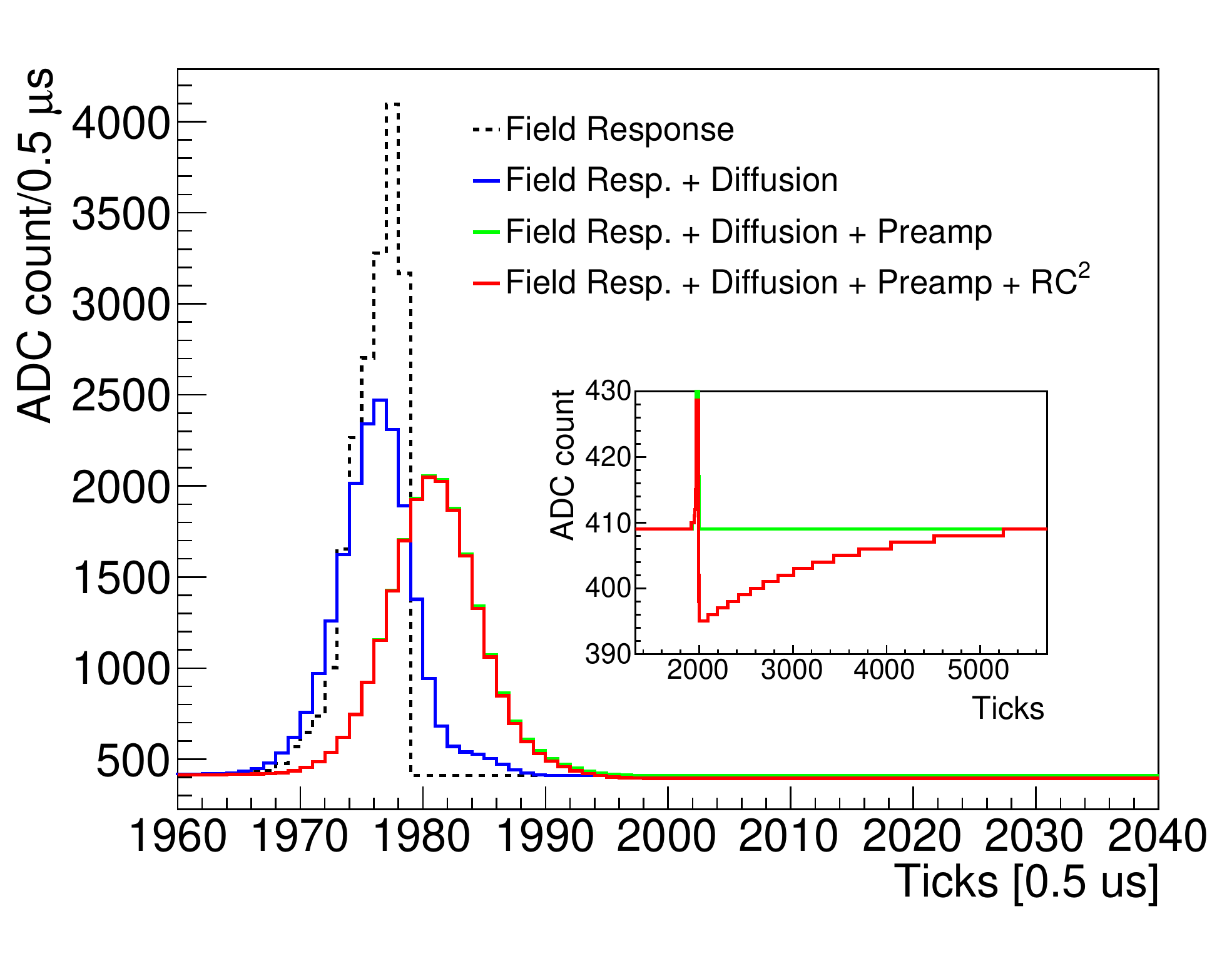}
  \caption{Breakdown of a full simulation of a point source of charge.
    The Y plane waveforms are taken as an example.
    The ADC baselines for Y plane are set to 409 ADC counts, \SI{200}{\mV}.
    An inset figure magnifies the long tail
    from the RC filters. The maximum magnitude of the negative tail is
    roughly \SI{0.7}{\percent} of the signal peak.
    Diffusion is a 2D smearing effect and the plot just shows the result 
    of the central wire.}
  \label{fig:stepbystep}
\end{figure}

Given a point charge, the changes of the waveform resulting from the
longitudinal Gaussian diffusion and electronics responses (preamplifier,
RC filters) are demonstrated in figure~\ref{fig:stepbystep}. Note that the
RC filters have an effect on the collection plane's signal,
though it is relatively small (less than \SI{1}{\percent}). The tail of
the RC filter also shows up in the Y plane waveform of figure~\ref{fig:simevent_wf}.

The simulation has the framework to incorporate multiple responses,
e.g., different electronics responses corresponding to normal channels
or mis-configured channels, as well as different field responses
corresponding to shorted wire regions. These specific responses are
relevant to real MicroBooNE data as mentioned in ~\cite{noise_filter_paper}
and important in data/MC comparisons as elaborated in ~\cite{SP2_paper}.

\subsubsection{Performing the convolution}

As described above, the drifted charge distribution and response
functions now cover a discrete two dimensional domain, where $\rho_j$ is
a measure in the pitch direction relative to a given wire.  That is,
$j$ spans the number of nearby drift paths.  The simulation
has drift paths which cover the central wire region and ten wire
regions to either side with ten paths per region.  This gives a total
of 210 paths per plane centered on each wire of interest.

This span in $\rho$ is used to identify all depositions that have any
portion of their total extent contained therein.  It is worth
remembering that these depositions are in a parameterized form
following a 2D Gaussian distribution.  These distributions must be
discretized in order to be convolved with the corresponding discrete response
function.

There are two aspects to this discretization which are important to
its performance and correctness. The first is due to the response
functions covering more transverse span than just a single wire
region.  All depositions falling in the nominal 21 wire regions must
be discretized for each wire.  Most of these same depositions will
also contribute to the neighboring wires.  Rediscretizing for each wire
would entail a factor of 20 redundant calculations.  Discretizing the
entire domain would require a prohibitive amount of computer
memory. To overcome computing limitations, the wires of
a plane are processed in order of their pitch location starting from
the smallest $\rho$.  As the calculation advances to the next wire, it
frees the memory associated with the wire region of the lowest $\rho$
and discretizes one new wire region at the high end.

\begin{figure}[htbp]
  \centering
    \begin{subfigure}[t]{0.49\textwidth}
        \centering
        \includegraphics[width=0.99\textwidth]{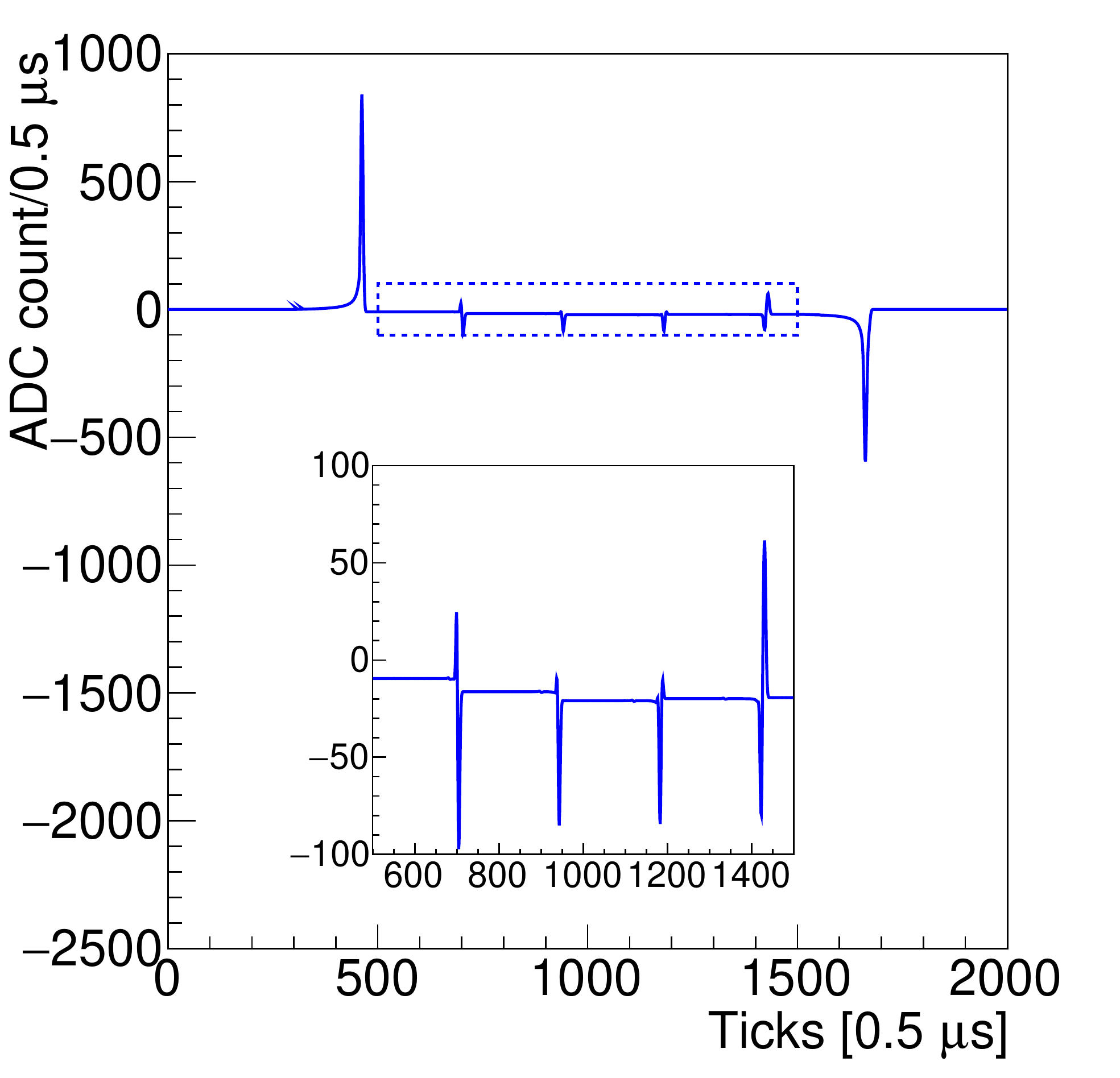}
        \caption{Average.}
        \label{fig:interp_1}
    \end{subfigure}
    \begin{subfigure}[t]{0.49\textwidth}
        \centering
        \includegraphics[width=0.99\textwidth]{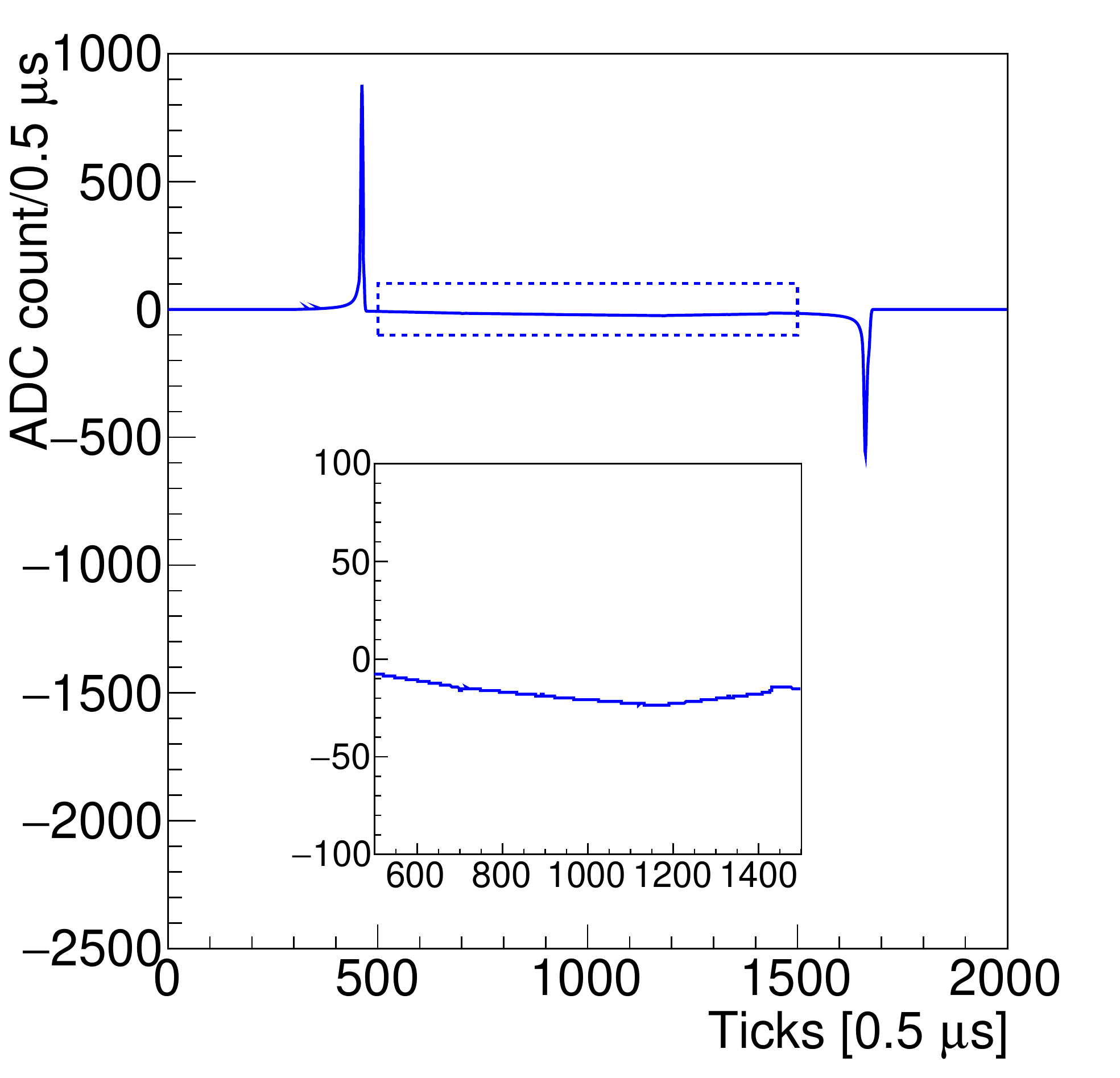}
        \caption{Linear interpolation.}
        \label{fig:interp_2}
    \end{subfigure}
  \caption{Simulated baseline-subtracted waveforms of an induction plane large angle track, near
    perpendicular to the wire plane. The track covers 6 paths / 5
    sub-pitches within a wire region. 
    The inset figure magnifies the waveform in the dashed box, displaying the fine
    grained field response as well as the cancellation of the bipolar
    field response at the paths of each calculated field response (boundaries of
    sub-wire regions). 
    (a) Average weighting, leading to improper cancellation of the bipolar response.  
    (b) Linear interpolation, leading to proper cancellation of the bipolar response.}
  \label{fig:interp}
\end{figure}

The second aspect relates to correctness.  The coverage of $\rho$ by
the response functions is very fined grained.  Neighboring drift paths
are separated by \SI{0.3}{\milli\meter} (1/10 of the wire pitch in MicroBooNE).
At this scale, the path-to-path response variation is typically small, within \SI{10}{\percent}.
However, particularly for the induction planes, at any scale of
such coverage of $\rho$ there is an aliasing effect that will occur for elongated charge distributions which fall in a line close to perpendicular to the wire planes. As shown in figure~\ref{fig:interp}, this effect is in a small but non-negligible phase space that must be mitigated.  
An interpolation is performed using the two nearest response functions at both $\rho_i$ and
$\rho_{i+1}$.

In order to calculate the interpolated response, for any
wire plane using its own Cartesian coordinate, the deposited
charge after drifting can be expressed as $q(t, z)$. This charge is
to be convolved with the field response $F(t,x,y,z)$ of which $x$ is
fixed at the starting location of the virtual response plane
and $y$ equivalently averages out.  Consider $\rho_j < z < \rho_{j+1}$
and the two field responses $F_1(t) = F_j(t, \rho_j)$ and $F_2(t) = F_{j+1}(t, \rho_{j+1})$,
the convolution is performed as follows:
\begin{equation}
    \label{eq:field_interp}
    \int^{\rho_{j+1}}_{\rho_j} \{ q(t, z)\otimes(F_1(t)\cdot u(z) + F_2(t)\cdot (1-u(z))) \} dz,
\end{equation}
where $u(z)$ is the weighting function for interpolation between two calculated paths of field response. 
Since transverse and longitudinal diffusion are independent processes,
$q(t, z)$ can be re-phased as $q(t)\cdot Gaus(z)$. The width and center of the
function $Gaus$ depend on the initial deposition location of the charge.
Therefore, expression~\ref{eq:field_interp} can be simplified as an integral
of $z$ and convolution of $t$, 
\begin{equation}
    \label{eq:field_interp2}
    \int^{\rho_{j+1}}_{\rho_j} \{Gaus(z)u(z)\} dz \cdot q(t) \otimes F_1(t) +  \int^{\rho_{j+1}}_{\rho_j} \{Gaus(z)(1-u(z))\}dz \cdot q(t) \otimes F_2(t)
\end{equation}
In practice, any charge deposition at $z$ between path $j$ and $j+1$ is
redistributed to the two positions $\rho_j, \rho_{j+1}$ against the
weight $ \int^{\rho_{j+1}}_{\rho_j} \{Gaus(z)u(z)\} dz$. The weighting
function $u(z)$ has two options at present,
\begin{align}
    \label{eq:weight}
    {\rm Average:} &~ u(z) = 0.5,{~\rm or}\\
    {\rm Linear:} &~ u(z) = \frac{z-\rho_{j}}{\rho_{j+1}-\rho_j}.
\end{align}
The weight integral can be analytically derived using a Gaussian
function and the error function. Linear interpolation more closely reflects
the underlying physics as illustrated in figure~\ref{fig:interp} because
linear interpolation lends to a continuous variation of the field response
and smooth waveforms at the paths of each calculated field response.

Following $Duct$, a final digitization step is applied, currently supplying a
simple ADC model.  It is parameterized by the ADC resolution, full scale
voltage range and baselines to apply to the input voltage waveforms. For MicroBooNE, a 2-MHz, 12-bit ADC is used with the dynamic voltage range up to \SI{2000}{\mV}.


\subsubsection{Truth values}\label{sec:MCtruth}

A new set of input quantities of the simulation (MC truth) is constructed to compare the ionization electrons
generated by Geant4 and propagated to the wire planes to those reconstructed
after signal processing. MC truth retains track ID and PDG code (particle
species) information (from the generator) to facilitate the assessment of the
efficacy of downstream reconstruction algorithms. As described in
section~\ref{sec:method_filter}, software filters are applied in signal
processing to extract the ionization signal and mitigate high frequency noise.
These filters smear the charge signal in the time and wire dimensions, or
longitudinally and transversely, respectively. For equitable treatment
of MC truth to that posed by signal processing, the same
high frequency time and wire software filters are treated directly to the
ionization electrons as a function of time and wire in construction of the MC
truth signal.

\subsubsection{Noise simulation}\label{sec:noise_model}

The signal simulation described above produces ADC-level,
time-domain waveforms in each channel. To simulate detector data,
waveforms arising from inherent electronics noise are added to the
signal waveform. An analytic method to simulate the noise is introduced
below and it is applicable to a wide range of noise simulations other
than the inherent electronics noise.

This method was motivated by~\cite{noise_Milind}.
The core of this method is to simulate the stochastic behavior of the noise
in the frequency domain. The stochastic effect originates from the random
occurrence of each noise pulse (e.g. due to thermal fluctuation) and the
method is applicable in the condition that the occurrence of each noise pulse
is uniformly distributed in time. With this condition, the noise stochastic
behavior in the frequency domain follows a random walk process and the
analytic mathematical description of the random walk can be used to model the
noise. A single parameter - the mean frequency amplitude - is required for
the noise simulation for each frequency. This parameter is extracted from
data. Details of the method are described below.

In general, the inherent electronics noise can be categorized into several
types: white noise, flicker noise (pink noise), and Brownian noise
(red noise)~\cite{op_amp_handbook, IEEE_noise, noise_webcast}. The
corresponding power spectrum densities are constant, proportional to
$1/f$, and proportional to $1/f^2$, respectively. The total noise divergence
(at infinite frequency for white noise and at infinitesimal frequency for the
latter two types of noise) does not occur because the detector devices
themselves have cutoffs at low and high frequencies. Meanwhile, the power
spectrum can be altered by the response of the dedicated electronics device.

Given any kind of inherent electronics noise pulse, assume its function in
the time domain is $i(t)$ and in the frequency domain is $I(\omega)$.  For
white noise $I(\omega)$ is $G(\omega) \cdot 1$  and for flicker noise
$G(\omega) \cdot \frac{1}{\sqrt{\omega}}$, where $G(\omega)$ includes the
normalization for a single noise pulse as well as the response of the
electronics device. 
 
Consider a train of noise pulses: the function in time domain is
\begin{equation}
    f(t) = \sum^{N}_{n=1}q_n \cdot i(t-t_n),     
\end{equation}
and in frequency domain is
\begin{equation}\label{eq:noisefreq}
    F(\omega) = \sum^{N}_{n=1}q_n \cdot I(\omega)\cdot e^{-j \omega t_n},     
\end{equation}
where $q_n$ is the sign (+ or -) of the noise pulse, $n$ is the index, $j$ is the imaginary unit, and $t_n$ is the occurrence time of each noise pulse, uniformly distributed in time.
        
Absorbing $q_n$ and the phase of $I(\omega)$ into the phase term, equation~\ref{eq:noisefreq} can be rewritten as
\begin{equation}\label{eq:randomwalk}
    F(\omega) = \sum^{N}_{n=1}|I(\omega)| \cdot e^{-j\theta_n(\omega)},
\end{equation}
where $\theta_n(\omega)$ is uniformly distributed in $[0, 2\pi)$ if\footnote{In practice, the discrete Fourier transform naturally discards the information if $\omega \cdot t_N < 2\pi$ and the analysis technique, e.g. region of interest (ROI), can further suppress the low frequency noise. The improper simulation of low frequency noise which does not meet the condition $\omega \cdot t_N \gg 2\pi$ can be ignored.} $\omega \cdot t_N \gg 2\pi$.
    
In the 2D complex plane, given a frequency $\omega$, equation~\ref{eq:randomwalk} follows a 2-dimensional random walk with the angle $\theta_n$ uniformly distributed over $[0, 2\pi)$. Since the number of steps, $N$, is large enough, the probability density distribution of the distance ($r$) from the origin to the end point can be analytically described by~\cite{randomwalk} 
\begin{equation}\label{eq:rayleigh}
    R(r; \sigma) = \frac{r}{\sigma^2}e^{-\frac{r^2}{2\sigma^2}},
\end{equation}
where $R(r; \sigma)$ is the Rayleigh distribution with the mean value of $\sigma\cdot \sqrt{\pi/2}$ and $\sigma^2 = 0.5\cdot N \cdot |I(\omega)|^2$. 

Then, we can represent equation~\ref{eq:randomwalk} in a polar form by
\begin{equation}\label{eq:noiserayleigh}
    F(\omega) = r(\omega) \cdot e^{-i\alpha_{\omega}},
\end{equation}
where $r(\omega)$ follows the Rayleigh distribution and $\alpha_{\omega}$ is uniformly distributed in $[0, 2\pi)$. If $N$ is large enough, the $\alpha_{\omega}$'s are mutually independent for different frequencies.

Equation~\ref{eq:rayleigh} is equivalent to two independent Gaussian distributions on the real and imaginary axes with the same standard deviation $\sigma$, which is the only parameter in the corresponding Rayleigh distribution.
This feature can be employed to simulate the random walk (noise) and to deduce that the summation of multiple different step length random walks can be described as in equation~\ref{eq:noiserayleigh}, because of the additive property of Gaussian distribution. 

\begin{figure}[htbp]
  \centering
  \includegraphics[width=0.6\textwidth]{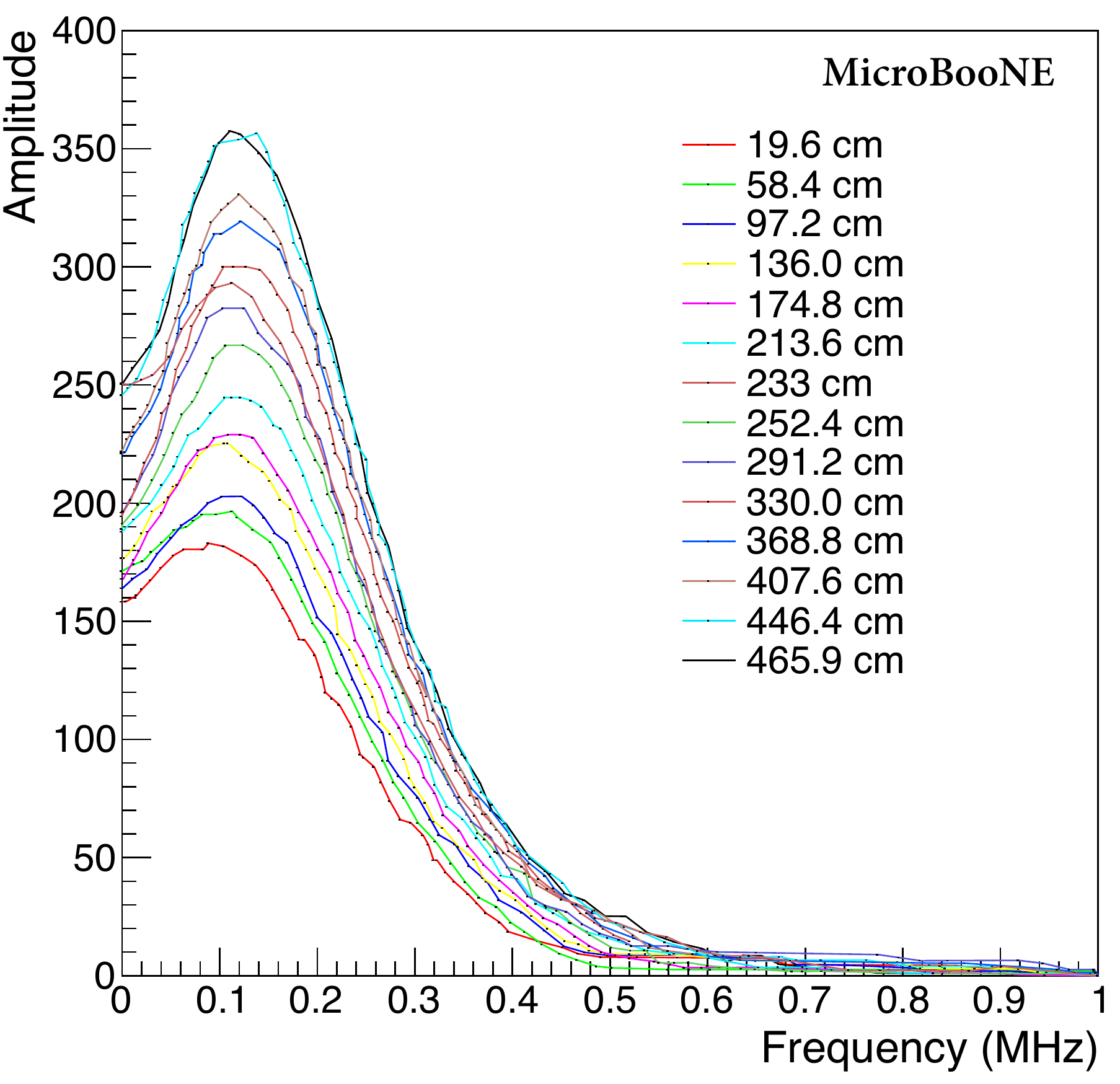}
  \caption{Mean frequency amplitude of the inherent electronics noise for 
    different lengths of wires are deduced from MicroBooNE data using ADC 
    waveforms after noise filtering and mis-configured channel  
    corrections~\cite{noise_filter_paper}. A small constant term 
    which is not associated with the pre-amplifier is subtracted.
    The length of \SI{233}{\cm} is associated with collection plane wires,
    and the rest are associated with induction plane wires. The mean      
    amplitude for intermediate wire lengths is obtained by interpolation.}
  \label{fig:noiseamp}
\end{figure}

As a result, all sources of noise can be summed, and for each frequency the simulation can be done by sampling the Rayleigh distribution, which has a single parameter $\sigma_{total}$ to randomize the frequency amplitude with a uniformly distributed phase from 0 to 2$\pi$. 
The $\sigma_{total}(\omega)$, i.e. the mean frequency amplitude divided by
$\sqrt{\pi/2}$, can be extracted from the data. In this paper, the mean frequency amplitude was extracted from MicroBooNE data as shown in figure~\ref{fig:noiseamp}. An inverse Fourier transformation of the randomized noise frequency spectrum will provide the final noise waveform in the time domain.

\subsection{Quantitative evaluation of the signal processing}\label{sec:quan_eva}
A qualitative demonstration of the signal processing performance is shown
in figure~\ref{fig:simevent}. Several features of the raw and reconstructed tracks can be seen: the fine structure of the charge depositions along the track, the short tracks off the main trajectory, and the high charge density of the track in the collection plane, mainly due to the small angle of the track with respect to the Y wires.

In this section, we present a quantitative evaluation of the signal processing.
This evaluation addresses the intrinsic charge smearing in the time and wire
dimensions, charge matching among the three wire planes, and reconstructed
charge resolution, bias, and inefficiency for large angle tracks.
Understanding these effects is critical for subsequent event reconstruction.
Below are the definitions of these quantities.
\begin{description}
    \item{\textit{Charge}} -- the number of ionization electrons. For a track (line charge), a constant charge density is used.
    \item{\textit{Time smearing}} -- the standard deviation of the Gaussian distribution of the deconvolved charge time spectrum on a wire, specifically associated with a point source of charge.
    \item{\textit{Wire smearing}} -- the fraction of the integrated deconvolved charge on a wire with respect to the total deconvolved charge over the entire range of fired wires, specifically associated with a point source of charge.
    \item{\textit{Charge bias}} -- the mean fractional difference between the total deconvolved charge and the true charge. In terms of a track (line charge), the mean value of the distribution of each wire's integrated deconvolved charges from a range of wires will be used to calculate the charge bias with respect to the true charge within one wire pitch.
    \item{\textit{Charge resolution}} -- the standard deviation of the total deconvolved charge relative to the mean deconvolved charge. In analogy to the definition of charge bias, in terms of a track (line charge), the distribution of each wire's integrated deconvolved charges from a range of wires will apply. 
    \item{\textit{Charge inefficiency}} -- specifically associated with tracks (line charges), the fraction of the wires which have ZERO deconvolved charge (no ROI found). Note that these wires will not be involved in the calculation of charge bias or charge resolution.

\end{description}

\subsubsection{Basic performance of the signal processing}\label{sec:basic_perf}
\begin{figure}[htbp]
  \centering
    \begin{subfigure}[]{1.0\textwidth}
    \includegraphics[width=0.95\textwidth]{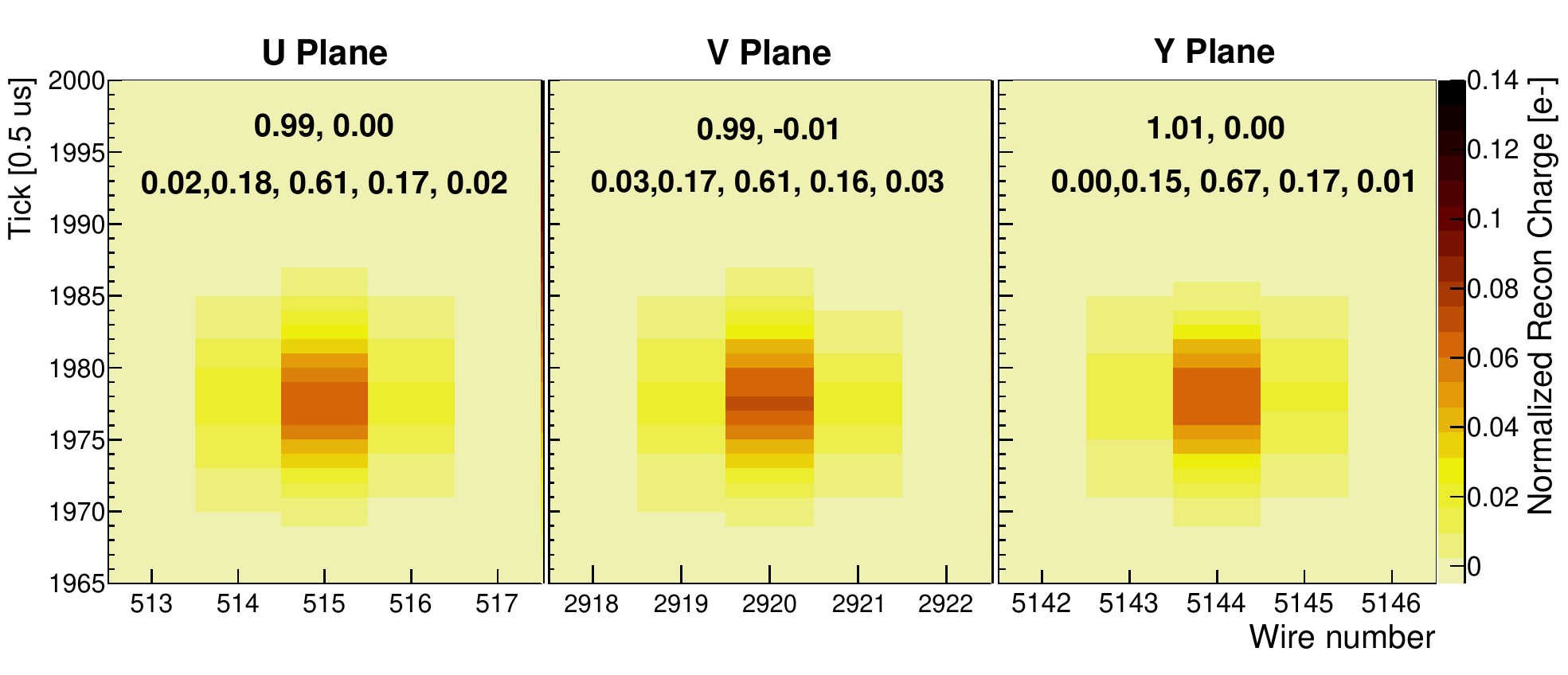}
        \caption{Point source at the wire (\SI{0.0}{\milli\meter} transverse position relative to the closest wire).}
    \label{fig:point-charge-wire}
    \end{subfigure}
    \begin{subfigure}[]{1.0\textwidth}
    \includegraphics[width=0.95\textwidth]{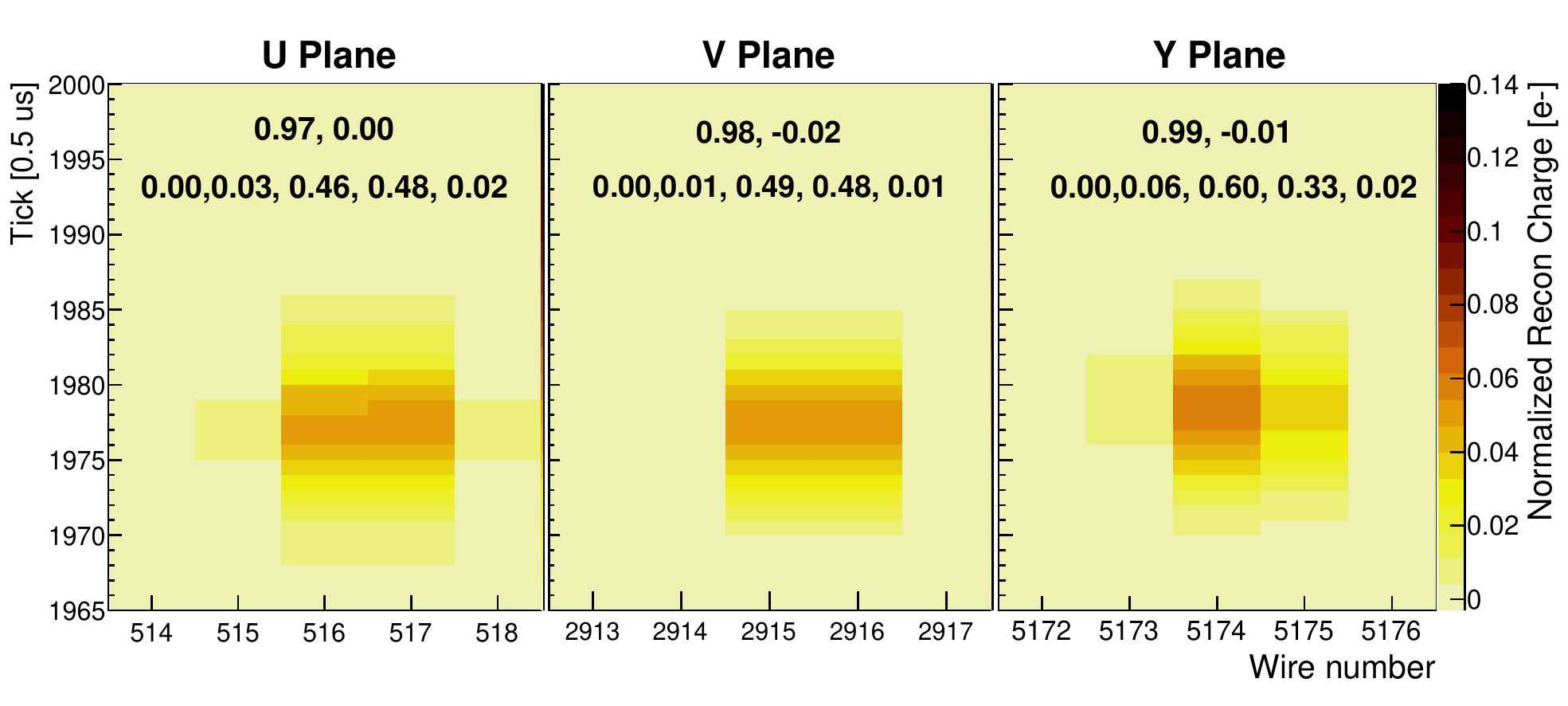}
        \caption{Point source at the wire region boundary (\SI{1.5}{\milli\meter} transverse position relative to the closest wire).}
    \label{fig:point-charge-bound}
    \end{subfigure}
  \caption[Simulated point charge.]{Point source charge after signal processing. Charge is shared across wires
    on a given plane. The fractional charge recovery relative to $1\times10^4$ electrons is indicated by the numbers in the top row, negative charge by the former number
    and the positive charge by the latter one, rounded to 0.01. Relative charge per wire is indicated in the lower row.}
  \label{fig:point-charge}
  \end{figure}

The deconvolved results in the three planes for a point source of charge are shown in figure~\ref{fig:point-charge}.
Ten thousand electrons were simulated one meter from the wire planes. No noise was included;
only diffusion and response functions were simulated. The electronics responses in simulation and deconvolution are identical.
The results for two extreme positions of the point source of charge were presented, at the wire and at the wire region boundary, respectively.

\textbf{Time smearing} is about 2.7 ticks and 2.3 ticks for induction and collection planes, respectively, largely dictated by the high frequency software filters in the signal processing. A one meter drift span provides a longitudinal/time
smearing of about 2 ticks.   

\textbf{Wire smearing} is indicated by the numbers in the second row
in figure~\ref{fig:point-charge}. More than \SI{60}{\percent} of charge
is extracted by the closest wire if the point source of charge is close to the wire. 
For the induction plane, about 50\% of the charge is extracted by both the closest wire and the adjacent wire if the point source of charge is close to the boundary of the wire region, due to the long range of induction as shown in figure~\ref{fig:fielddemo}. For the collection plane, this effect is smaller due to the predominant collection signal on the closest wire. 

\textbf{Charge bias} is at the \SI{1}{\percent} level. In this case, it originates from the mismatch of the field response in signal simulation (fine-grained) and deconvolution (average response). Positive charge (a fraction of distorted deconvolved charge spectrum below zero) can also be reconstructed due to this mismatch. A 1-m drifting diffusion is applied which smears the charge deposition and reduces the mismatch of the average response and position-dependent response within one wire pitch. 
As a result, the extracted charge across the three wire planes is reasonably
well-matched. Charge bias arising from small position-dependency
within the wire region is inevitable. 

The deconvolved charge spectrum of any event topology can be obtained through a convolution with the point-charge ``response'', i.e. the deconvolved charge spectrum of point-charge as shown in figure~\ref{fig:point-charge}. For instance, it is anticipated that the charge bias should also be at the \SI{1}{\percent} level for a track of charge.  
However, including the electronics noise has a significant impact on the performance, as explained in the following section.

\subsubsection{Charge resolution due to electronics noise}\label{sec:cr_elecnoise}
Given an energy deposition, the overall charge resolution originates from the fluctuation of the number of ionization electrons in the signal formation. Several effects are considered, including the statistical fluctuation in the production of ionization electrons, recombination with argon ions, and absorption in drifting due to impurities of liquid argon.
The total statistical fluctuation can be ignored relative to the impact of electronics noise. For instance, a MIP track
corresponds to $\sim10^4$ ionization electrons within one wire pitch. The
statistical fluctuation in production is $\sqrt{10^4\cdot 0.1} = 32$ and in drifting is $\sqrt{10^4 \cdot 0.7 \cdot (1-0.7)} = 50$, where $0.1$ is the typical Fano factor~\cite{fano_factor} and $0.7$ is the typical survival probability of ionization electrons considering recombination and absorption~\cite{recombination1, recombination2, recombination3}. By contrast, the equivalent noise charge (ENC) after deconvolution is $\sim$1k electrons.

\begin{figure}[htbp]
  \centering
  \begin{subfigure}[]{0.49\textwidth}
    \includegraphics[width=0.95\textwidth]{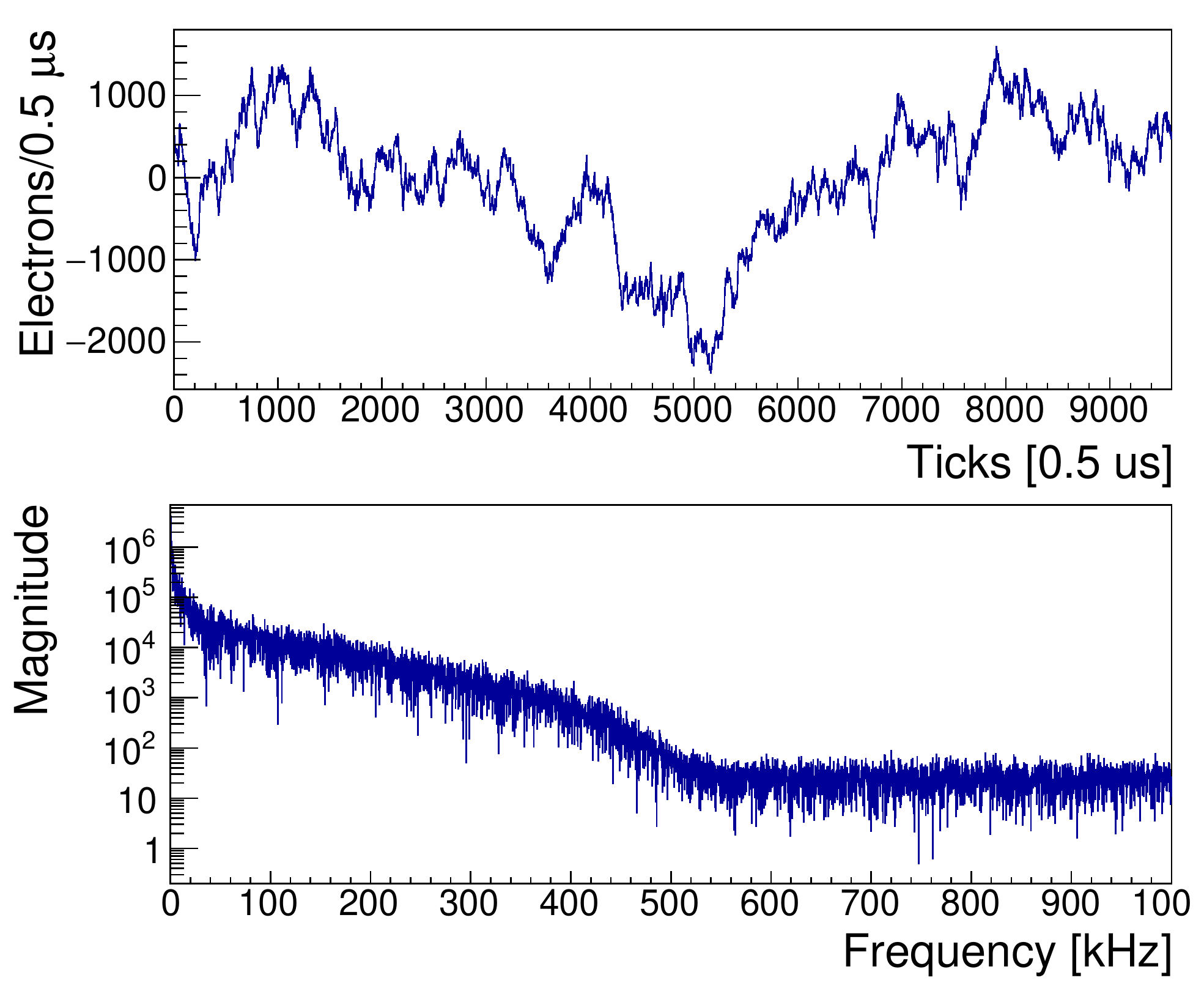}
    \caption{Induction plane.}
    \label{fig:noisecharge_a}
  \end{subfigure}
  \begin{subfigure}[]{0.49\textwidth}
    \includegraphics[width=0.95\textwidth]{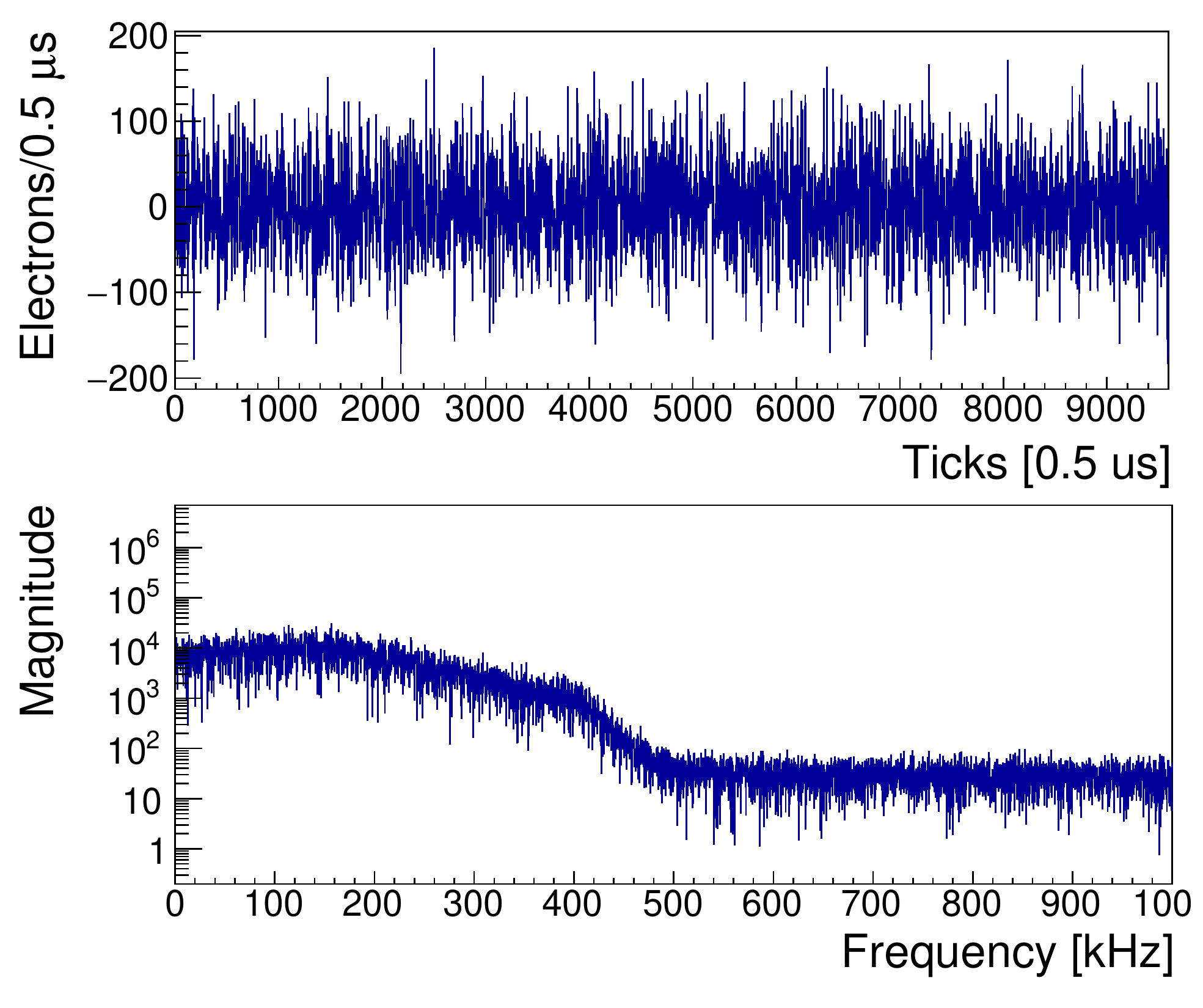}
    \caption{Collection plane.}
    \label{fig:noisecharge_b}
  \end{subfigure}
  \caption{Noise-induced charge distribution after signal
    processing for the full waveform. The top row is in the time domain.
    The bottom row is in the frequency domain after FFT.}
  \label{fig:noisecharge}
\end{figure}

Thus, the electronics noise is the main contributor to the charge resolution.
Before deconvolution, the ENC is roughly the same for all three wire planes
(about 300$\pm$50 electrons irrespective of wire length) as shown
in ~\cite{noise_filter_paper} and illustrated in
figure~\ref{fig:noiseamp}. After deconvolution, the noise-induced charge
is magnified from the ENC value, especially by the bipolar field response
of the induction planes. When performing the deconvolution in the frequency
domain for the induction planes, the induction plane bipolar response is
in the denominator. This response is suppressed at low frequencies (see
figure~\ref{fig:induction_field}), thus amplifying the low frequency noise.
Figure~\ref{fig:noisecharge} presents the noise-induced charge after
signal processing for the full waveform. For the induction plane, there is a very large contribution from the lowest frequencies.
By contrast, the collection plane
has a similar deconvolved charge distribution as the original noise
waveform, apart from a unit conversion from electrons to ADC
counts ($\sim$182 $e^-$/ADC).   

\begin{figure}[htbp]
  \centering
  \begin{subfigure}[]{0.49\textwidth}
    \includegraphics[width=1\textwidth]{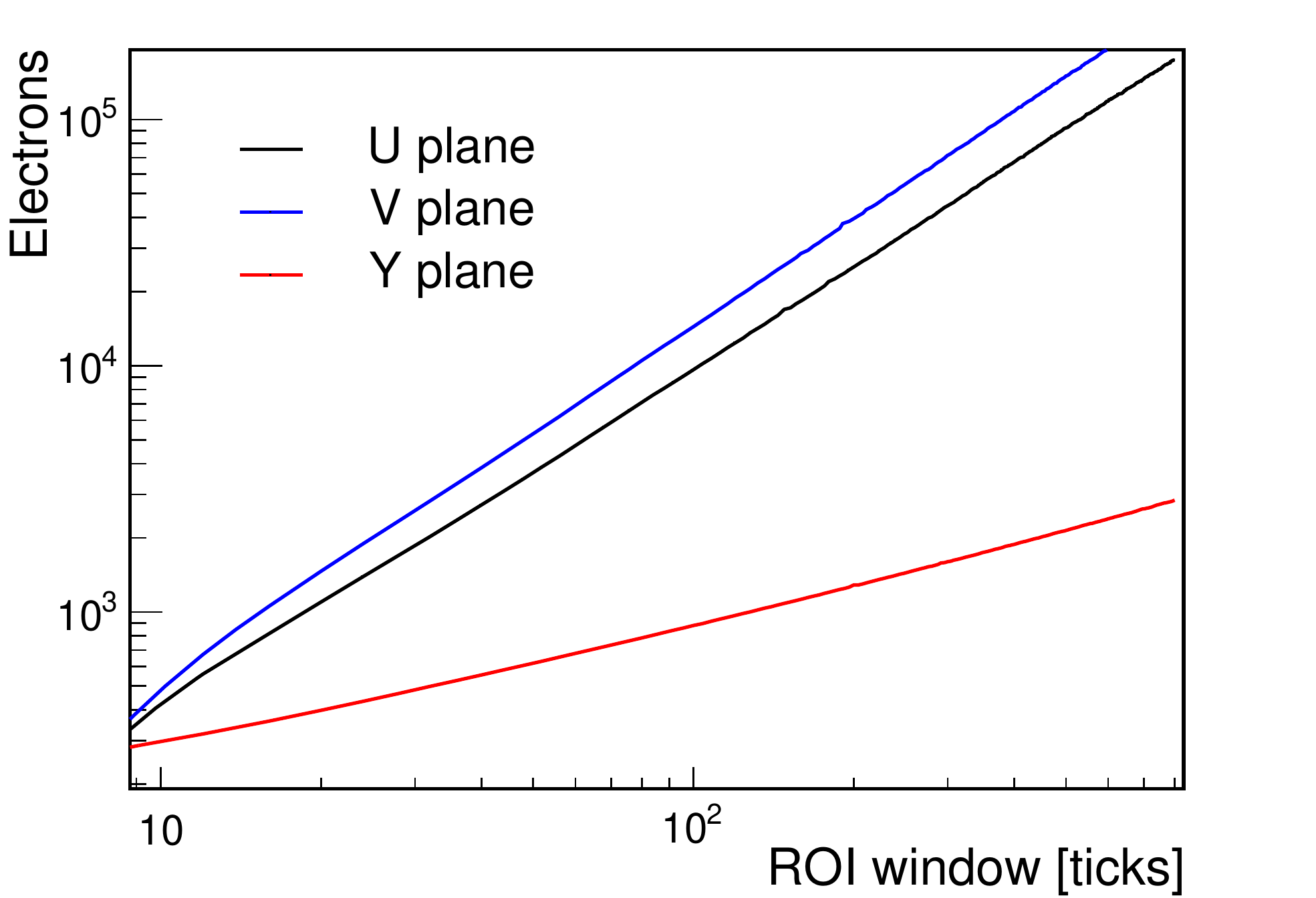}
    \caption{Total charge resolution.}
    \label{fig:charge_res_a}
  \end{subfigure}
  \begin{subfigure}[]{0.49\textwidth}
    \includegraphics[width=1\textwidth]{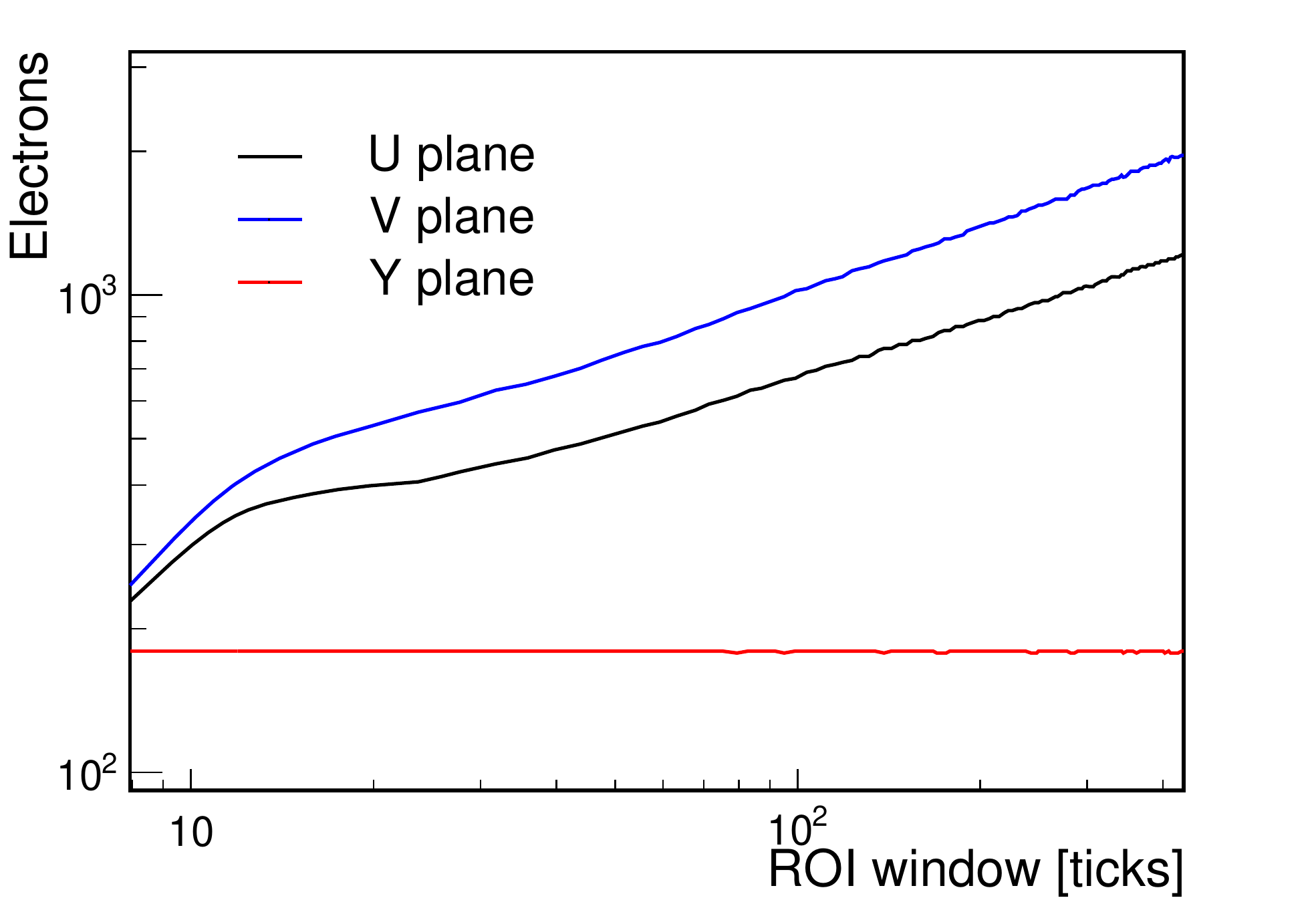}
      \caption{Single bin charge resolution (1 bin = 4 ticks).}
    \label{fig:charge_res_b}
  \end{subfigure}
  \caption{Charge resolution (in units of electrons) due to the electronics
    noise. 1 tick = \SI{0.5}{\micro\second}. 
    (a) Total charge resolution within the entire ROI window.
    (b) Single bin charge resolution within the minimum time unit for subsequent event reconstruction. 
    See text for additional explanation of the variation of the single bin charge resolution for the induction plane.}
  \label{fig:charge_res}
\end{figure}

ROI finding aims to identify a ROI window based on the `true' signal;
therefore, the charge resolution due to noise can be obtained by analysis
of various ROI windows. Two types of charge resolution will be studied
here. The first type is the total charge resolution within the entire ROI
window, which is related to the energy reconstruction for each wire,
i.e. $dE/dx$. The second type is the single bin charge resolution, where
one bin corresponds to the minimum time unit for the subsequent event
reconstruction (e.g. 1 bin may account for multiple time ticks). Figure
~\ref{fig:charge_res} illustrates the total charge resolution and bin
charge resolution for each wire plane as a function of ROI window length.

A linear baseline subtraction is needed for the induction planes to
remove the distortion introduced from the large amount of low frequency
components as discussed in section~\ref{sec:method}. For the collection
plane, this linear baseline subtraction is unnecessary. In fact, if the
baseline subtraction were performed, it would contribute to additional
smearing. The underlying mathematics is implied by the charge spectrum
in the frequency domain as shown in figure~\ref{fig:noisecharge}. The
induction plane has many more low frequency components and stronger
bin-to-bin correlations, whereas the collection plane has smaller
bin-to-bin correlations.

The single bin charge resolution is constant for the collection plane.
For the induction plane, the single bin charge resolution increases 
since the impact of the baseline correction decreases with an increase 
of ROI window length. In addition, the single bin charge resolution 
for the induction plane depends on the location of the
bin within the ROI window. The bin resolution at the boundaries of the
ROI window is zero and increases rapidly for the intermediate region
like a step function. The ``flat top'' of the step has a shallow valley
at the center of the ROI window which varies in flatness up to
\SI{20}{\percent}. For bins ranging from 1 to 6 ticks, the single bin
charge resolution is roughly proportional to the number of time ticks
in one bin.

Based on figure~\ref{fig:charge_res}, given the signal length, the ROI window should be as small as possible while covering the signal in order to improve the resolution. In general, for a point source of charge or a line charge close to parallel to the wire plane (small $\theta_{xz}$), the ROI length is typically $\sim$20 ticks, driven by the time smearing.

Due to the charge smearing, the ability to identify the ROI boundaries is impacted by a low signal-to-noise ratio locally. The variation of the ROI window length in turn alters the charge bias (see section~\ref{sec:charge_bias}) and noise-induced charge (leading term in charge resolution as shown in figure~\ref{fig:charge_res}), introducing an additional smearing of
the total deconvolved charge as part of the charge resolution. This additional smearing comprises up to \SI{10}{\percent} of the total charge resolution. 



\subsubsection{Charge bias due to thresholding in ROI finding}\label{sec:charge_bias}
Waveform ROI finding is required due to noise, especially that magnified by deconvolution in the induction planes. As a consequence, a threshold (see section~\ref{sec:finding_ROIs}) based on RMS noise on
the signal strength dictates the ROI window size on the charge spectrum. The ``true'' charge is smeared after signal processing and a
fraction of deconvolved charge falls below the ROI threshold. Though extension
of the original ROI window is performed, signal loss can occur due to its
exclusion from the final ROI window. This is the major source of charge bias given that the use of the average field response in signal processing has limited impact (see section~\ref{sec:basic_perf}). 
For instance, if noise is included, the deconvolved charge in the distant adjacent wires
in figure~\ref{fig:point-charge}, i.e. a few percent of the total charge
shared by the $\pm$2 wires, will be overwhelmed by noise and entirely below
the ROI threshold; thus, the charge is left absent from the signal processing. 
Additional results in the context of line charges can be seen
in figure~\ref{fig:trackcharge_rbe}.

\subsubsection{Inefficiency of line charge extraction} \label{sec:charge_inefficiency}
Because of the charge bias and charge resolution, 
signal processing can be inefficient for specific event topologies, which means the deconvolved charge spectrum is entirely below the ROI threshold and no ROI window is created.  
This effect is especially pronounced in the induction planes for prolonged tracks with large $\theta_{xz}$, where the bipolar response causes a suppression of the signal, while at the same time increasing the noise. 

The hits on a track tend to share the same $\theta_{xz}$, and this can lead to gaps in the identified track or even the complete disappearance of the reconstructed track. 
Disconnected or absent tracks pose a
challenge for pattern recognition with regard to the track reconstruction as well as vertex identification.

\subsubsection{Performance of line charge extraction} \label{sec:evaluation_result}
To evaluate the signal processing performance of the line charge extraction,
various topologies were simulated. As
described in section~\ref{sec:topology_signal}, the nominal coordinate
system is the detector (collection plane's) Cartesian coordinate
system, with the $y$-axis in the vertical up direction, the $y$-axis in
the drifting field direction, and the $z$-axis in the beam direction.
The line charge is located on this $x-z$ plane with $\theta_y=90^{\circ}$
and a set of varying $\theta_{xz}$'s which define the shape of
the signal. The simulated charge density is $1.6\times10^4$ electrons per
\SI{3}{\milli\meter} (wire pitch) along the trajectory from a 1-meter MIP track centered one meter from the wire plane. Diffusion and inherent electronics noise are applied in the simulation.

\begin{figure}[htbp]
    \centering
    \includegraphics[width=0.7\textwidth]{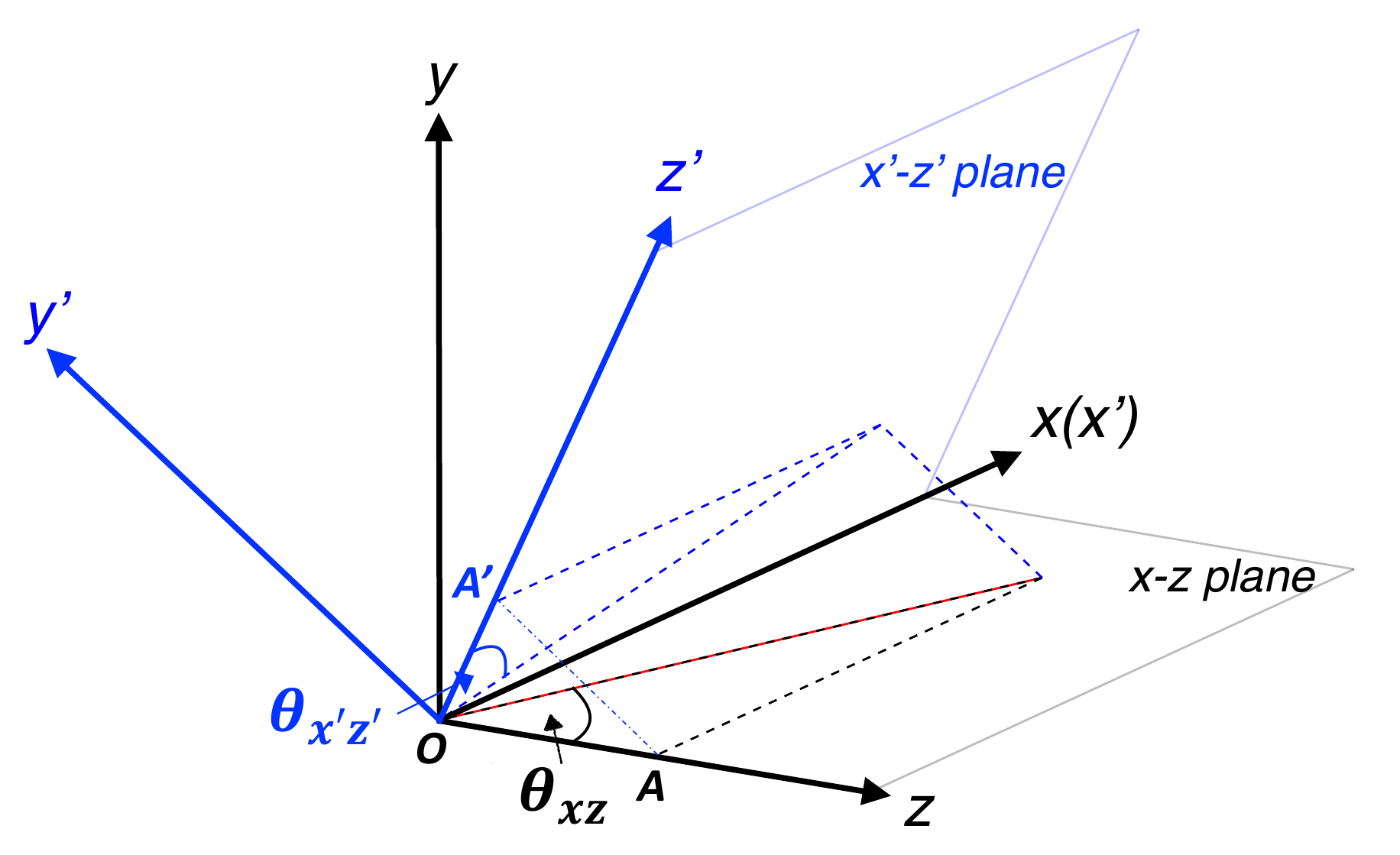}
    \caption{Illustration of the line charge employed to evaluate
        the signal processing performance. Coordinates and angles are defined in 
        figure~\ref{fig:geometry}.
        The line charge (red line) is located
      on the $x-z$ plane (collection plane coordinate, black axes).
      The corresponding projection on the $x'-z'$ plane (induction plane
      coordinate, blue axes), which is a 60$^{\circ}$ rotation of
      the $y(z)$-axis around $x$-axis, is plotted as well. Projections on  
      either wire pitch direction ($\overline{OA}$ and $\overline{OA'}$) 
      are also indicated and $\overline{OA}   = 2\cdot \overline{OA'}$.}
    \label{fig:tracktopo}
\end{figure}

Here one line charge corresponds to two $\theta_{xz}$ angles,
as in $\theta_{x'z'}$ and $\theta_{xz}$ for the induction
planes (same value for U, V planes) and collection plane,
respectively. In figure~\ref{fig:tracktopo}, $\overline{OA}$ and $\overline{OA'}$ are the projections of the track on wire pitch direction for collection plane and induction plane, respectively. In this case ($\theta_y=90^{\circ}$), because $\overline{OA} = 2\cdot \overline{OA'}$ as a result of the $60^{\circ}$ rotation of induction wires, ${\rm tan}\theta_{x'z'} = 2\cdot {\rm tan}\theta_{xz}$ and the total deposited charge within one wire pitch which is inversely proportional to the length
of projection on wire pitch direction is scaled up by a factor of two for the induction planes. 

\begin{figure}[htbp]
  \centering
  \includegraphics[width=0.9\textwidth]{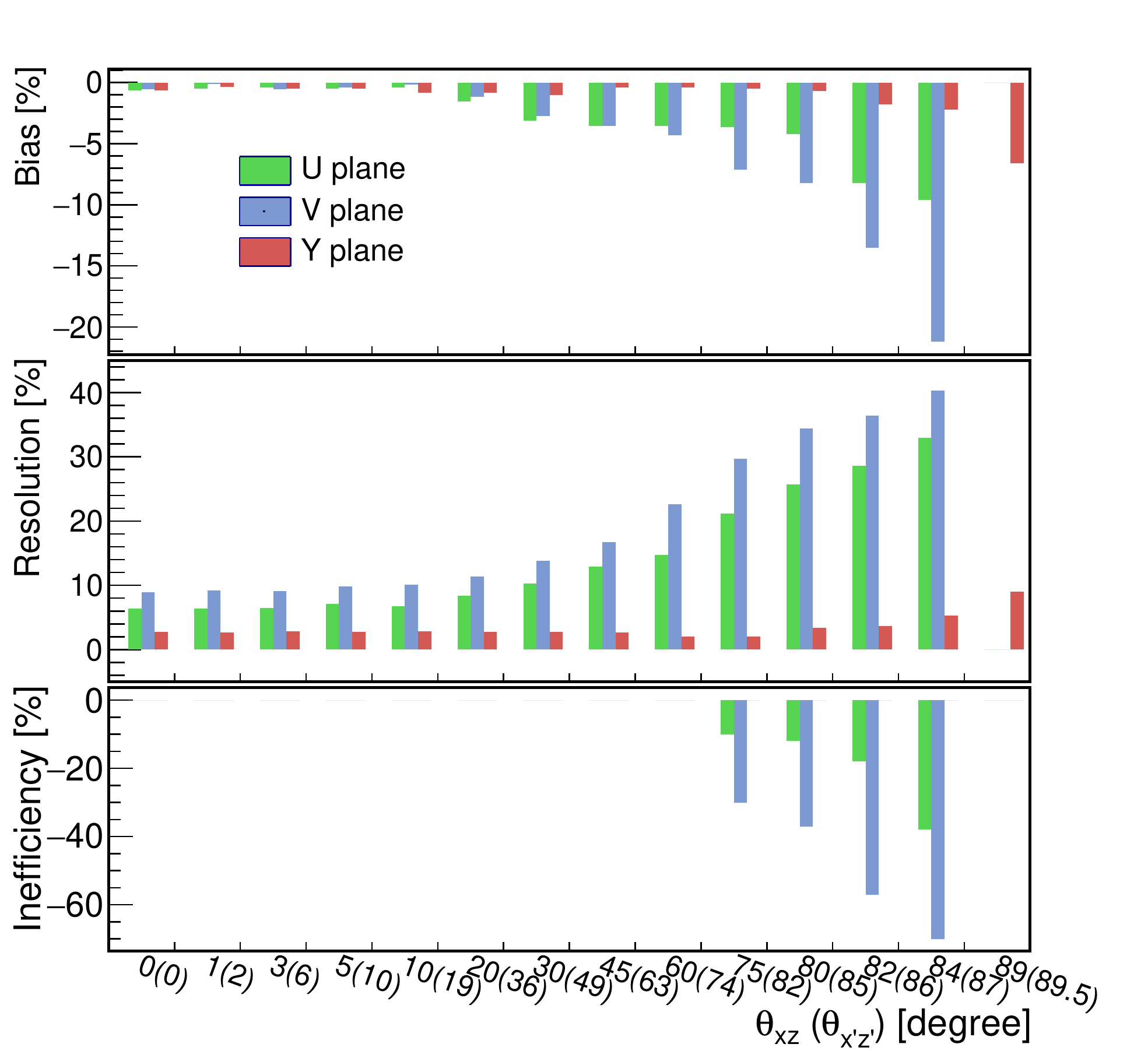}
  \caption{Charge bias, resolution, and inefficiency of the
    reconstructed charge in one wire pitch as a function of
    $\theta_{xz}$ ($\theta_{x'z'}$). The line charge is
    simulated assuming a 1-meter MIP track. Any wire/channel with no
    charge reconstructed is included in the inefficiency
    calculation and ignored in the resolution and bias calculations.
    The last bin (89.5$^{\circ}$) 
    for induction plane corresponds to full inefficiency of charge extraction 
    and has no input for the resolution and bias calculations.
    The Y-axis is relative to the `true' charge deposition
    within one wire pitch before drifting.} 
  \label{fig:trackcharge_rbe}
\end{figure}

\begin{figure}[htbp]
  \centering
  \includegraphics[width=0.55\textwidth]{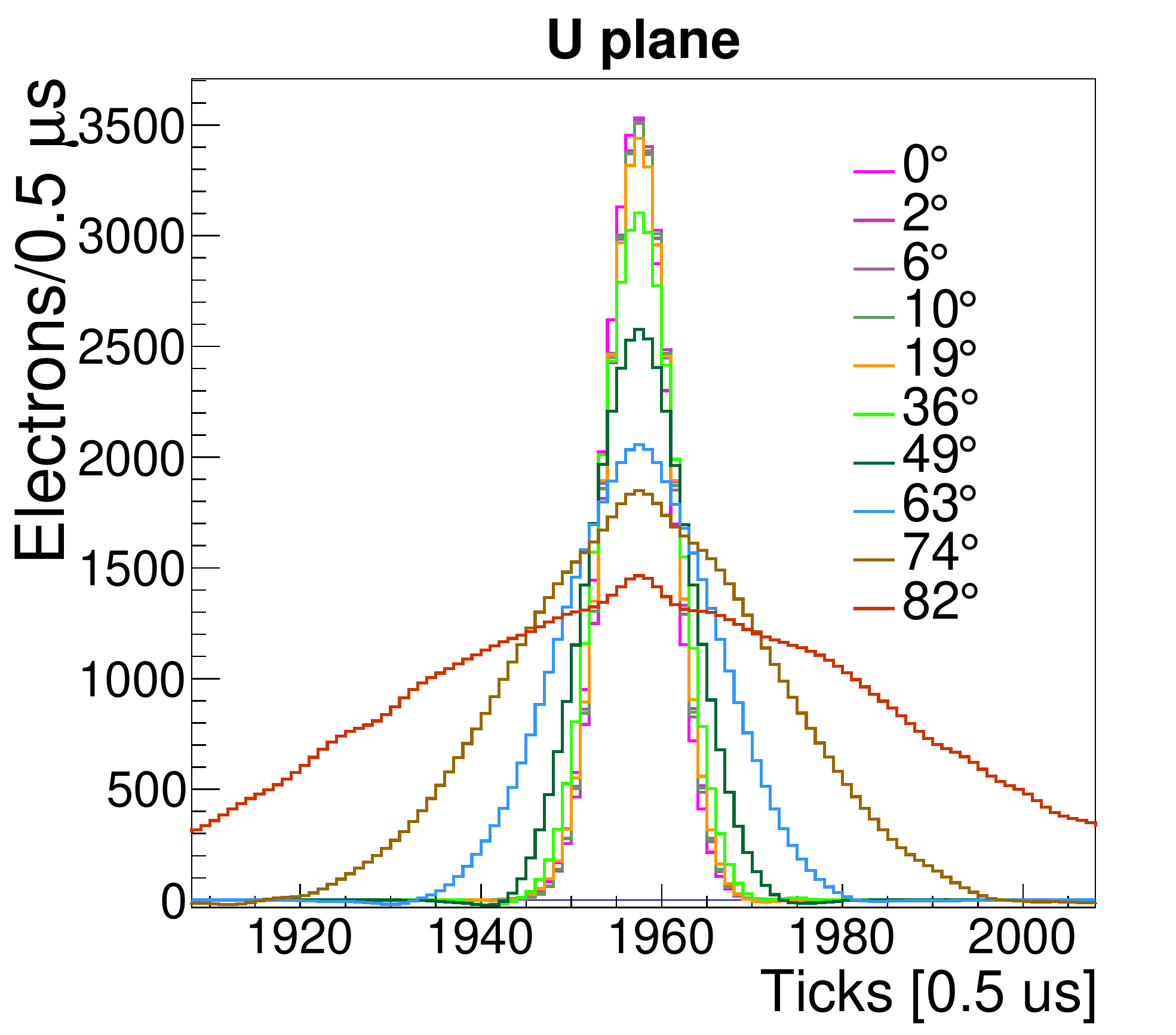}
  \includegraphics[width=0.55\textwidth]{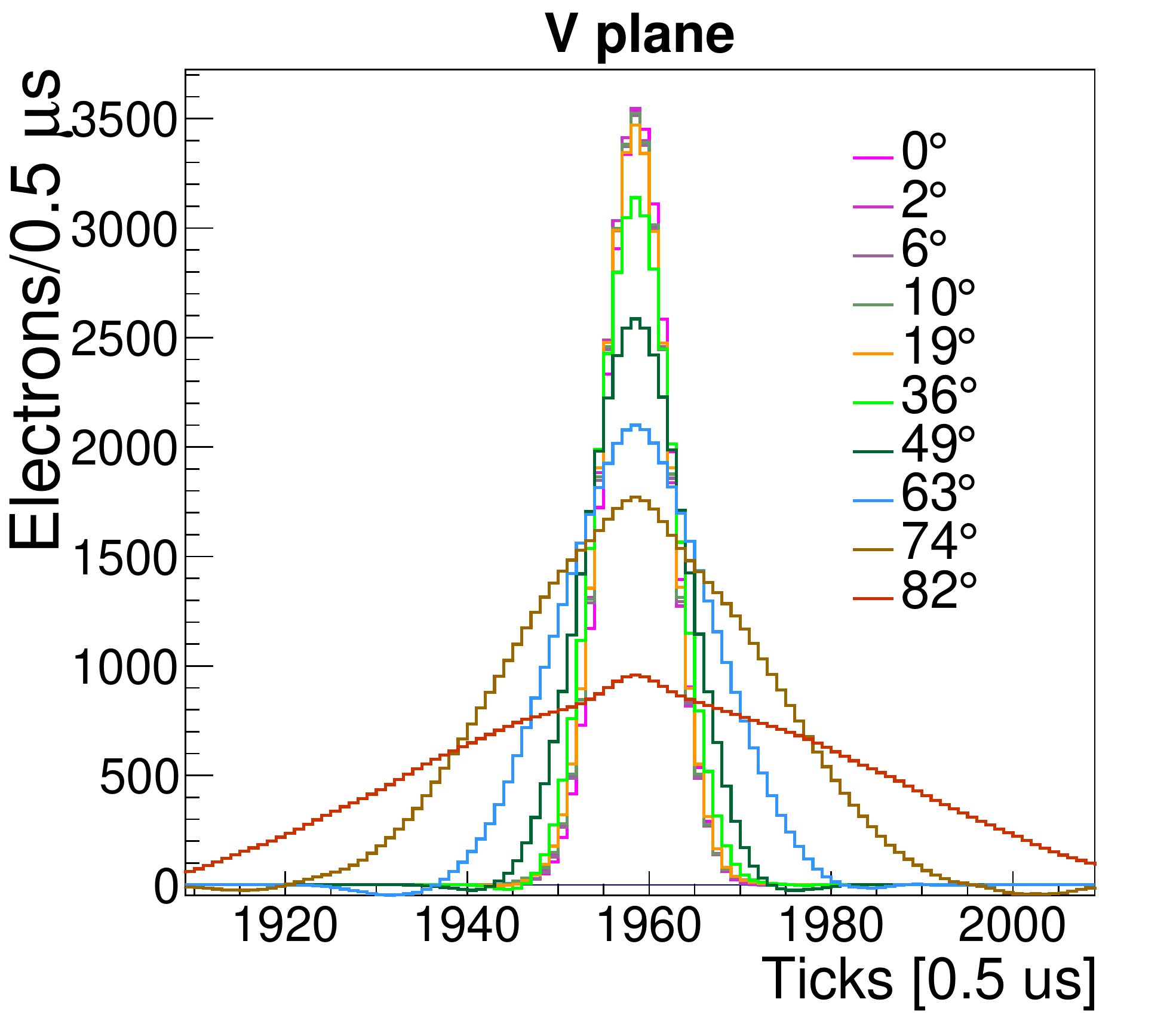}
  \includegraphics[width=0.55\textwidth]{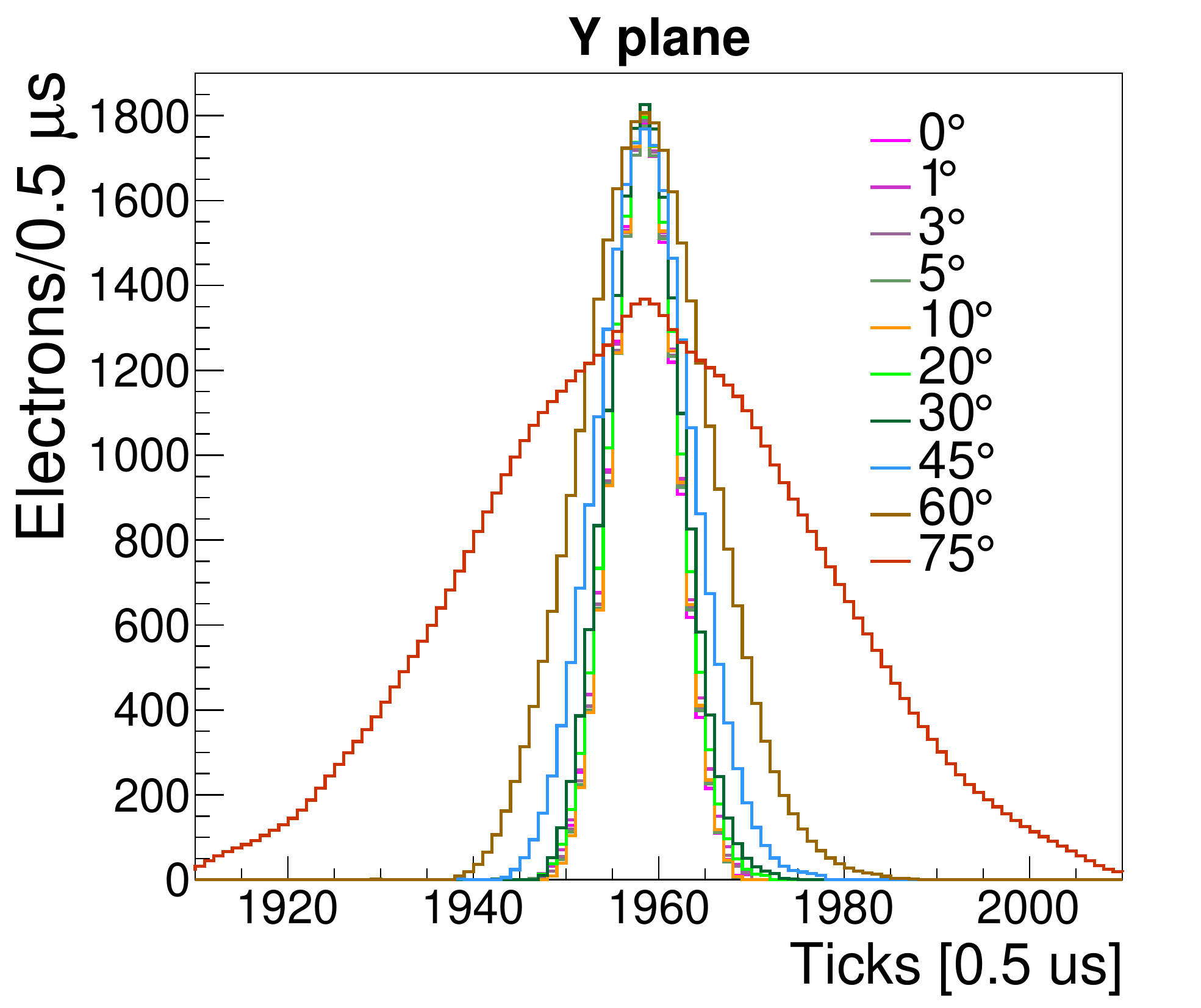}
  \caption{Average reconstructed charge distribution
    within one wire region for the three wire planes
    at various $\theta_{xz}$ values corresponding to figure~\ref{fig:trackcharge_rbe}.}
  \label{fig:trackcharge_wf}
\end{figure}

In general, the performance of the line charge extraction deteriorates with increasing $\theta_{xz}$ as shown in
figure~\ref{fig:trackcharge_rbe}, where 0$^{\circ}$ corresponds
to a track parallel to the wire plane (isochronous) and
90$^{\circ}$ to a track perpendicular to the
wire plane.
The shape of the reconstructed charge spectrum as shown in figure~\ref{fig:trackcharge_wf} is dominated by the Gaussian smearing for small $\theta_{xz}$, or by the track topology for large $\theta_{xz}$ in which case the contributions from the charge in neighboring wires are stretched in time and produce a triangular-like shape.  
A larger $\theta_{xz}$ corresponds to a wider (larger ROI window) and flatter (smaller signal-to-noise ratio) reconstructed charge distribution, therefore associated with a larger charge resolution and bias.

As explained in section~\ref{sec:charge_inefficiency}, due to the bipolar response cancellation (see figure~\ref{fig:sim_perpendicularmip}) and magnified noise (see figure~\ref{fig:noisecharge}), the induction plane has considerably worse performance than the collection plane and is more sensitive to large $\theta_{xz}$.
Specifically, the induction plane has a significant inefficiency of charge extraction for large $\theta_{xz}$. This inefficiency is related to diffusion which introduces an additional smearing. In figure~\ref{fig:trackcharge_rbe}, the result corresponds to the average performance considering diffusion along the track.

The extreme case of a long perpendicular track ($\theta_{xz}=90^{\circ}$)
is not shown in figure~\ref{fig:trackcharge_rbe}. For the induction
planes, similar to the last point in this figure, a long perpendicular track has zero efficiency to extract charge,
with the exception of the ends of the track which have non-zero efficiency.
However, for physics analyses, the phase space around this extreme case, e.g. 89$^{\circ}$-90$^{\circ}$, can be ignored, because, in practice, multiple scattering of the particle along a trajectory most likely results in a few degree path deflection. 
Moreover, for a high energy neutrino beam experiment with the predominance of forward-going event topologies, this impacted phase space is even smaller.
For the collection plane, there is still a high efficiency to reconstruct the charge if the track is perpendicular to the wire plane\footnote{If the track is prolonged in time, the waveform may be identified as noise baseline shifting and filtered.}. Similar to a point source
as shown in figure~\ref{fig:point-charge}, the reconstructed charge is
shared by adjacent wires. 



\section{Discussion}\label{sec:discussion}

The signal processing and TPC simulation described and evaluated
here allows for a detailed appraisal of cosmic and neutrino-induced
TPC event reconstruction in MicroBooNE. 
For example, with this work, the study of cosmic ray reconstruction efficiencies in MicroBooNE,
as shown in ~\cite{uboone-mucs}, can be extended to address the
reconstruction efficacy of tagged cosmic events at the raw waveform level in MicroBooNE. 
However, the current state of signal processing and
TPC simulation is not without deficiencies even though we have ideal implementations of the TPC and electronics designs. 
Some of these shortcomings and methods to remedy them are discussed in the following section.

\subsection{Limitations of 2D field responses}\label{sec:3D_field}

As described above, this work relies on calculating field response
functions using a simplified 2D model of the MicroBooNE detector.  While
these calculations are an improvement over previous ones in terms of including both
fine-grained variations and more correct long-range induction, there exists
some uncertainty and concern about any residual limitations.  This section
lists the potential limitations, discusses some remediation and
enumerates the technical challenges to overcome.

The current model assumes three parallel wire planes which extend to
infinity in both transverse directions and are limited transversely to
ten wires on either side of the central wire of interest.
As such, the model cannot
accommodate detector edges nor variations in individual wire location,
angle, or bias voltage.  The 2D nature disallows accounting for any
possible variation along the direction of the wires and in particular
requires a somewhat arbitrary choice to be made for the
transverse location of wires in one plane relative to the
others.

Also as described above, the induced current is appreciable over a
greater range than just the wire region nearest to a given element of
drifting charge. At the same time, after the distance of ten wire
regions, the strength is reduced to a negligible level in most cases. This means
that multiple sets of 2D field response functions can be used to model
variations on a patchwork basis.  Such an approach is developed for
modeling the MicroBooNE wires with bias voltages consistent with a
short to ground.  This improvement still suffers from the above limitations,
and in particular must have the parameters governing bias voltage and
relative wire positions tuned to match real data from the detector.

One approach to address these limitations is to develop a new model
which spans all three spatial dimensions. This approach goes beyond what the
current Garfield 2D analytic calculation provides.

The Finite Element Method can be applied to 3D geometry of arbitrary
construction.  The required computation time for this method naively
scales as the volume of the geometry relative to the feature size.
Calculations spanning tens of centimeters with wires of \SI{0.15}{\mm}
diameter become computationally challenging.  This feature can be
somewhat mitigated by applying adaptive grid techniques.  Development
of FEM based 3D field calculations is an active area of
investigation but currently they only span a
subset of the total required volume.

Another promising approach to calculate the electrostatic fields in 3D is to
use the Boundary Element Method~\cite{bem}.   Its computation
cost scales as the surface area instead of the volume of the geometry.
An initial pursuit of this approach has been investigated~\cite{bempp}
and some examples of its results are included below.

Regardless of the method employed to solve the Laplace equation in 3D,
there are various technical computing challenges related to the
increase in the size of the resulting field response data set compared
to that from 2D calculations.  For example, the equivalent of defining
a 1D linear grid of six drift path starting points as illustrated in
figure~\ref{fig:garfield_schem} is to calculate paths starting
on some 2D planar grid.  Continuing to require discrete translational
symmetry reduces the problem substantially. The minimum set of unique
drift paths on the MicroBooNE wire crossing pattern (a hexagonal lattice) is
an order of magnitude larger than the 1D linear grid case. For each drift path,
one must still calculate one field response function for each wire within range
for each plane.
The simulation must convolve the drifting charge distributions now in
3D partly by doing a 2D interpolation to the nearest drift paths.

\begin{figure}[htbp]
  \centering
  \includegraphics[width=0.6\textwidth]{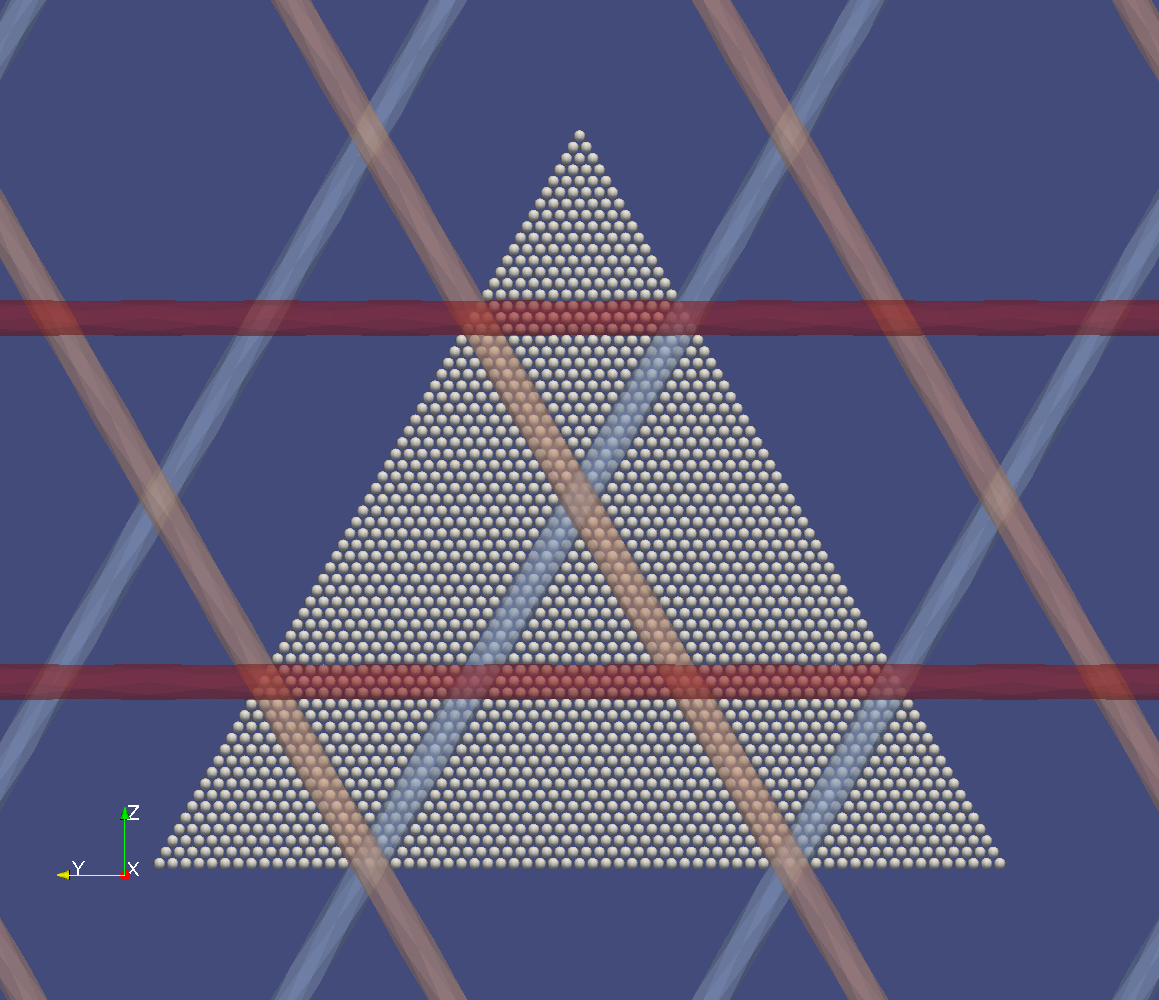}
  \caption{A region of starting points for electron drift paths given a 3D
    model for MicroBooNE.  Blue wires are in the U plane, red are in the Y
    plane.  Potential starting points are represented by white balls.  In
    this example, their separation is roughly \SI{0.1}{\mm} and they span
    a region about four times as large as the absolute minimum given the wire
    crossing symmetry.}
  \label{fig:larf-triangle}
\end{figure}

\begin{figure}[htbp]
  \centering
  \includegraphics[height=3.5cm,clip,trim=0cm 5cm 0cm 17cm]{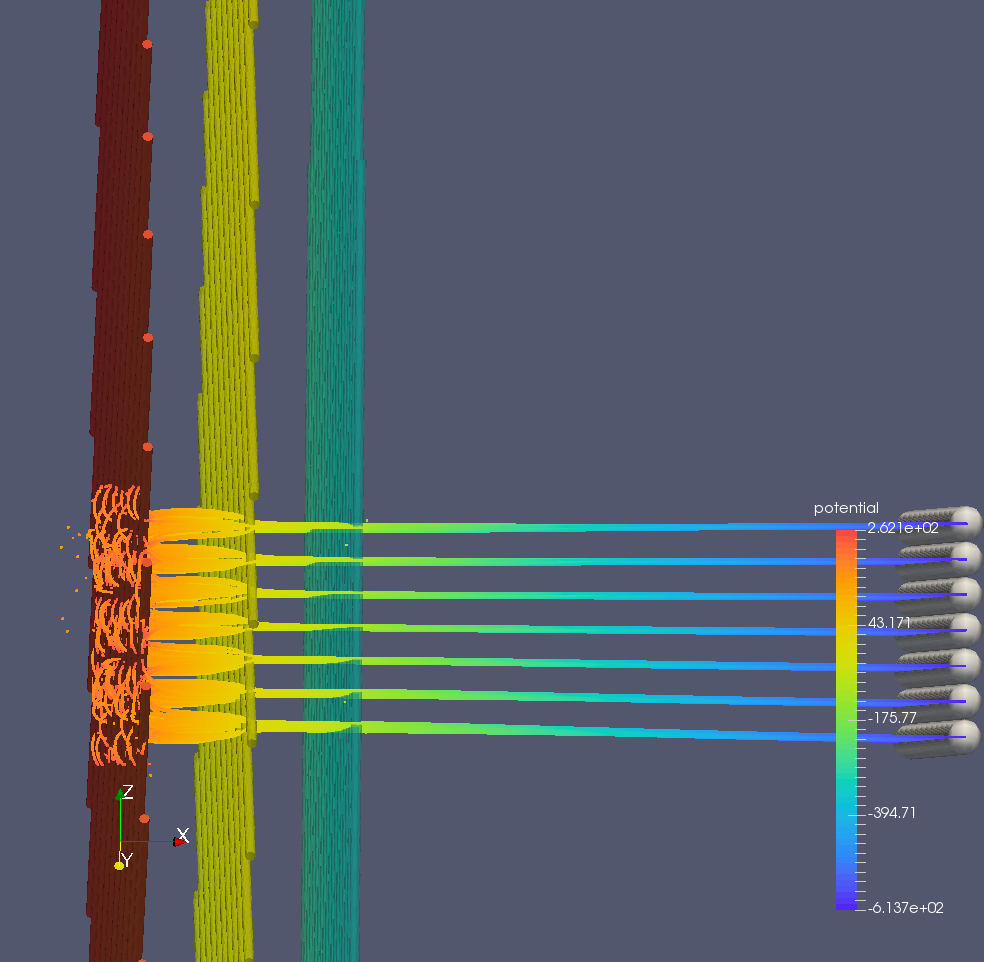}

  \includegraphics[height=4.5cm,clip,trim=10cm 0cm 10cm 0cm]{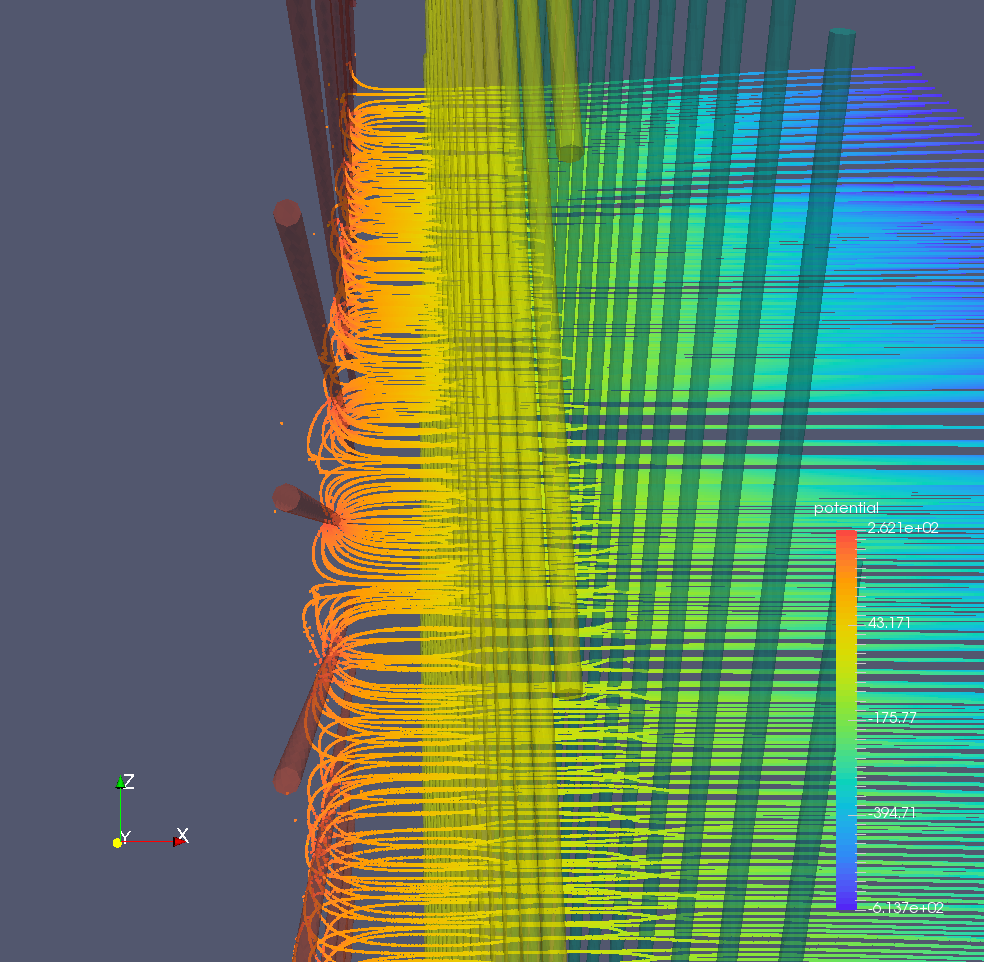}
  \includegraphics[height=4.5cm,clip,trim=4cm 0cm 10cm 0cm]{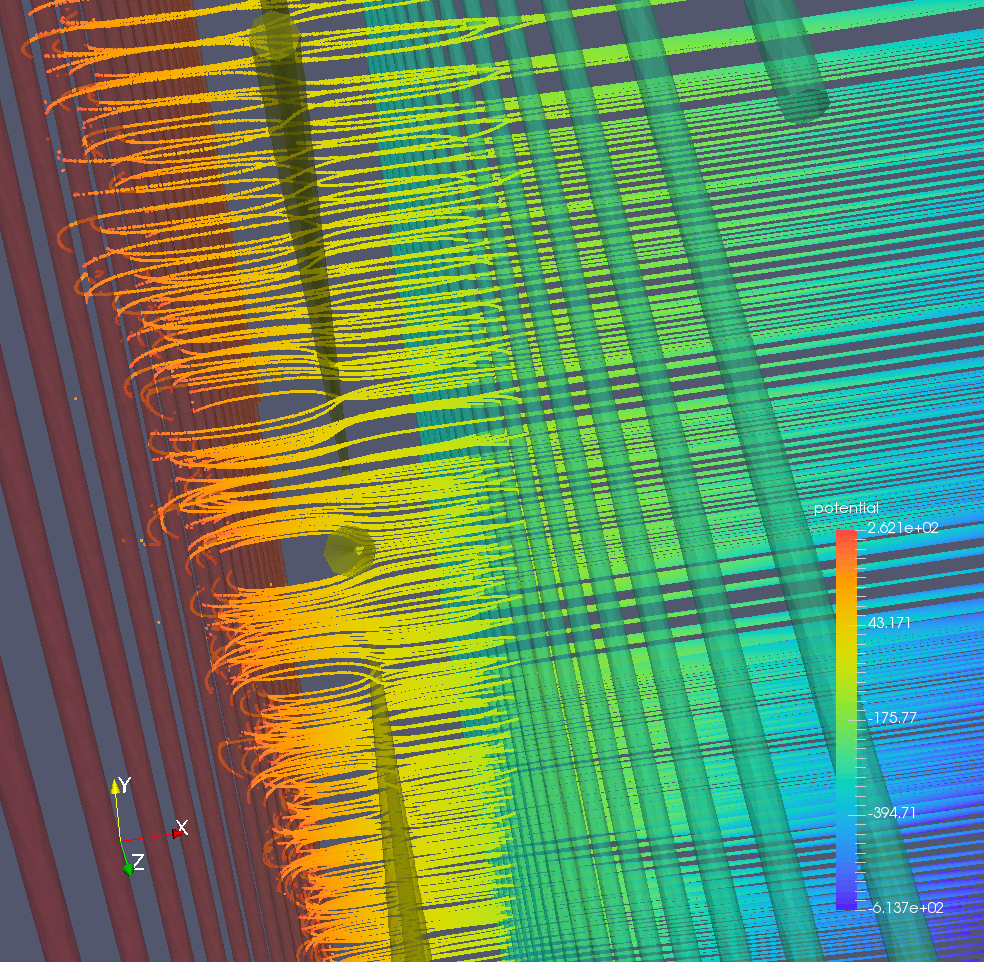}
  \includegraphics[height=4.5cm,clip,trim=4cm 0cm 10cm 0cm]{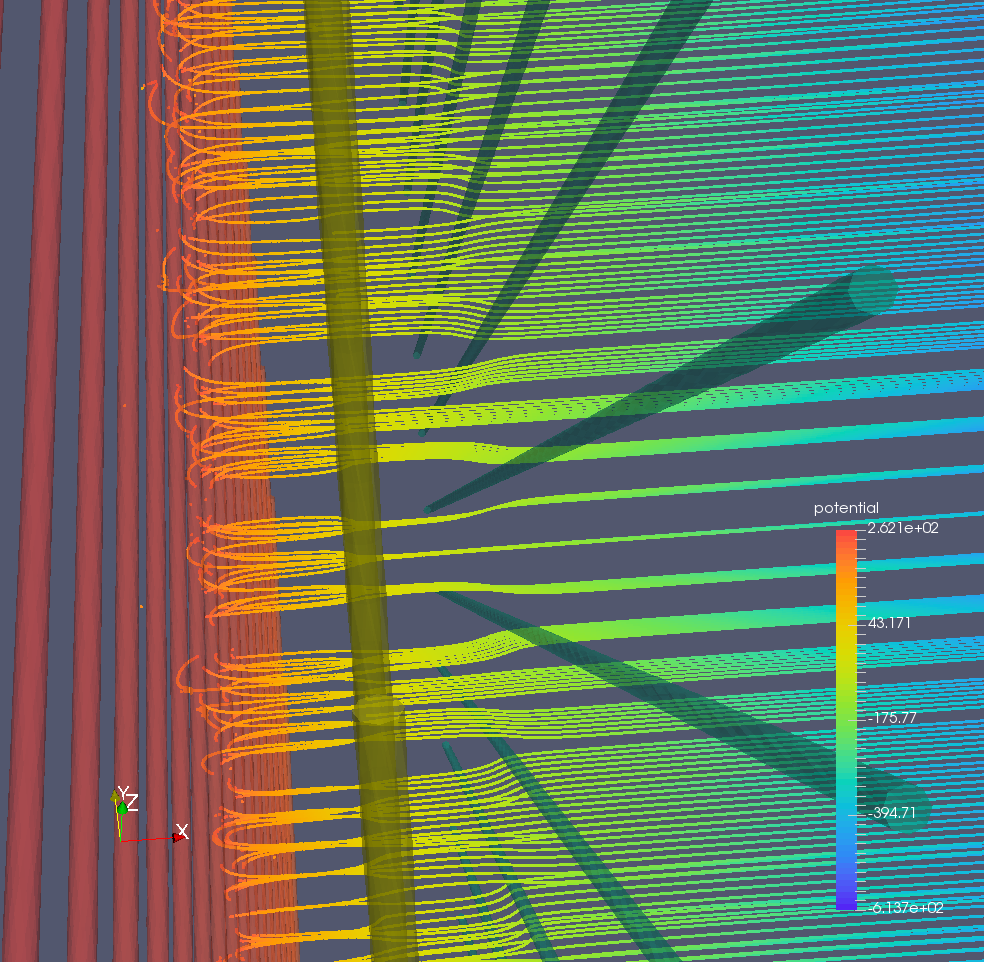}
  \includegraphics[height=4.5cm,clip,trim=10cm 0cm 0cm 10cm]{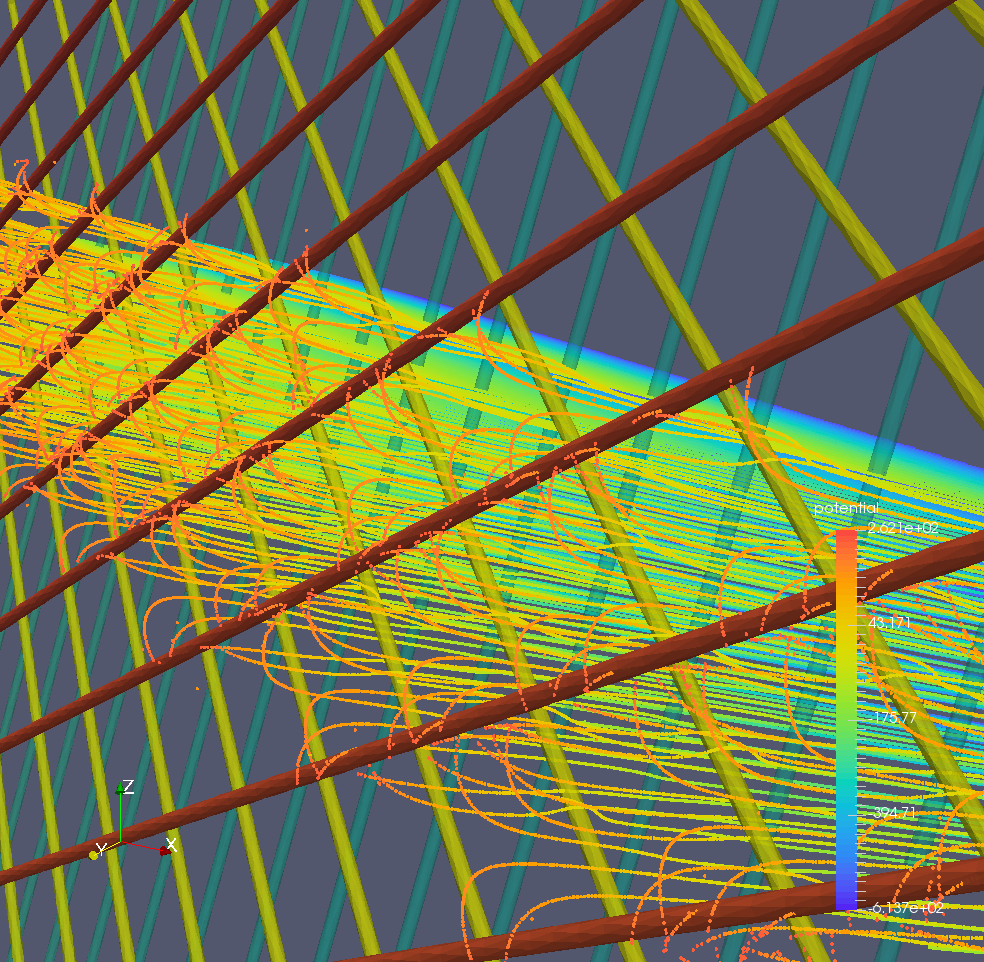}
  \caption{Various views of a family of possible drift paths through a
    3D model of MicroBooNE wires.  Paths are colored based on the
    local electrostatic potential.  Path starting points are shown as
    white balls.  Red wires make up the collection plane.  The top
    image shows a view looking approximately across the detector,
    perpendicular to the collection wires.  The bottom row shows
    views, in order of left to right: along a Y, V and U wire.
    The final view on the bottom is from behind the wire planes.}
  \label{fig:larf-paths}
\end{figure}

Figure~\ref{fig:larf-triangle} illustrates one possible grid
which spans about four times the minimum triangle and has roughly
\SI{0.1}{\mm} spacing.  Detectors lacking the symmetry of the MicroBooNE
wire crossing pattern will require a much larger minimum region.  To
model a region that lacks translational symmetry requires calculating
a family of drift paths which span the entire region.
Figure~\ref{fig:larf-paths} shows an example family of drift paths
somewhat equivalent to those shown in figure~\ref{fig:efield} but
calculated in 3D using the BEM.

In addition to the residual discrepancy between the 2D approximation
and the real 3D case, the current 2D Garfield calculation has
limitations due to its own configuration. Close proximity of the
cathode to the wire planes (in our calculation the
cathode is \SI{20}{\cm} from the V plane) results in a squeezing of the
weighting potential. The effect of this squeezing is most pronounced on
the U plane, which has no shielding. Nevertheless, this squeezing effect
is only considerable in the case of a long track at small $\theta_{xz}$.
In this circumstance, the field response from a larger range of wires
(greater than the $0\pm10$ wire in the current calculation) is required
because the coherent summation of long-range induction from distant wires
is non-negligible.

A dedicated test-stand to calibrate the single-phase LArTPC field response
to a point source charge would greatly aid in validating the residual 3D
effect to a 2D field response calculation.




\subsection{Limitations of the current ROI finding}\label{sec:limitation_ROI}
\begin{figure}[htbp]
  \centering
  \includegraphics[width=0.8\textwidth]{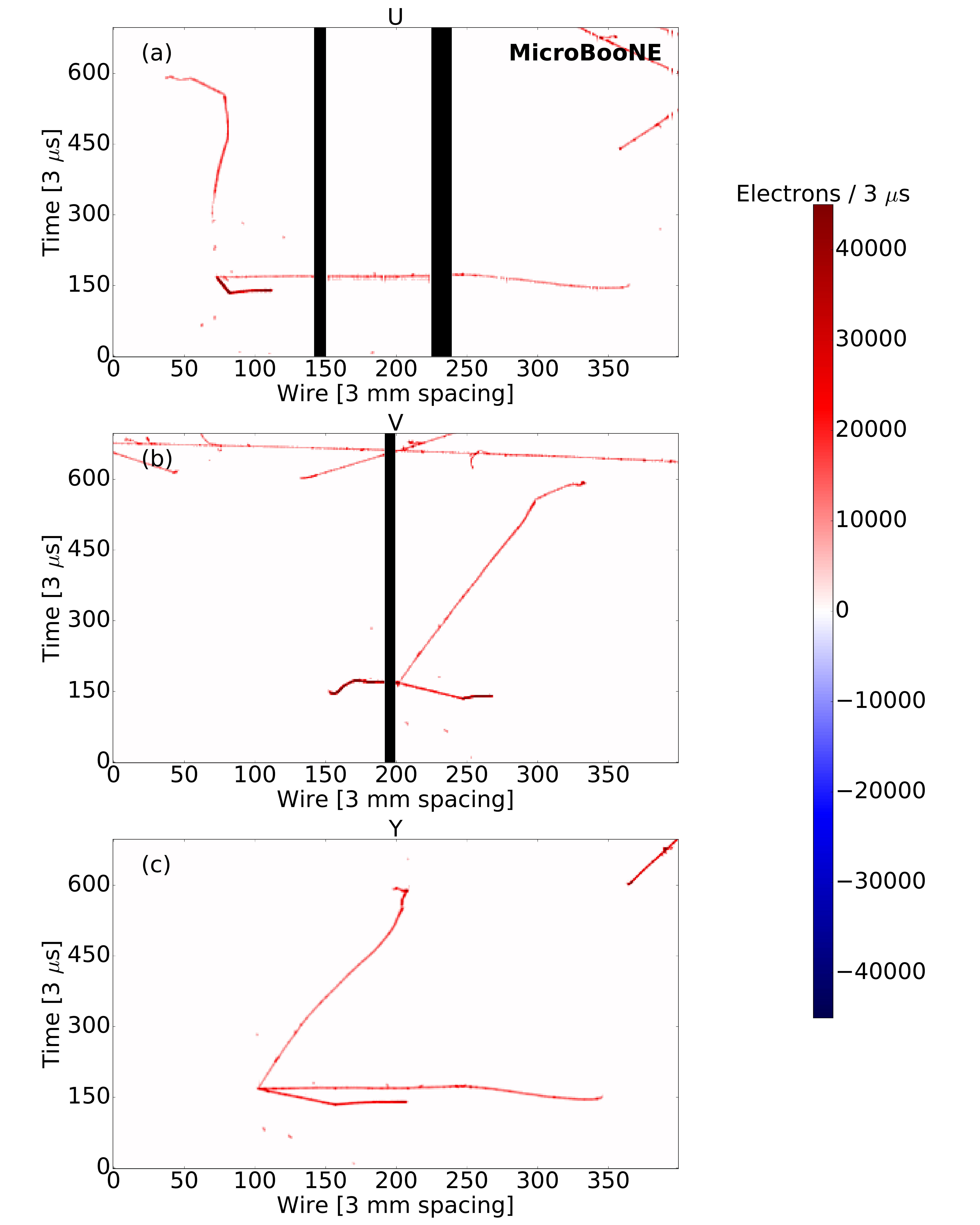}
  \caption{An example topology from MicroBooNE data (event 31,
    run 5366) for which the signal processing
    failed to find an ROI on the U plane. The signal after
    2D deconvolution is displayed. Black rectangles
    denote inactive TPC wires. (a) U plane view. (b) V
    plane view. (c) Y plane view}
  \label{fig:roi-failure}
\end{figure}

\begin{figure}[htbp]
  \centering
  \includegraphics[width=0.8\textwidth]{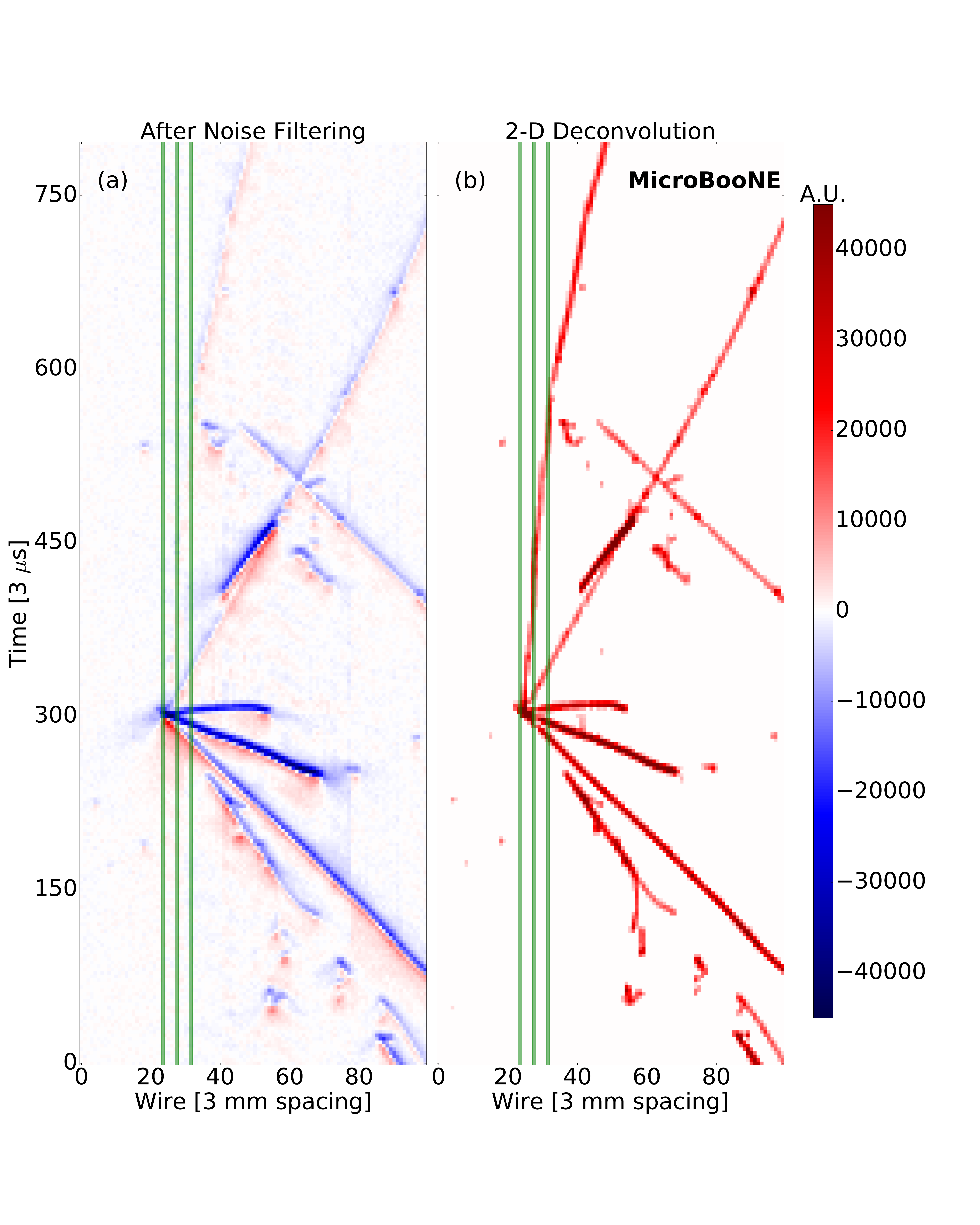}
  \caption{A neutrino candidate from MicroBooNE data (event
    41075, event 3493) measured on the U plane after (a) noise
    filtering (in units of average baseline subtracted ADC scaled
    by 250 per \SI{3}{\us}) and (b) 2D deconvolution (in units
    of electrons per \SI{3}{\us}). The vertical green
    lines denote the wires corresponding to the waveforms and
    charge spectra, respectively, in figure~\ref{fig:triangle}.}
  \label{fig:triangleEVD}
\end{figure}

\begin{figure}[htbp]
  \centering
  \includegraphics[width=0.8\textwidth]{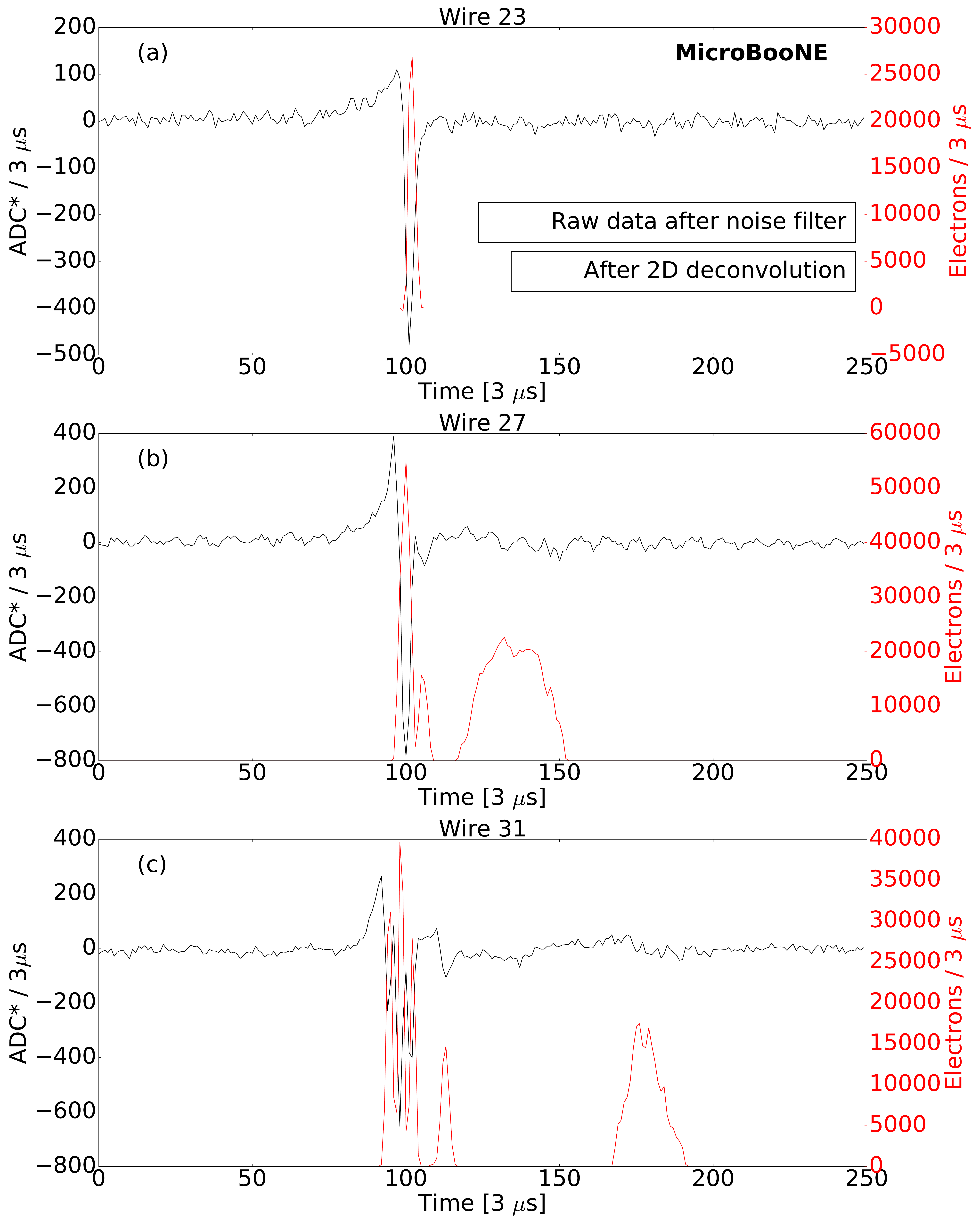}
  \caption{Noise-filtered waveforms (black, in units of average baseline
    subtracted ADC per \SI{3}{\us}) and charge spectra (red, in units of
    electrons per \SI{3}{\us}) corresponding to a U plane track
    with $\theta_{xz}\approx84^{\circ}$.
    Note, the quasi-triangular shaped reconstructed signal from the
    large-angle track. This signal is indistinguishable from the
    baseline in the noise-filtered waveform.}
  \label{fig:triangle}
\end{figure}

ROI finding is a critical step in charge extraction and has direct
consequences for charge bias and inefficiency. For the induction
planes, a prolonged (large $\theta_{xz}$) track will have a large
bias and inefficiency in charge extraction. An example of such a track
is presented in figure~\ref{fig:roi-failure}.
The ROI failure occurs on the U plane. The V and Y planes provide an
indication of how the topology should have been reconstructed in the
U plane view. 
Given that these highly inclined tracks are extended in time and have small amplitude,
the associated ROIs tend not to be found. This effect is more serious at large $\theta_{xz}$ 
and for tracks with long drift time, where the effects of diffusion further reduce the amplitude.
For short drifting distance with small diffusion, ROI finding for large
angle tracks has found some success. Figures~\ref{fig:triangleEVD}
and~\ref{fig:triangle} demonstrate
the successful recovery of a large $\theta_{xz}$, time prolonged signal on the U
plane. In the raw data waveform after noise filtering, notice
that the signal is buried within the resolution of the baseline.
In this case, the 2D deconvolution
procedure excels at extracting this low amplitude signal from
noise. This example also highlights the intra- and inter-wire
dependence of the weighting field on the signal shape, which is
quasi-triangular as discussed in section~\ref{sec:evaluation_result}. 
A web-based interactive tool to explore the raw waveform, waveform
after noise filtering, and the charge spectra after deconvolution
can be found in~\cite{magnify}.

At present, the ROI finding considers the amplitude of the deconvolved
signal compared with a certain predefined noise threshold as
well as the connectivity of the ROI windows in wire and
time dimensions. However, the shape of the reconstructed charge
distribution may be helpful to predict or extrapolate the ROI
window to fully cover the signal range, and the start drifting position
would be used to deconvolve the corresponding diffusion smearing. An
adaptive ROI finding threshold would be useful to balance the noise level
and charge extraction efficacy. In short, a second-pass signal processing
may be attempted with more knowledge of the charge deposition that
can only be obtained after downstream event reconstruction, e.g., start
time matching, 3D object matching/imaging, clustering, tracking, etc.
These considerations are the subject of ongoing work.

\subsection{Impact of excess noise in real data}\label{sec:impact_noise}
Excess noise, as introduced in~\cite{noise_filter_paper},
can be characterized and largely removed by noise filtering prior to
signal processing. However, the noise filter changes the noise
spectrum as well as the signal spectrum. For the MicroBooNE detector,
the coherent noise removal in the noise filter has a large
impact on signal processing. The coherent noise originates
from a low-voltage regulator which serves a number of
channels/wires and manifests itself at the low end of
the frequency spectrum, e.g. $\leq$\SI{30}{\kHz}. For an
isochronous track, the noise filter can mistake the signal as
part of the coherent noise and distort the waveforms from a
range of wires/channels. Unfortunately, signals from the
induction plane, particularly the U plane, are especially affected
and have a large fraction of the power spectrum interfered
with by the coherent noise. This gives rise to an additional
bias for the reconstructed charge for the U plane, a -10 to
-\SI{20}{\percent} effect. As the angle of the track increases
with respect to the wire plane, this effect is mitigated
dramatically, and can typically be ignored for tracks with angle greater than 5$^\circ$.

\section{Summary}\label{sec:summary}
This paper presents the principle of single phase LArTPC signal
formation and a method to perform ionization electron signal
analysis and processing for wire readout schemes. Several components of the signal
processing are described. Most notably, the implementation
of a 2D deconvolution (in time and wire dimensions) is
introduced as a natural complement to the field ranges
and contours inherent to LArTPC signals. Also, the
careful selection and refinement of signal regions of interest
are emphasized for a successful extraction of the charge distribution
from all sense wire planes.

A full TPC signal simulation with long-range and fine-grained
field responses has been developed and employed to evaluate the signal processing.
Also, an analytic method has been
introduced to simulate inherent electronics noise, 
which is a driving factor in signal processing performance. 
This motivates the design decision to develop and employ cold
electronics~\cite{cold} for LArTPC applications.

With the signal and noise simulation, a quantitative assessment of the
signal processing method has been performed. For track-like topologies, 
the signal processing for the collection plane achieved
better than \SI{5}{\percent} charge resolution, 
full efficiency, and negligible charge bias for track angle
$\theta_{xz}\lesssim80^{\circ}$;
the signal processing for the induction plane achieved
better than \SI{10}{\percent} charge resolution,
full efficiency, and negligible charge bias with
$\theta_{xz}\lesssim40^{\circ}$. For the induction
planes, both the charge bias and charge resolution 
increase significantly with an increase of track angle
$\theta_{xz}$ greater than $40^{\circ}$.
The signal processing has a significant increase in
inefficiencies/failures for induction plane tracks
with $\theta_{xz}\gtrsim75^{\circ}$. 
The simulation is meant not only
for the evaluation of signal processing but also as a
fundamental tool to understand the LArTPC detector by
comparison with real data. The data-to-simulation
comparison will be demonstrated in a forthcoming paper~\cite{SP2_paper}.

Weak points of the current signal processing method are also presented.
These shortcomings are strongly dependent on signal topology.
Improvements to these deficiencies are informed by the full
chain of event reconstruction and are an active area of development.

Signal processing is the foundation of LArTPC event reconstruction
paradigms such as Pandora multi-algorithm
pattern recognition~\cite{uboone-pandora}, deep learning with convolutional neural networks~\cite{uboone-dl}, Wire-Cell tomography~\cite{wirecell}, and therefore the foundation of
physics analyses. We present a procedure to extract signal from all planes,
taking special precautions to counteract difficult circumstances
innate to the induction plane signal. This technique makes a more robust input to the subsequent 3D reconstruction. The signal processing
method outlined here brings closer the realization of the
promised capability of LArTPC detector technology.

\acknowledgments

This document was prepared by the MicroBooNE collaboration using the
resources of the Fermi National Accelerator Laboratory (Fermilab), a
U.S. Department of Energy, Office of Science, HEP User Facility.
Fermilab is managed by Fermi Research Alliance, LLC (FRA), acting
under Contract No. DE-AC02-07CH11359.  MicroBooNE is supported by the
following: the U.S. Department of Energy, Office of Science, Offices
of High Energy Physics and Nuclear Physics; the U.S. National Science
Foundation; the Swiss National Science Foundation; the Science and
Technology Facilities Council of the United Kingdom; and The Royal
Society (United Kingdom).  Additional support for the laser
calibration system and cosmic ray tagger was provided by the Albert
Einstein Center for Fundamental Physics, Bern, Switzerland.

\appendix



\bibliographystyle{JHEP}

\bibliography{Signal_Processing_part1}{}

\end{document}